\documentclass[aps,prb,showpacs,twocolumn]{revtex4-1}

\usepackage{amsmath,amssymb,stmaryrd} 

\usepackage[caption=false]{subfig} 

\usepackage[draft,marginclue,margin,author=]{fixme}
\fxsetup{theme=color,marginface=\small}

  \usepackage[pdftex,breaklinks=true]{hyperref}
  \usepackage{graphicx} 

\graphicspath{{./FigsSubmit2002/}}

\renewcommand{\vec}[1]{\boldsymbol{#1}}
\renewcommand{\Re}{\mathop{\mathrm{Re}}\nolimits} 
\renewcommand{\Im}{\mathop{\mathrm{Im}}\nolimits}

\begin{document}

\title{Quantum control of nonlinear thermoelectricity at the nanoscale}
 
\author{Nobuhiko Taniguchi}\email{taniguchi.n.gf@u.tsukuba.ac.jp}
\affiliation{Physics Division, Faculty of Pure and Applied Sciences,
  University of Tsukuba, Tennodai Tsukuba 305-8571, Japan} 
\date{\today}

\begin{abstract}
  We theoretically study how one can control and enhance nonlinear
  thermoelectricity by regulating quantum coherence in nanostructures
  such as a quantum dot system or a single-molecule junction.  In
  nanostructures, the typical temperature scale is much smaller than
  the resonance width, which largely suppresses thermoelectric
  effects. Yet we demonstrate one can achieve a reasonably good
  thermoelectric performance by regulating quantum coherence. Engaging a
  quantum-dot interferometer (a quantum dot embedded in the ring
  geometry) as a heat engine, we explore the idea of thermoelectric
  enhancement induced by the Fano resonance.  We develop an analytical
  treatment of fully nonlinear responses for a dot with or without
  strong interaction. Based on the microscopic model with the
  nonequilibrium Green function technique, we show how to enhance
  efficiency and/or output power as well as where to locate an optimal
  gate voltage.  We also argue how to assess nonlinear
  thermoelectricity by linear-response quantities.

\end{abstract}



\begin{widetext}
\maketitle
\end{widetext}













\section{Introduction}

Thermoelectricity is a phenomenon that can directly convert between
heat and electric
power~\cite{RoweBook06,Dubi11,WangBook14,ZlaticBook14,GoldsmidBook16} to make
heat engines or refrigerators possible.  Though the effect has long
been known, it has recently been attracting wider interest by its
capacity to realize thermoelectric generators or energy harvesters
that convert waste environmental heat into electric energy.  Despite a
few decades of extensive studies, materials suitable for these
applications are still scarce, and the demand for new
materials with better thermoelectricity is ever-increasing.
Thermoelectric ability is commonly characterized by 
the figure of merit 
\begin{align}
& ZT = S^{2}\, GT/K,
\label{def:ZT}
\end{align}
where $T$ is the temperature, $S$
is the thermopower (or Seebeck coefficient), $G$ is electric
conductance, and $K$ is thermal conductance. The index $ZT$
is a linear-response quantity that comes in handy for quantifying
thermoelectricity. The larger value anticipates the better
thermoelectric performance. 
The linear-response estimate of the
achievable maximal efficiency $\eta_{L}^{\max}$ regarding the Carnot
efficiency $\eta_{C}$ is given by
\begin{align}
& \frac{\eta^{\max}_{L}}{\eta_{C}} =
\frac{\sqrt{ZT+1}-1}{\sqrt{ZT+1}+1}. 
\label{eq:linear-eta}
\end{align}
In addition to electrons, phonons (or photons) may also contribute to
the thermal conductance $K$ without causing any charge
conduction. Because of it, heat conduction due to phonons or photons
is always harmful to high efficiency.
Most thermoelectric materials available today exhibit
$ZT\approx 1\sim 2$. Yet a larger value of $ZT$ is preferable for
viable thermoelectric applications~\cite{Majumdar04,Vining09,He17b}.

Since the ratio $GT/K$ empirically remains almost constant as stated
by the Wiedemann-Franz law, a common strategy to enhance $ZT$ is to
find a material with a large $S$, and nanoscale or low-dimensional
materials have been seen as a promising
candidate~\cite{Dresselhaus07}.
Typically, one places a nanostructure (a quantum dot or
a single-molecule junction) between two thermal reservoirs with
different temperatures and electrochemical potentials.  The sharp resonance
by their discrete levels provides energy filtering effects, which
makes nanostructures work as either a heat engine or a refrigerator by
exchanging particles between the reservoirs~\cite{Humphrey02}. 
Without any moving parts, one could easily scale down such solid-state
machines. They are suitable for applications where ``cost and
energy efficiency were not as important as energy availability,
reliability, predictability and the quiet operation of
equipment''~\cite{Dresselhaus07}.  With bio-compatible quantum
dots~\cite{Zhou15}, they could possibly act as an on-chip micro-power supply for
medical applications in the future.
Moreover, it has been theoretically suggested that if the DOS has an
extremely sharp peak width like the $\delta$-function, the figure of
merit $ZT$ may get huge almost
unlimitedly~\cite{Hicks93,Hicks93b,Mahan96}.
One may argue, however, that such a situation is rather unrealistic
and unphysical because the peak width of nanostructures usually
exceeds the temperature scale, largely suppressing $ZT$ much smaller
than unity.
Nevertheless, by taking advantage of great freedom in fabrication, we
expect that wisely and effectively designed nanoscale devices may
overcome those difficulties to realize much improved thermoelectric
performance. 

\begin{figure}
\subfloat[\label{fig:setting-ring}]{
  \includegraphics[width=0.45\linewidth]{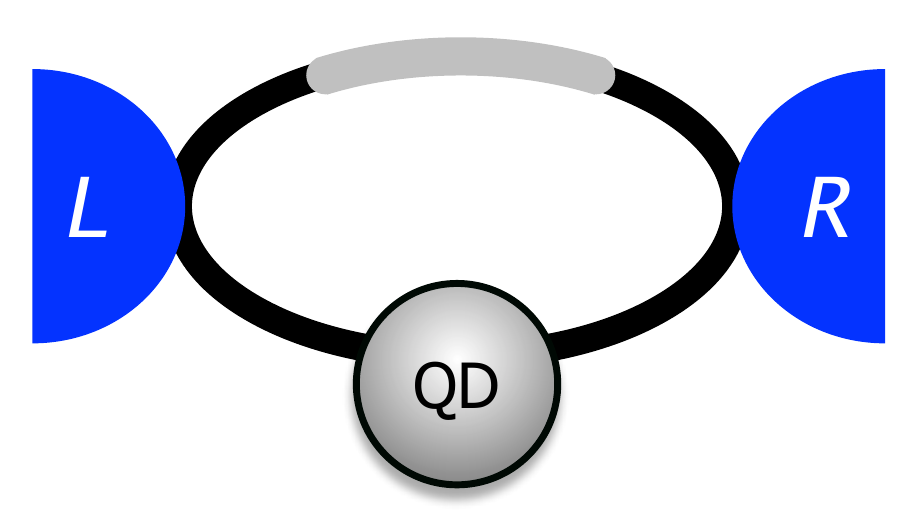}} \quad
\subfloat[\label{fig:setting-fig}]{
  \includegraphics[width=0.45\linewidth]{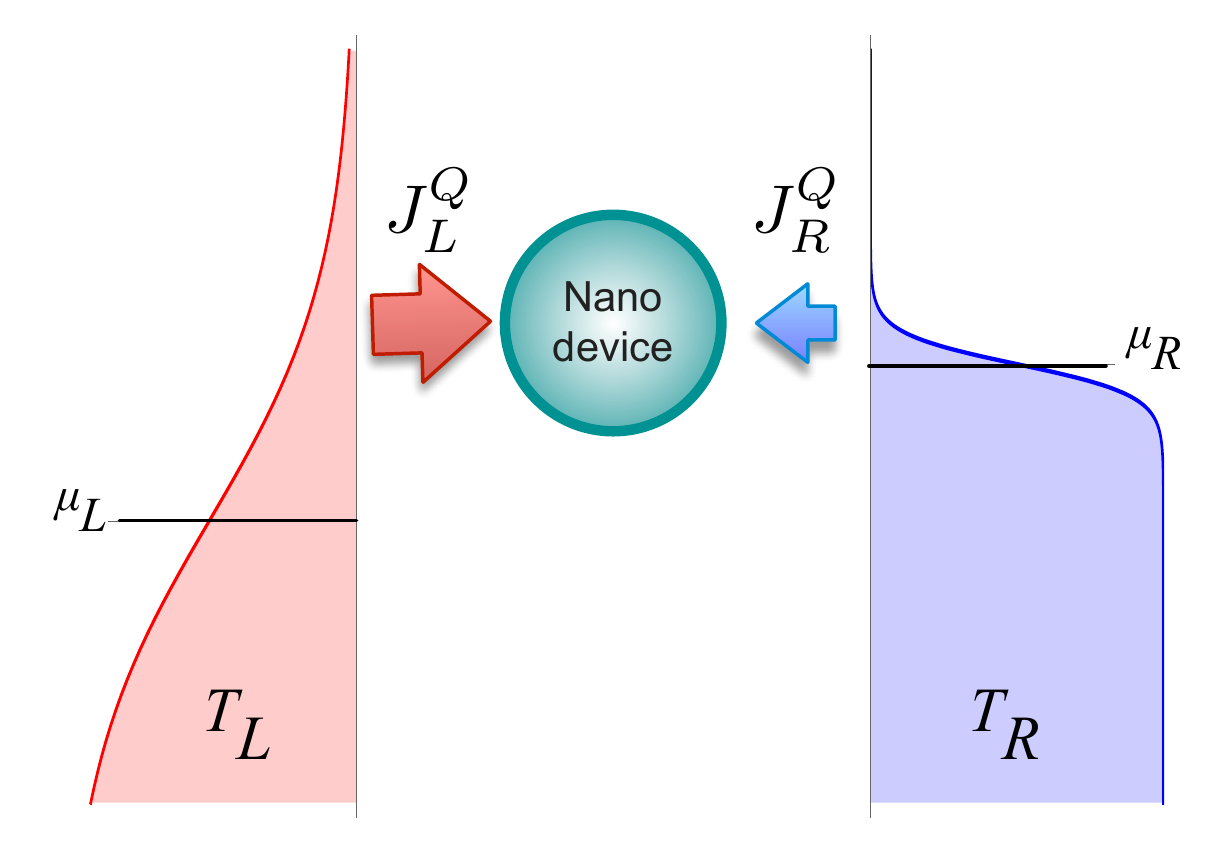}}
\caption{(a)~A schematic of a quantum-dot
  interferometer. The hopping along the direct conducting channel
  between the two reservoirs (the gray region) can be
  adjusted. (b)~The setting of electrical and thermal voltages for a
  heat engine: $T_{L}>T_{R}$ and $\mu_{L}<\mu_{R}$.}
\end{figure}

The purpose of the paper is to examine and demonstrate how one can
control thermoelectric transport through a nanostructure by regulating
quantum coherence.  Unlike intrinsic properties in bulk materials,
properties of coherent transport at the nanoscale are largely
determined by the type of a junction, which one can engineer.  In this
paper, taking a quantum dot interferometer
(Fig.~\ref{fig:setting-ring}), we explore the idea of thermoelectric
enhancement by the Fano resonance, focusing on the nonlinear transport
regime.  The idea has been pushed forward within the linear-response
theory both
theoretically~\cite{Finch09,Gomez-Silva12,Trocha12,Garcia-Suarez13,Bevilacqua16,Wojcik16,Menichetti18}
and experimentally (see Ref.~\cite{Cui17b} and references therein), as
well as regulating quantum
coherence~\cite{Karlstrom11,Lambert16} or the effect caused
by the transmission
node~\cite{Bergfield09b,Bergfield10,Abbout13,Yamamoto17}.
In nanostructures, however, thermoelectric phenomena usually occur in
the nonlinear regime, where the reliability of linear-response
estimates such as Eq.~\eqref{eq:linear-eta} remains
uncertain~\cite{Meair13,Azema14}.  To examine and demonstrate the
viability of enhanced thermoelectricity, we need to investigate
nonlinear transport.  Based on the microscopic model of a quantum-dot
interferometer, we analyze nonlinear flows using nonequilibrium Green
functions, by ignoring phonon or photon contributions to the heat
conduction.  Unlike the scattering theory approach, the method allows
us to incorporate strong correlation effect on the dot.
By developing analytical treatments, we will demonstrate how one can
control both linear and nonlinear transport for better
thermoelectricity.
To our knowledge, this is the first of showing such effect in
nonlinear transport.  Furthermore, by comparing linear and nonlinear
results, we will argue what kind of criteria is appropriate to
adjust optimal parameters, for achieving better efficiency or power in
the nonlinear regime.

The Fano resonance has been revealed in many nanostructure
systems~(see Ref.~\cite{Miroshnichenko10} and references therein).
Experimental realizations of tunable Fano resonances include
semiconductor quantum
dots~\cite{Gores00,Kobayashi02,Kobayashi03,Johnson04} or molecular
junctions~\cite{Papadopoulos06} as well as engineered graphenes or
nanoribbons~\cite{Kim03b,Kim05,Babic04b,Gong13,Briones-Torres17,Zhou18b}.
It is noteworthy that quantum coherence in some single-molecule
junctions remains not only at low temperatures but also at room
temperature
\cite{Vazquez12,Ballmann12,Guedon12,Prins11,Arroyo13,Aradhya12,Aradhya13}.
Our microscopic model of the quantum-dot interferometer
(Fig.~\ref{fig:setting-ring}) can serve as an effective description
for those systems with asymmetric resonances.
Many aspects of how the Fano resonances enhance linear-response
thermoelectricity have been theoretically investigated in
the literature~\cite{Finch09,Gomez-Silva12,Trocha12,Garcia-Suarez13,Saiz-Bretin15,Wojcik16,Bevilacqua16,Menichetti18}.
We develop an analytical treatment in both linear and
nonlinear thermoelectric responses, to provide a simple picture of how
to improve thermoelectric performance.

It is possible to manipulate quantum coherence to
achieve a better performance by amplifying either positive or negative
thermopower,  but we need a different discipline of optimization. 
As a concrete illustration, we focus on a setup for a heat
engine where temperature gradient drives the current flow
against the bias voltage (Fig.~\ref{fig:setting-fig}).
For this case, we improve thermoelectricity by amplifying negative
thermopower by tuning the bias voltage up to the stopping bias voltage. 

In addition, we are particularly concerned with the situation where the
temperature scale is much smaller than the resonant peak width of the
dot. Although this is a typical situation for semiconductor quantum
dots, it makes the system nonthermoelectric with $ZT \sim 0.1$ if it
is simply coupled with the reservoirs, disconnecting the gray line in
Fig.~\ref{fig:setting-ring} (see Fig.~\ref{fig:ZT-Beta5-xi00} below) .
We demonstrate that with a small tweaking in designing a
nanostructure, one can turn such a non-thermoelectric system into
being thermoelectric, \textit{i.e.}, by making and adjusting a direct
conducting channel between the reservoirs. This thermoelectric
enhancement occurs in both efficiency and output power and remains
effective in the nonlinear regime. 

The paper is organized as follows.  We start in
Sec.~\ref{sec:phenomenology} by giving a phenomenological discussion
of how the Fano resonance can enhance the figure of merit.
Section~\ref{sec:model-theory} introduces the microscopic model of a
quantum-dot interferometer in nonequilibrium. The correspondence
between phenomenological parameters introduced in
Sec.~\ref{sec:phenomenology} and microscopic parameters is
presented. In Sec.~\ref{sec:nonlinear-flows}, we review how to
obtain exactly Landauer-type formulas of nonlinear flows using the
nonequilibrium Green function approach, as well as how to incorporate
strong Coulomb interaction on the dot in a simple, analytical way.
Sections~\ref{sec:analytical-expressions} and \ref{sec:results}
constitute our main results: Sec.~\ref{sec:analytical-expressions}
summarizes analytical expressions of nonlinear flows of particle,
energy, and heat; Sec.~\ref{sec:results} demonstrates and discusses
how the efficiency and the power output get enhanced by the Fano
resonance in both linear and nonlinear transport for a quantum dot
with or without strong correlation.  We discuss further in
Sec.~\ref{sec:criteria} the criteria on where in the parameters we
should look for better efficiency and output power.  Finally we
conclude in Sec.~\ref{sec:summary}.  In Appendix A and B, we present
one integral formula and summarize linear-response quantities for
convenience.

\section{Phenomenology of thermoelectric enhancement}
\label{sec:phenomenology}

We start by making a pedagogical exposition of how transport affects
and possibly enhances thermoelectric performance, especially the
figure of merit $ZT$.  One can gain an invaluable insight into the Fano
effect by examining the linear response theory at low
temperature~\cite{Garcia-Suarez13}, ignoring phonon contribution and
Coulomb interaction.  In that case, the scattering matrix theory can
connect various transport quantities with the transmission spectrum
$\mathcal{T}(\varepsilon)$ at the electrochemical potential $\mu$ (see
also Appendix \ref{sec:linear-response}).  Among them, the Cutler-Mott
formula~\cite{Cutler69} gives an estimate of thermopower via the
temperature-dependent conductance $G(\mu,T)$ by
$S \approx - (\pi^{2} k_{B}^{2}T/{3e}) d \log G(\mu,T)/d\mu$.
Therefore, in the low temperature, we can directly connect the figure
of merit $ZT$ with the transmission spectrum as
\begin{align}
& ZT \approx \frac{\pi^{2}}{3} \bigg( k_{B}T\frac{d \ln
  \mathcal{T}(\mu)}{d\mu} \bigg)^{2}.
\label{eq:ZT-by-Cutler-Mott}
\end{align}
The formula shows that to enhance $ZT$, one needs to
find materials with the transmission spectrum $\mathcal{T}(\varepsilon)$ 
whose logarithmic slope is sufficiently large --- with strongly and
asymmetrically energy-dependent resonances around $\mu$.  

When a single level $E_{d}$ of the dot coupling with reservoirs
acquires finite resonance width $\Gamma$, the transmission spectrum
usually takes a Breit-Wigner form,
\begin{align}
& \mathcal{T}_{\text{BW}} (\varepsilon) 
=\frac{\alpha \, \Gamma^{2}}{ \left|\varepsilon - E_{d} +
   i \Gamma \right|^{2} }. 
\end{align}
where $\alpha$ is the asymmetric factor [see Eq.~\eqref{eq:alpha}]. 
Then, putting it into
Eq.~\eqref{eq:ZT-by-Cutler-Mott}, we find $ZT$ bound from above by
$(\pi^{2}/3) (k_{B}T/\Gamma)^{2}$ which occurs at
$\mu = E_{d} \pm \Gamma$.  Since a typical resonant width $\Gamma$ in
nanostructure is larger than the temperature scale, it is quite
hard to achieve $ZT$ of the order of unity, except for a system with an
extremely sharp peak like the Kondo resonance.

One finds the situation drastically different if
$\mathcal{T}(\varepsilon)$ has a node such as
$\mathcal{T}(\varepsilon) \propto (\varepsilon -
\epsilon_{\text{node}})^{k}$ (for
$k>0$)~\cite{Bergfield09b,Bergfield10}. Then
Eq.~\eqref{eq:ZT-by-Cutler-Mott} suggests the figure of merit diverges
at low temperature when the electrochemical potential crosses the node energy
$\epsilon_{\text{node}}$ as
$ZT \propto (k_{B}T)^{2}/(\mu - \epsilon_{\text{node}})^{2}$. Such
divergence turns out to be cut off by finite temperature effect so
that this leads to a universal mechanism of providing an
order-of-unity $ZT$, even if $\Gamma$ is much larger than $k_{B}T$.

A great advantage of nanoscale systems is that one can make such a
transmission node or \emph{nodify} the spectrum
$\mathcal{T}(\varepsilon)$ of any materials by manipulating how a
nanostructure connects with the surroundings, \textit{i.e.}, by the
Fano
effect~\cite{Finch09,Garcia-Suarez13,Bevilacqua16,Menichetti18}. One
can control the effect by changing gate voltages along the direct
conducting channel of a quantum-dot interferometer, or rotating the
side group of a single-molecule junction. This contrasts with bulk
materials whose $\mathcal{T}(\varepsilon)$ is seen as intrinsic, being
always proportional to the local DOS.
The transmission spectrum subject to the Fano effect is expressed by
the so-called Fano formula~\cite{Fano61} (see also
Ref.~\cite{Miroshnichenko10} and references therein),  
\begin{align}
& \mathcal{T}_{\text{Fano}}(\varepsilon) 
= \mathcal{T}_{0} \left| \frac{\varepsilon - E_{d} +
    q\Gamma}{\varepsilon - E_{d} + i\Gamma}  \right|^{2},
\label{eq:Fano-T}
\end{align}
where constant $\mathcal{T}_{0}$ describes the transmission of
conducting channel and the Fano parameter $q$, which may be either
complex or real, accounts for the asymmetry of the transmission
profile.  The Fano effect is an outcome of the quantum interference
between the two transport channels via discrete and continuum
levels.  
The node of $\mathcal{T}(\varepsilon)$ is located at
$E_{d}-\Gamma \Re q$, so that we expect an order-of-unity $ZT$ when we
set $\mu$ around this Fano resonance dip even for $k_{B}T \lesssim
\Gamma$.

Although the above phenomenological argument is quite useful to draw
a rough picture of how we may expect the Fano effect to improve
thermoelectric performance, we should bear in mind that the above is
based on the linear response theory, not to mention on the
low-temperature expansion.  To gauge thermoelectric performance in
nanostructures, we need to take account of two additional factors:
nonlinear transport and
interaction~\cite{Kim03,Lopez13b,Azema14,Sierra14,Whitney14}.
In the next section, we will work out the microscopic model of a
quantum-dot interferometer and show how these phenomenological
parameters constituting $\mathcal{T}_{\text{Fano}}(\varepsilon)$ can
be controlled in practice.

\section{Microscopic model}
\label{sec:model-theory}

\subsection{Microscopic Hamiltonian}

As a concrete microscopic realization of a system with the Fano
resonance effect, we consider a quantum-dot interferometer (see
Fig.~\ref{fig:setting-ring}), a quantum dot embedded in the ring
geometry and coupling with two reservoirs (the left and right
reservoirs $a = L, R$) with different electrochemical potentials $\mu_{a}$
and temperatures $T_{a}$.  To operate it as a heat engine, we
arrange $T_{L} > T_{R}$ and $\mu_{L} < \mu_{R}$ to make the
temperature voltage drive the heat flow
against the potential bias (Fig.~\ref{fig:setting-fig}).
We assume a single spin-degenerate discrete level of the dot
predominantly contributes to transport. With interaction on the dot,
the model is essentially the single-impurity Anderson model augmented
by the direct hopping between the
reservoirs~\cite{Hofstetter01,Kim03}.  A similar model has also been
studied in examining the Rashba spin-orbit interaction
effect~\cite{Sun05b,Crisan09,Taniguchi12}.

The total Hamiltonian is
$H = H_{D} + H_{T} + H_{A} + \sum_{a=L,R} H_{a}$, where $H_{D}$
represents the dot Hamiltonian; $H_{T}$, the hopping between the dot
and the leads; and $H_{A}$, the direct hopping between the left and
right leads.  They are given by
\begin{align}
& H_{D} = \sum_{\sigma} \epsilon_{d}\, \hat{n}_{\sigma} + U \hat{n}_{\uparrow}
\hat{n}_{\downarrow},\\ 
& H_{T} =  \sum_{a=L,R} \sum_{\vec{k},\sigma}\left(
    V_{d\sigma a}\, d^{\dagger}_{\sigma}  c_{a\vec{k}\sigma} + V_{a
      d\sigma}\, c^{\dagger}_{a\vec{k}\sigma} d_{\sigma} \right),\\
&   H_{A} = \sum_{\vec{k},\sigma} \left( V_{LR}\, c^{\dagger}_{L\vec{k}\sigma}
    c_{R\vec{k}\sigma} + V_{RL}\, c^{\dagger}_{R\vec{k}\sigma}
    c_{L\vec{k}\sigma} \right),
\end{align}
where $\hat{n}_{\sigma} = d^{\dagger}_{\sigma} d_{\sigma}$ is the dot
electron number operator.  The Hamiltonian $H_{a}$ describes
noninteracting electrons on the lead $a=L,R$, which can be characterized by
the DOS $\rho_{a}$ in the wide-band approximation. 
One may incorporate the Aharonov-Bohm effect by introducing the phase
factor $\phi$ of
$V_{Rd\sigma} V_{d\sigma L} V_{LR} = \big| V_{Rd} V_{dL} V_{LR} \big|
\, e^{i\phi}$.  We will present all of our analytical results including
this effect, but we will choose $\phi=0$ for numerical
presentation.

\subsection{Connection with the Fano formula}

To effectively regulate quantum coherence via the Fano resonance
effect, we need to identify phenomenological parameters introduced in
Sec.~\ref{sec:phenomenology} in terms of microscopic parameters of the
Hamiltonian~\cite{Taniguchi12}.  We here summarize those connections
without waiting for details that we will present in the
next section.

Our convention of the relaxation rates $\gamma_{a}$ due to the
leads $a=L,R$ is
\begin{align}
& \gamma = \gamma_{L} + \gamma_{R}; \qquad \gamma_{\ell} = \pi
|V_{da}|^{2} \rho_{a}, 
\label{eq:gamma} \\ 
& \alpha = 4 \gamma_{L} \gamma_{R}/\gamma^{2}, 
\label{eq:alpha}
\end{align}
where $\alpha$ is the asymmetric factor regarding the dot-lead
couplings. 
The most important parameter we utilize to control quantum coherence
is the dimensionless parameter $x$, defined by
\begin{align}
& x = 4\pi^{2} \rho_{L} \rho_{R} |V_{RL}|^{2}.
\label{eq:x}
\end{align}
The parameter $x$ describes how much the direct conducting channel
contribute to transport.
In the absence of the electron correlation on the dot, we can exactly
evaluate transmission spectra $\mathcal{T}(\varepsilon)$ (see
Sec.~\ref{sec:GR-without-int} below), which confirms the Fano
formula~\eqref{eq:Fano-T}.
This enables us to identify phenomenological parameters in
Eq.~\eqref{eq:Fano-T} as
\begin{align}
& q = \sqrt{\frac{\alpha}{4x}} \left( e^{i\phi} - x
  e^{-i\phi} \right), 
\label{eq:Fano-q} \\
& E_{d} = \epsilon_{d} - \Gamma \sqrt{\alpha x} \cos\phi,
\label{eq:Ed} \\
& \Gamma = \gamma/(1+x), 
\label{eq:Gamma} \\
& \mathcal{T}_{0} = 4x/(1+x)^{2}.
\label{eq:T0}
\end{align}
The limit $x \to 0$ corresponds to a Breit-Wigner resonance where
$E_{d} \to \epsilon_{d}$ and $\Gamma \to \gamma$.  Another interesting
limit is when the dot-lead couplings are extremely asymmetric.  In
this case, transmission $\mathcal{T}(\varepsilon)$ becomes an
anti-resonance, which means the dot is side-coupled to the
conducting channel.

In the presence of strong interaction on the dot, the role of the Fano
parameter as characterizing the asymmetric transmission profile become
obscure, because $\mathcal{T}(\varepsilon)$ gets deformed also by the
interaction. Accordingly, we make a point of using
Eqs.~\eqref{eq:Fano-q}--\eqref{eq:T0} as the definitions of our
controlling parameters (see also the argument in
Sec.~\ref{sec:GR}).

\section{Nonlinear flows of particle, energy and heat}
\label{sec:nonlinear-flows}

\subsection{Current formulas}

One can evaluate nonlinear flows of the particle, the energy, and the
heat (denoted by $I$, $J^{E}$, and $J^{Q}$), using nonequilibrium
Green functions techniques along the standard line of
treatment~\cite{Hofstetter01,Kim03,Meir92,HaugBook08}.
Expressions of these flows usually involve the lesser Green function
of the dot as well as the retarded one.  However, in the case where
the dot-reservoir couplings are proportional to each other, one can
safely eliminate the lesser Green function by using the conservation
laws of particle and energy.~\cite{Meir92}.  Accordingly, we can
express those flows only by using the retarded Green function even if
the strong interaction is present on the dot.
Choosing the left reservoir as the reference, we can write those
nonlinear inflows (per spin) to the dot as
\begin{align}
& I_{L} = \int \frac{d\varepsilon}{h}
\mathcal{T} (\varepsilon) \left[ f_{L}(\varepsilon) -
  f_{R}(\varepsilon) \right], 
\label{eq:I-by-LB}
\\
& J_{L}^{E} = \int \frac{d\varepsilon}{2h}
\mathcal{T} (\varepsilon) \varepsilon \left[ f_{L}(\varepsilon) -
  f_{R}(\varepsilon) \right], 
\label{eq:JE-by-LB}\\ 
& J_{L}^{Q} =  \int \frac{d\varepsilon}{h}
\mathcal{T} (\varepsilon) (\varepsilon - \mu_{L} )
\left[ f_{L}(\varepsilon) - f_{R}(\varepsilon) \right],
\label{eq:JQ-by-LB}
\end{align}
where $h$ is the Planck constant and
$f_{a}(\varepsilon) = [e^{(\varepsilon-\mu_{a})/k_{B}T_{a}}+1]^{-1}$
is the Fermi distribution on the lead $a$ with $T_{a}$ and $\mu_{a}$.
They take exactly the same forms as the Landauer-B\"{u}ttiker formulas
by using the effective transmission function
$\mathcal{T}(\varepsilon)$, which is defined in terms of the exact
retarded Green function $G^{R}(\varepsilon)$.  For the present case of
a quantum-dot interferometer, we find $\mathcal{T}(\varepsilon)$
as~\cite{Taniguchi12}
\begin{align}
& \mathcal{T} (\varepsilon) = \mathcal{T}_{0}
- \Im \left[ \mathcal{T}_{q}\, G^{R} (\varepsilon) \Gamma \right],
\label{eq:T-by-G}\\
& \mathcal{T}_{q} = \mathcal{T}_{0} \, (q-i) (q^{*}-i).
\label{eq:Tq}
\end{align}

We emphasize that while one usually derives the Landauer-B\"{u}ttiker
formula assuming the one-particle scattering
theory~\cite{Sivan86,Butcher90,Sanchez16,Benenti17}, the above
Landauer-like description of nonlinear flows is exact, whether with or
without the interaction on the dot.  All the correlation effect is
encoded in $\mathcal{T}(\varepsilon)$, and its validity goes beyond
the one-particle approximation.

\subsection{Efficiency and output power}

To assess the nonlinear thermoelectric performance as a heat engine,
we mainly use two benchmarks: the output power $\mathcal{P}$ and the
thermal efficiency $\eta$.  Because of our configuration
$\Delta \mu =\mu_{R}-\mu_{L}>0$ and $\Delta T=T_{L} - T_{R}>0$, the output power $\mathcal{P}$
and the efficiency $\eta$ are defined by
\begin{align}
& \mathcal{P} = \sum_{a} J_{a}^{Q} 
= (\mu_{R} - \mu_{L}) I_{L}, 
\label{def:P}\\
& \eta = \frac{\mathcal{P}}{J^{Q}_{L}} = \frac{(\mu_{R}-\mu_{L})
  I_{L}}{J_{L}^{Q}}.
\label{def:eta}
\end{align}
The system works as a heat engine for a positive output power
$\mathcal{P}>0$ with a positive heat inflow from the left reservoir
$J_{L}^{Q}>0$.

Since the nonlinear flows are expressed by the Landauer-like
formulas~\eqref{eq:I-by-LB}--\eqref{eq:JQ-by-LB}, its nonlinear
transport is fully consistent with thermodynamics, namely, the
positive entropy flow (the second law of thermodynamics).  This means
the efficiency $\eta$ is bound from above by the Carnot efficiency
$\eta_{C}= \Delta T/T_{L}$.  Moreover, quantum mechanics also bounds
the power output $\mathcal{P}$ from above~\cite{Whitney14}:
$\mathcal{P}$ should be smaller than
$A_{0} (\pi k_{B} \Delta T)^{2}/h$ (with $A_{0} \approx 0.0321$) for
the two-terminal single-level dot.
Later in Sec.~\ref{sec:results}, we make a point of normalizing
the output power $\mathcal{P}$ by $\mathcal{P}_{\Delta T} = k_{B}^{2}
(T_{L} - T_{R})^{2}/4h$.  In this 
unit, this quantum upper bound corresponds to $4\pi^{2} A_{0} \approx 1.267$.

\subsection{Evaluating the retarded Green function}
\label{sec:GR}

To find nonlinear flows of a quantum-dot interferometer according to
Eqs.~\eqref{eq:I-by-LB}--\eqref{eq:JQ-by-LB}, we need the dot's
retarded Green function $G^{R}(\varepsilon)$ out of equilibrium,
connecting with the leads with different temperatures and electrochemical
potentials.  While we can obtain it exactly for a noninteracting dot,
we can no longer do so if the dot involves strong interaction.  For
the latter case, we will make a simple yet effective analytical
approximation suitable to describe charge-blocking physics that the
strong correlation induces.

\subsubsection{Noninteracting dot}
\label{sec:GR-without-int}

For a noninteracting dot, one finds exactly the retarded
Green function~\cite{Hofstetter01} to be
\begin{align}
& G^{R}(\varepsilon) = \frac{1}{\varepsilon - E_{d} + i\Gamma}.
\end{align}
where $E_{d}$ and $\Gamma$ were given in
Eqs.~\eqref{eq:Ed} and \eqref{eq:Gamma} (see also derivations by the
diagram approach~\cite{Kim03}, the equation of motion~\cite{Crisan09}
or the Kelshy path integral~\cite{Taniguchi12}).  Indeed, such
connections were established by using the above into
Eq.~\eqref{eq:T-by-G}. The effective transmission
$\mathcal{T}(\varepsilon)$ becomes the Fano formula,
\begin{align}
& \mathcal{T}(\varepsilon) = \mathcal{T}_{\text{Fano}}(\varepsilon),
\label{eq:effective-T-without-int}
\end{align}
with parameters given in Eqs.\eqref{eq:Fano-q}--\eqref{eq:T0}.

\subsubsection{Strong correlation on the dot and Coulomb blockade}
\label{sec:GR-with-int}

It has been known that the strong correlation out of equilibrium is
quite hard to treat systematically.  One-particle approximations such
as the Hartree-Fock theory are valid only for weak interaction,
failing to explain strong correlation effects. Moreover, it is
somewhat embarrassing to find that a nonequilibrium perturbation
calculation regarding the interaction sometimes gives results that
disrespect fundamental laws such as the current
conservation~\cite{Hershfield92,Taniguchi14}. To make a sensible
assessment of the efficiency, it is crucial to abide by the
conservation laws.
Below, we will use a simple yet effective analytical approximation
that conforms to the conservation of the particle and the energy as
well as the spectral sum rule of the dot spectral function
$-\pi^{-1}\Im \int d\varepsilon\, G^{R}_{\sigma}(\varepsilon) = 1$.

We focus on the strongly correlated case where the interaction $U$ is
much larger than the resonant peak width or temperature.  This is a
typical situation of a quantum dot where charge blocking physics (the
Coulomb blockade effect) dominates.  
Due to the strong Coulomb repulsion on the dot, the dot energy
increases by the presence of another electron and depends on its
occupation. Therefore, we may well view its energy as
$\epsilon_{d} + U \hat{n}_{\bar{\sigma}}$ (with
$\bar{\sigma}=-\sigma$).  When we ignore dynamical fluctuations of the
dot number operator, this leads to the following approximation of the
retarded Green function~\cite{Taniguchi17}:
\begin{align}
& G_{\sigma}^{R}(\varepsilon) \approx 
\left\langle \frac{1}{\varepsilon -
    E_{d} - U \hat{n}_{\bar{\sigma}} + i\Gamma} \right\rangle 
 \\ & \quad
 = \frac{1-\langle \hat{n}_{\bar{\sigma}} \rangle}{\varepsilon -E_{d} +
  i\Gamma}
+ \frac{\langle \hat{n}_{\bar{\sigma}} \rangle}{\varepsilon -E_{d} -U +
  i\Gamma}. 
\label{eq:GR-int}
\end{align}
The treatment corresponds to the Hubbard I approximation, which one
can also derive within the equation-of-motion method by decoupling
higher-order correlators~\cite{Hubbard63,Hewson66}. Unlike the
treatment of Ref.~\cite{Meir91}, it ignores the correlation effect on
the resonant width and the Kondo correlation that become prominent at
extremely low temperature (below the Kondo temperature).  The
approximation has been shown to capture quite well the essence of
correlation effects in nonlinear responses above the Kondo
temperature, and it was recently used to successfully explain strongly
nonlinear thermal voltage observed in interacting quantum
dots~\cite{Sierra14}.

To complete the approximation, we still have to determine the average
occupation $\langle \hat{n}_{\bar{\sigma}}\rangle$. This is done by
requiring self-consistently the particle conservation out of
equilibrium, which we can solve analytically.  For a quantum-dot
interferometer, one can write the
particle conservation as~\cite{Taniguchi12,Taniguchi14}
\begin{align}
& \langle \hat{n}_{\sigma} \rangle = -\frac{1}{\pi} \int d\varepsilon\,
\bar{f} (\varepsilon) \Im G_{\sigma}^{R}(\varepsilon),
\label{eq:nd-by-GR}
\end{align}
where $\bar{f}$ is the weighted Fermi distribution defined by 
\begin{align}
& \bar{f} (\varepsilon) = \sum_{a=L,R} \frac{\Gamma'_{a}}{\Gamma}
f_{a} (\varepsilon), \\
& \Gamma'_{a}
= \frac{\gamma_{a} + x \gamma_{\bar{a}} +
  2\sqrt{x\gamma_{a}\gamma_{\bar{a}}}
  \sin\phi_{a}}{(1+x)^{2}},
\end{align}
with the convention $\phi_{R} = -\phi_{L} = \phi$ and $\bar{L}=R$ etc.
Since our problem is spin-independent, by putting
Eq.~\eqref{eq:GR-int}, we can readily find the
solution $n_{d} = \langle \hat{n}_{\uparrow} + \hat{n}_{\downarrow}
\rangle$ of the self-consistent 
Eq.~\eqref{eq:nd-by-GR}. (One can easily extend the treatment to
the spin-dependent case as well.)  We prefer organizing the solution as
\begin{align}
& n_{d} = \frac{2 n_{0}(E_{d})}{1 + n_{0}(E_{d}) - n_{0}(E_{d} + U)},
\label{eq:nd}
\end{align}
where $n_{0}(E_{d})$ is the average occupation of a noninteracting dot
per spin as a function of $E_{d}$. Using
Eq.~\eqref{eq:integral-formula}, we can obtain its explicit form as
\begin{align}
& n_{0} (E_{d}) = \sum_{a} \frac{\Gamma'_{a}}{\Gamma}
\left[ \frac{1}{2} - \frac{1}{\pi} \Im
\psi(\tfrac{1}{2}+z_{a}) \right],
\end{align}  
where $\psi(\tfrac{1}{2}+z_{a})$ is the digamma function with the argument
$z_{a} = z(\beta_{a}, \mu_{a}-E_{d})$ defined by 
\begin{align}
& z(\beta, \zeta) = \frac{\beta}{2\pi i} (\zeta + i\Gamma); \quad \zeta
= \mu - E_{d}.
\label{def:z-by-beta-mu}
\end{align}
Combining all the above enables us to obtain a closed analytical
approximation for a quantum-dot interferometer with strong
interaction.

Let us briefly discuss the immediate consequence of this approximation. Using
$n_{d}$ of Eq.~\eqref{eq:nd} in Eq.~\eqref{eq:T-by-G}, we find that
the effective transmission $\mathcal{T}(\varepsilon)$ in the Coulomb
blockade regime becomes essentially a superposition of the two Fano
resonances $\mathcal{T}_{\text{Fano}}$ of Eq.~\eqref{eq:Fano-T},
around $E_{d}$ (with weight $1-n_{d}/2$) and $E_{d}+U$ (with weight
$n_{d}/2$): 
\begin{align}
& \mathcal{T}  (\varepsilon) = \left( 1 - \frac{n_{d}}{2} \right)
\mathcal{T}_{\text{Fano}}(\varepsilon) + \frac{n_{d}}{2}\, 
\mathcal{T}_{\text{Fano}} (\varepsilon - U).
\label{eq:effective-T-with-int}
\end{align}
One needs to choose phenomenological parameters according to
Eqs.~\eqref{eq:gamma}--\eqref{eq:Fano-q}.  The form indicates
that the effective transmission $\mathcal{T}(\varepsilon)$ depends on
temperatures and electrochemical potentials of the leads through $n_{d}$.
Numerically speaking, however, the dependence of $n_{d}$ on the thermal
and electrical bias, $k_{B} \Delta T$ and $\Delta \mu$, is often
negligible when the system is set up for a heat engine.  This is
because to make it work as a heat engine, the bias $\Delta \mu$ must
be of the same order of $k_{B} \Delta T$ and usually much smaller
than $\Gamma$.  The major role of strong interaction in a
quantum-dot interferometer is to make the effective transmission split
into two Fano resonances with reduced weights.

\section{Analytical expressions of nonlinear flows}
\label{sec:analytical-expressions}

Having obtained the effective transmission
$\mathcal{T}(\varepsilon)$ for either a noninteracting or interacting
dot [as in Eq.~\eqref{eq:effective-T-without-int} or
Eq.~\eqref{eq:effective-T-with-int}], we
are now in a position to write down nonlinear flows defined by
Eqs.~\eqref{eq:I-by-LB}--\eqref{eq:JQ-by-LB}. We can reach their
explicit forms by completing the energy integrals 
by using the formula~\eqref{eq:integral-formula} in
Appendix \ref{sec:integral}.

\subsection{Noninteracting quantum dot}

By applying Eq.~\eqref{eq:integral-formula} to
Eqs.~\eqref{eq:I-by-LB}--\eqref{eq:JQ-by-LB}, it is straightforward to
obtain analytical formulas for $I_{L}$, $J^{E}_{L}$ and $J^{Q}_{L}$.
With $\zeta_{a} = \mu_{a} - E_{d}$, we organize the results as
\begin{align}
& I_{L} = \mathcal{I} (\beta_{L}, \zeta_{L}) - 
\mathcal{I}(\beta_{R}, \zeta_{R}), \\
& J_{L}^{E} =E_{d} I_{L} +  \mathcal{J}(\beta_{L}, \zeta_{L}) - 
\mathcal{J}(\beta_{R}, \zeta_{R}),
\end{align}
and heat flow is $J_{L}^{Q} = J_{L}^{E} - \mu_{L} I_{L}$. 
Functions $\mathcal{I}(\beta,\mu)$ and $\mathcal{J}(\beta,\mu)$
describe contributions from each leads, and involve 
Euler's digamma function:
\begin{align}
& h\, \mathcal{I} (\beta,\zeta)
= \mathcal{T}_{0} \zeta - \Gamma \Im \left[ \mathcal{T}_{q}
  \left( \psi(\tfrac{1}{2}+z) - \log \beta \Gamma  \right) \right], \\
& h\, \mathcal{J} (\beta, \zeta) 
= \mathcal{T}_{0} \Big( \frac{\zeta^{2}}{2} +
  \frac{\pi^{2}}{6\beta^{2}} \Big)
- \Gamma \Im \Big[ \mathcal{T}_{q} \zeta
\notag \\ & \qquad \qquad \qquad
- i\Gamma \mathcal{T}_{q}
  \left\{  \psi(\tfrac{1}{2}+z) - 
    \log \beta\Gamma \right\}  \Big],
\end{align}
with $z=z(\beta, \zeta)$ in Eq.~\eqref{def:z-by-beta-mu}.
When one disconnect the direct hopping between the reservoirs
($x \to 0$), they reduce (with $E_{d} \to \epsilon_{d}$ and $\Gamma
\to \gamma$) to 
\begin{align}
& h\, \mathcal{I} (\beta,\zeta) 
= - \alpha \Gamma \Im \left[ \psi(\tfrac{1}{2}+z) \right], \\ 
& h\, \mathcal{J}(\beta,\zeta) 
=  \alpha \Gamma^{2} \left[\Re  \psi(\tfrac{1}{2}+z) - \ln \beta
  \gamma \right].
\end{align}


\subsection{Quantum dot with interaction}

In the presence of strong correlation on the dot, 
Eq.~\eqref{eq:effective-T-with-int} tells us that  the effective transmission
$\mathcal{T}(\varepsilon)$ becomes a superposition of the two Fano
resonances, around $E_{d}$ and $E_{d}+U$. Hence nonlinear flows are
also expressed by a superposition of these two contributions.
\begin{align}
& I_{L} = \left( 1-\frac{n_{d}}{2} \right) \left[ \mathcal{I}(\beta_{L},
\zeta_{L}) - \mathcal{I}(\beta_{R}, \zeta_{R}) \right]
\notag \\ & \qquad \quad
+ \frac{n_{d}}{2} \left[ \mathcal{I}(\beta_{L},
\zeta_{L}-U) - 
\mathcal{I}(\beta_{R}, \zeta_{R}-U) \right], 
\label{eq:particle-flow-with-int} \\
& J_{L}^{E} = E_{d} I_{L}
+  \left( 1-\frac{n_{d}}{2} \right) \left[ \mathcal{J}(\beta_{L},
\zeta_{L}) - \mathcal{J}(\beta_{R}, \zeta_{R}) \right]
\notag \\ & \qquad \quad
+ \frac{n_{d}}{2} \left[ \mathcal{J}(\beta_{L},
\zeta_{L}-U) - \mathcal{J}(\beta_{R}, \zeta_{R}-U) \right].
\label{eq:energy-flow-with-int}
\end{align}
Using these $I_{L}$ and $J_{L}^{E}$, the heat current
becomes $J_{L}^{Q} = J_{L}^{E} - \mu_{L} I_{L}$.

\section{Numerical results and discussion}
\label{sec:results}

In this section, we present numerical results to support how one can
greatly improve nonlinear thermoelectric performance by regulating
quantum coherence via the Fano resonance effect.  Based on the
analytical results in Sec.~\ref{sec:model-theory}, we now make a fully
nonlinear analysis, focusing on the thermal efficiency and the output
power as benchmarks. For a better heat engine, we will intend to
amplify a negative thermopower, which makes us choose $q >0$ and
$\phi=0$ [see Eq.~\eqref{eq:Fano-q}].  If we aim to enhance a
positive thermopower, we have to choose $q<0$ instead.

We choose to fix the temperature and electrochemical potential of
the left reservoir as a reference while changing those of the right
reservoir, assuming the symmetric dot-reservoir couplings $\alpha=1$.
In addition, for most calculations, we take the temperature much smaller
than the resonant width (setting $k_{B}T_{L} = 0.2 \gamma$).  Such a
situation common in nanostructures is certainly unfavorable to achieve
high thermoelectricity ($ZT \sim 0.1$ without the Fano effect as in
Fig.~\ref{fig:ZT-Beta5-xi00} below). Nevertheless, we will demonstrate
that we can improve the thermoelectric performance 10 times as much,
by adjusting the Fano resonance effect.

We deliberately present all the results in a way that one can easily
compare between linear and nonlinear responses.  For convenience, we
summarize the explicit forms of linear-response quantities as well as
the Onsager coefficients in Appendix~\ref{sec:linear-response}.

\subsection{Noninteracting quantum dot}


\begin{figure}
  \centering
\subfloat[\label{fig:ZT-Beta5-xi00}]{
  \includegraphics[width=0.45\linewidth]{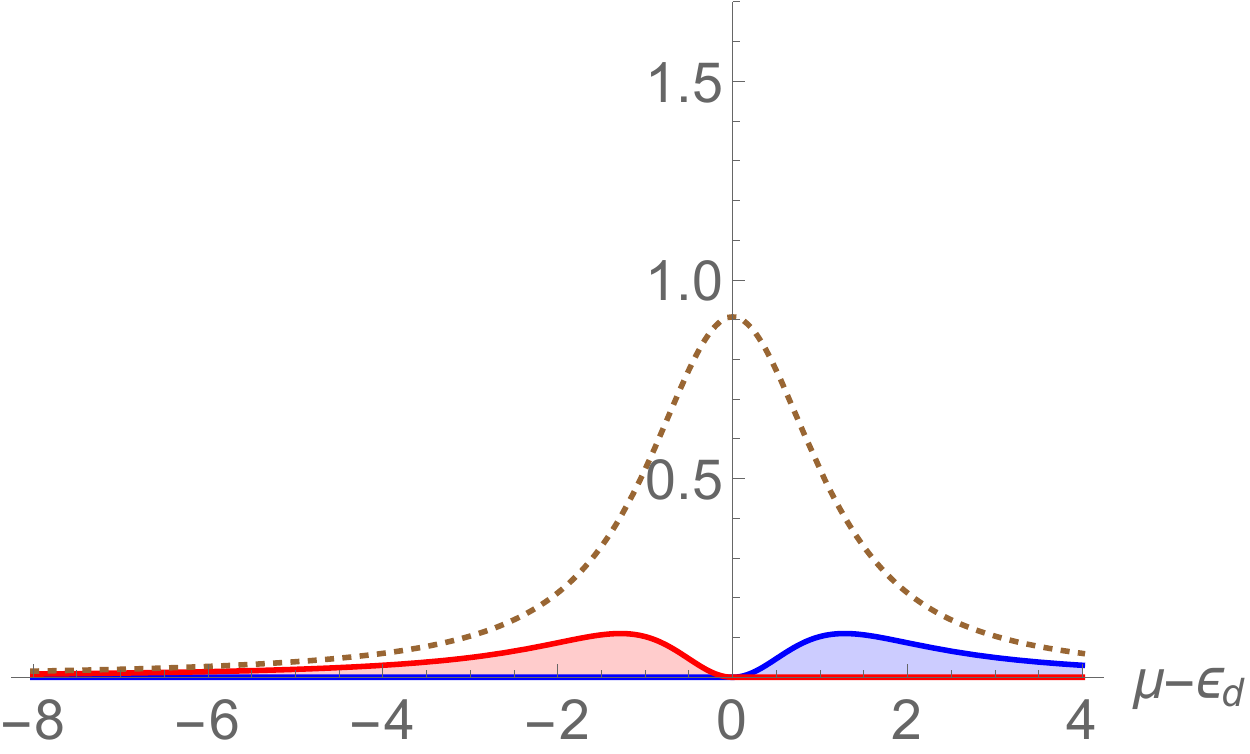}} \quad
\subfloat[\label{fig:ZT-Beta5-xi01}]{
  \includegraphics[width=0.45\linewidth]{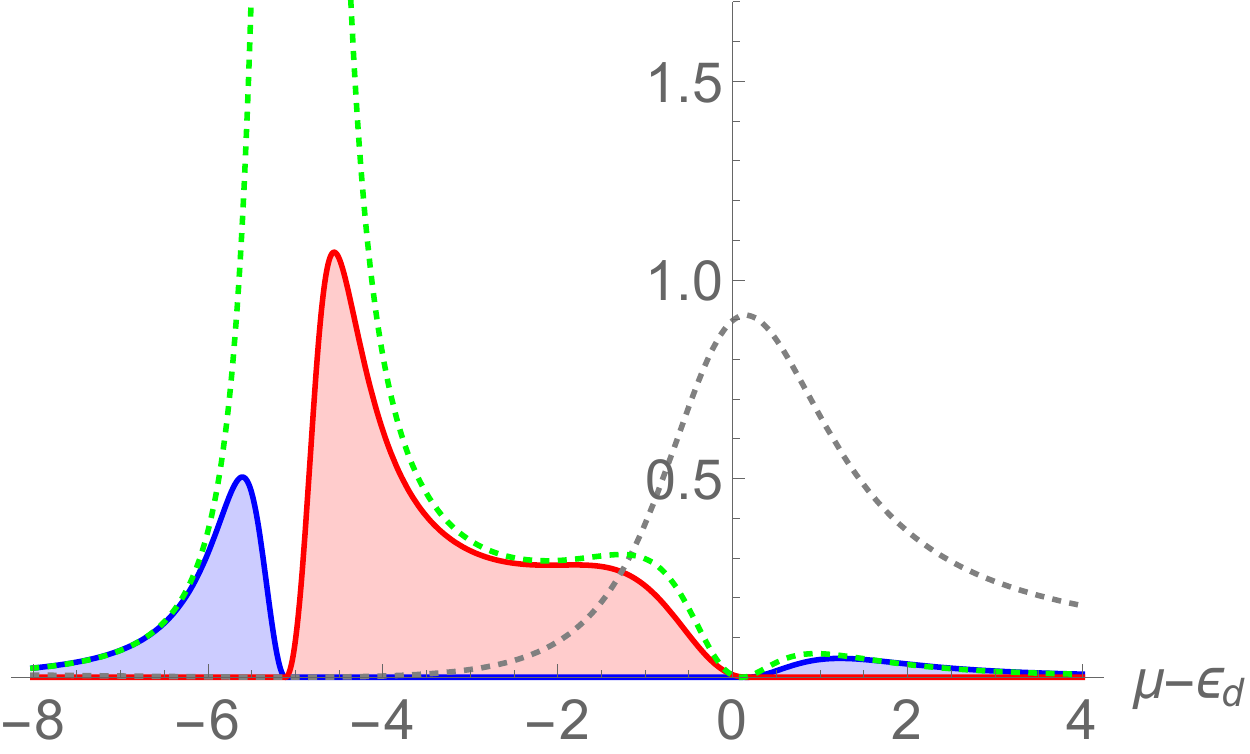}} \\
\subfloat[\label{fig:ZT-Beta5-xi10}]{
  \includegraphics[width=0.45\linewidth]{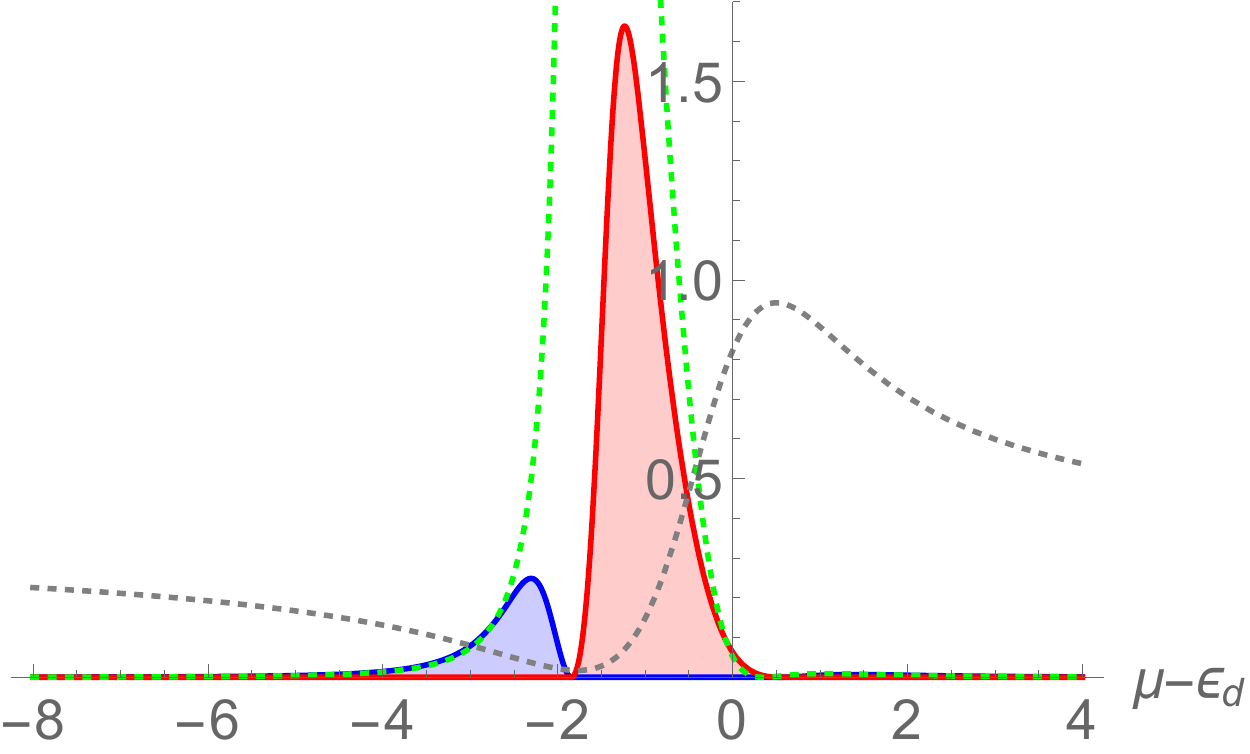}} \quad
\subfloat[\label{fig:ZT-Beta5-xi30}]{
  \includegraphics[width=0.45\linewidth]{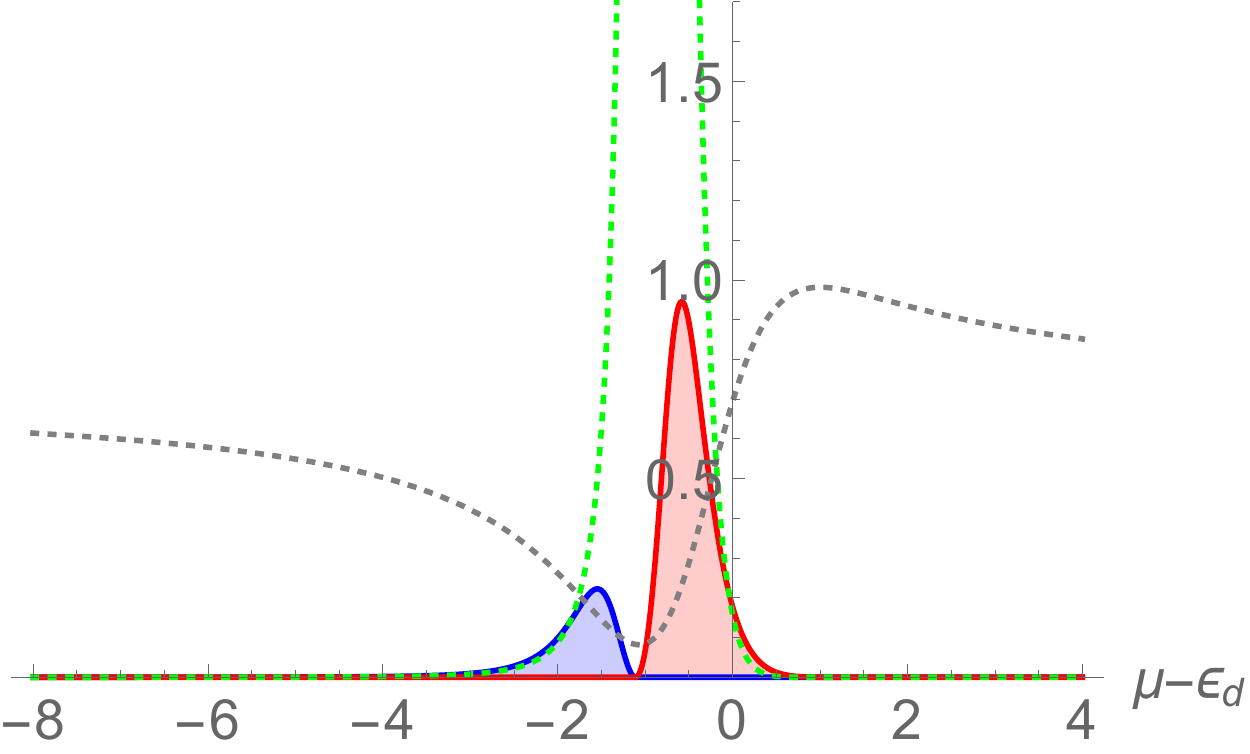}}
\caption{The figure of merit $ZT$ at $k_{B}T = 0.2\gamma$, as a
  function of $\mu - \epsilon_{d}$ by changing (a) $x=0$ ($q=\infty$),
  (b) $x=0.01$ ($q=4.95$), (c) $x=0.1$ ($q=1.42$) and (d) $x=0.3$ ($q=0.64$).
  Red and blue indicate corresponding thermopower is positive
  or negative.  Green lines show $ZT$ at the zero-temperature limit.
  Dotted lines correspond to the normalized conductance. }
\label{fig:ZT}
\end{figure}

\subsubsection{Linear responses}

\begin{figure}
  \centering 
\subfloat[\label{fig:ZT-Beta5}]{
  \includegraphics[width=0.45\linewidth]{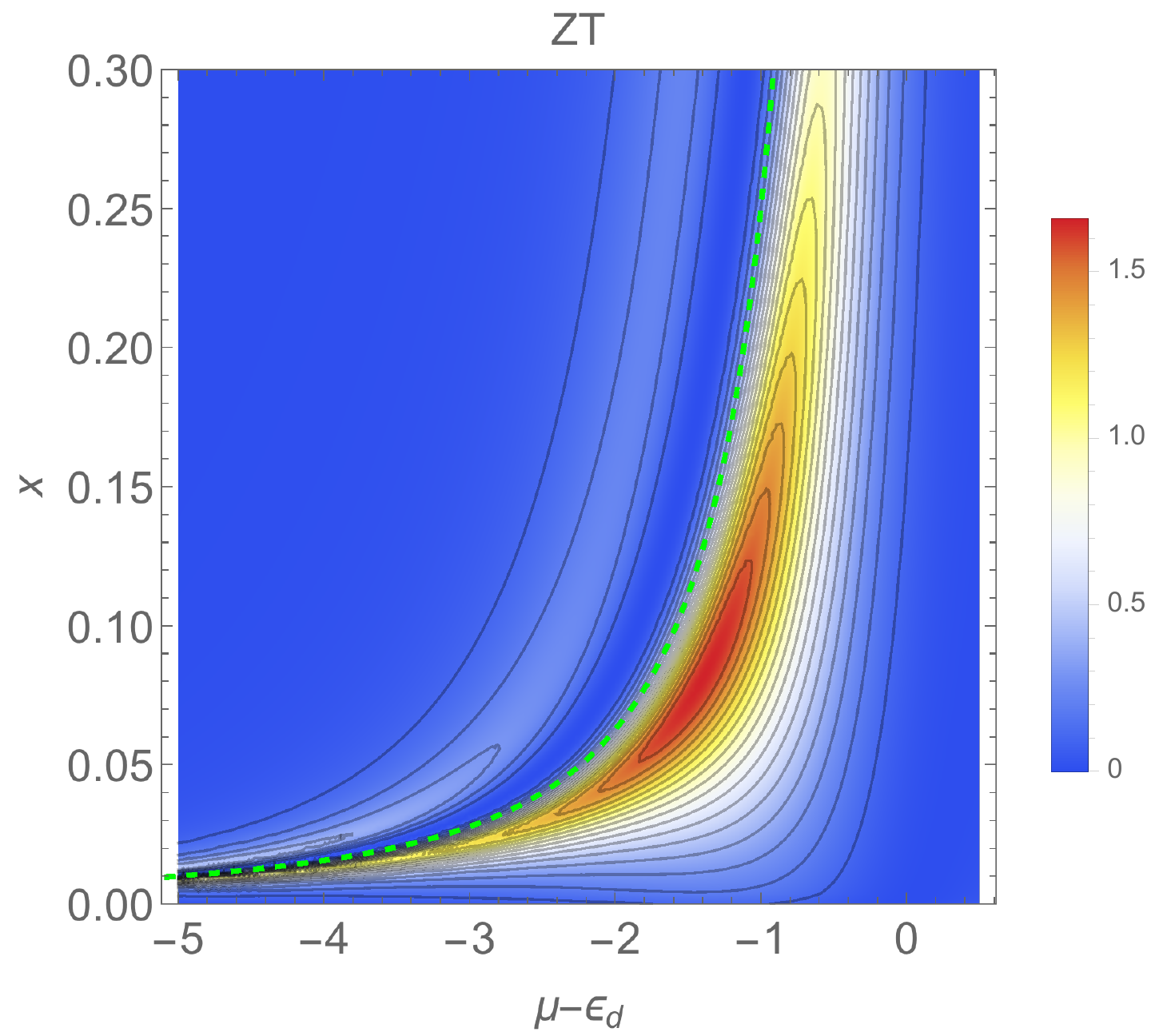}} \quad
\subfloat[\label{fig:ZT-Beta1}]{
  \includegraphics[width=0.45\linewidth]{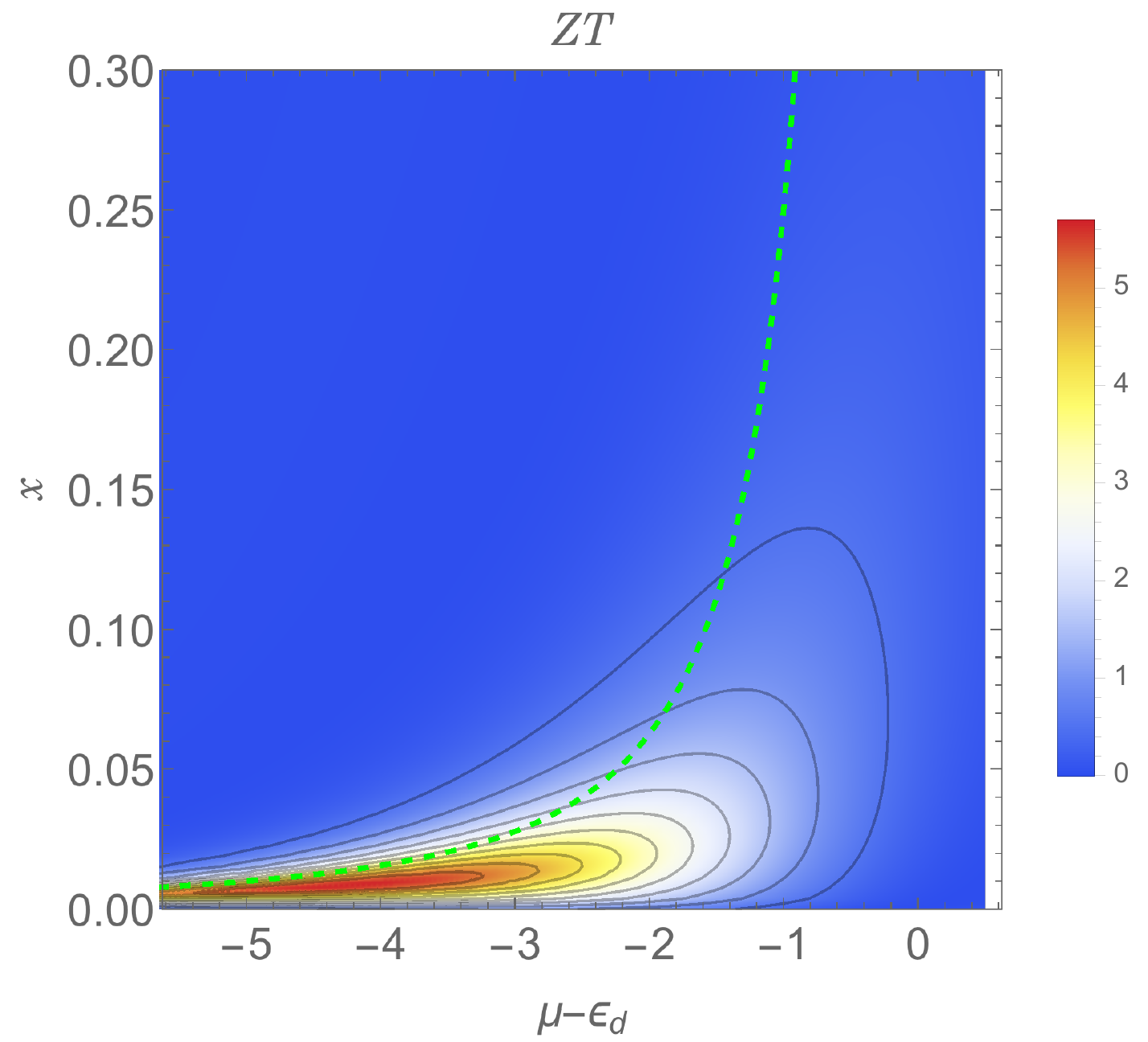}} \\
\subfloat[\label{fig:Pmax-Beta5}]{
  \includegraphics[width=0.45\linewidth]{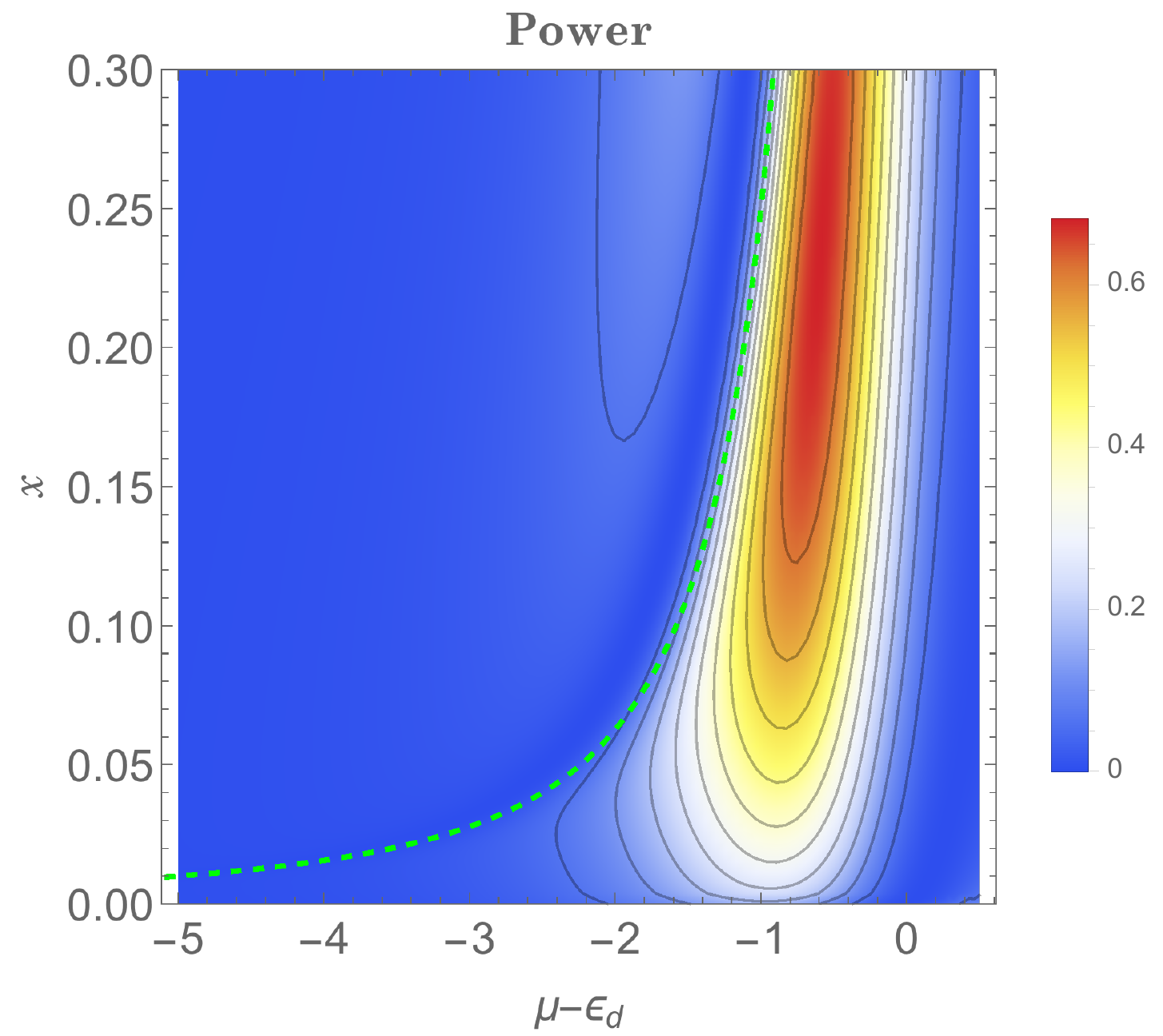}} \quad
\subfloat[\label{fig:Pmax-Beta1}]{
  \includegraphics[width=0.45\linewidth]{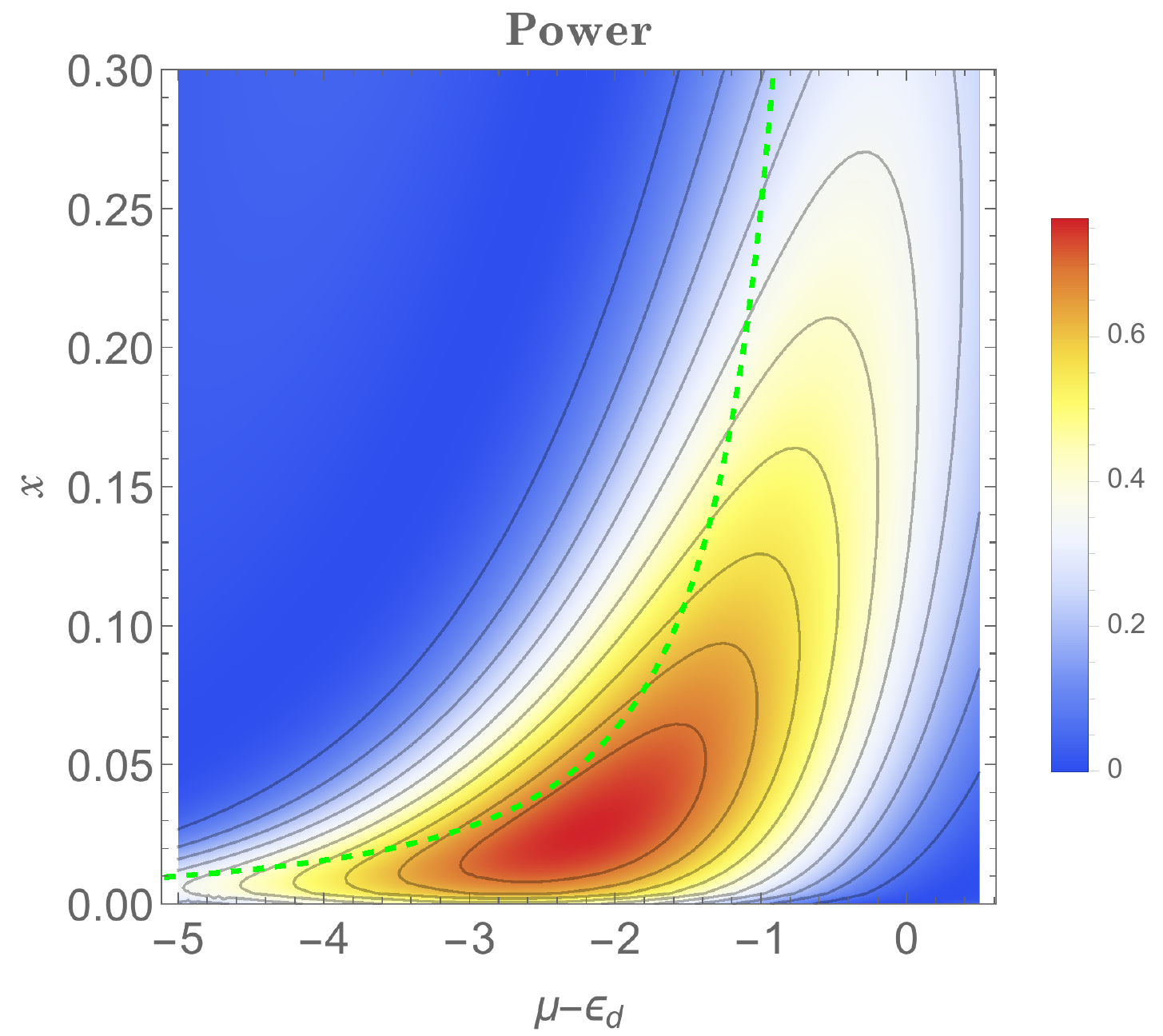}}
\caption{Density plots of the figure of merit $ZT$ and the
  linear-response estimate of the maximal power $\mathcal{P}_{\max}$,
  as a function of $\mu - \epsilon_{d}$ and $x$.  (a) $ZT$ at
  $k_{B}T = 0.2\gamma$, (b) $ZT$ at $k_{B}T = \gamma$, (c)
  $\mathcal{P}_{\max}$ at $k_{B}T = 0.2\gamma$, (d) $\mathcal{P}_{\max}$ at
  $k_{B}T = \gamma$.  Green dashed lines specify the location of the
  Fano node.}
\label{fig:density-plot-linear}
\end{figure}

We start by examining linear-response quantities, focusing the figure
of merit $ZT$ of Eq.~\eqref{eq:ZT}.  Figure \ref{fig:ZT} shows the
figure of merit $ZT$ at $k_{B}T = 0.2\gamma$ as a function of
$\mu - \epsilon_{d}$ by changing the parameter $x$ (or the Fano
parameter $q$).  Red (blue) shade corresponds to a negative (positive)
thermopower region.  The zero-temperature limit of $ZT$ (green lines) as
well as the normalized conductance (dotted lines) is also shown in the
same figure. 
Because $k_{B} T$ is much smaller than the peak width $\gamma$, $ZT$
is small ($\approx 0.11$) at $x=0$, but it quickly reaches more than
unity by introducing a small amount of $x$ ($ZT=1.07$ for $x=0.01$,
$1.64$ for $x=0.1$, and $0.94$ for $x=0.3$). Figures
\ref{fig:ZT-Beta5} and \ref{fig:ZT-Beta1} show the density plot of
$ZT$ as a function of $\mu - E_{d}$ and $x$. We see, if we choose a
larger value of $k_{B}T$, $ZT$ can get even larger.  We find the value
of $ZT$ exceeds 5 at $k_{B}T = \gamma$ (not shown), though such a
situation may be hard to realize in nanostructure systems except for
extremely low-temperature Kondo regime.
As argued in Sec.~\ref{sec:phenomenology}, we anticipate such
enhancement around the Fano resonant node
$\mu - E_{d} \approx - q\Gamma$, which we depict by the green dashed
lines in Fig.~\ref{fig:density-plot-linear}.
The optimal $x$ that achieves the largest $ZT$ (hence the efficiency)
is around $x = 0.1$ for $k_{B}T = 0.2\gamma$, and around $x = 0.01$
for $k_{B}T=\gamma$.  The optimal value of $x$ depends on 
$k_{B}T/\gamma$ and decreases with increasing $k_{B}T/\gamma$.

\begin{figure}
  \centering
\subfloat[\label{fig:PL-etaL-Beta5}]{
  \includegraphics[width=0.45 \linewidth]{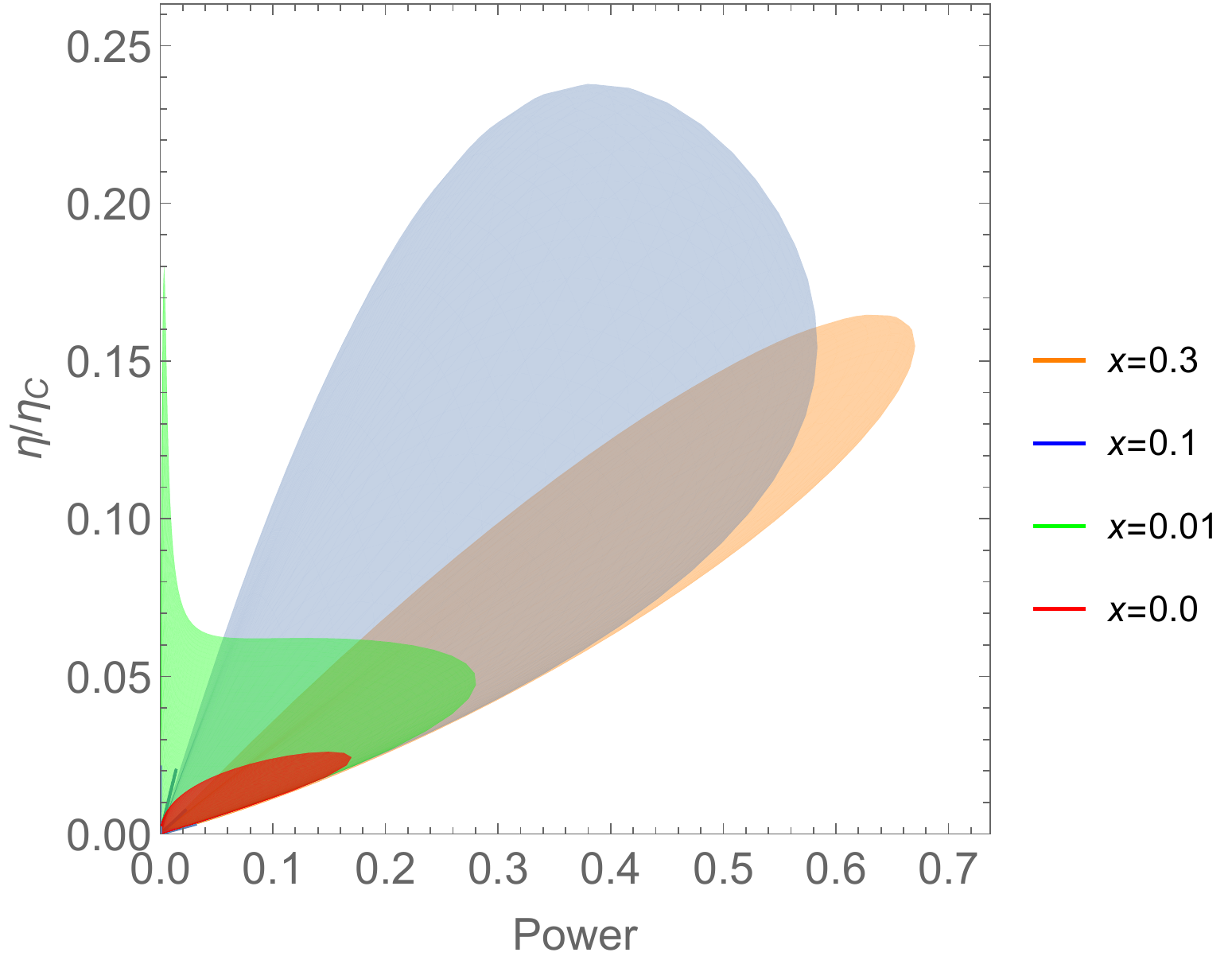}} \quad
\subfloat[\label{fig:PL-etaL-Beta1}]{
  \includegraphics[width=0.45 \linewidth]{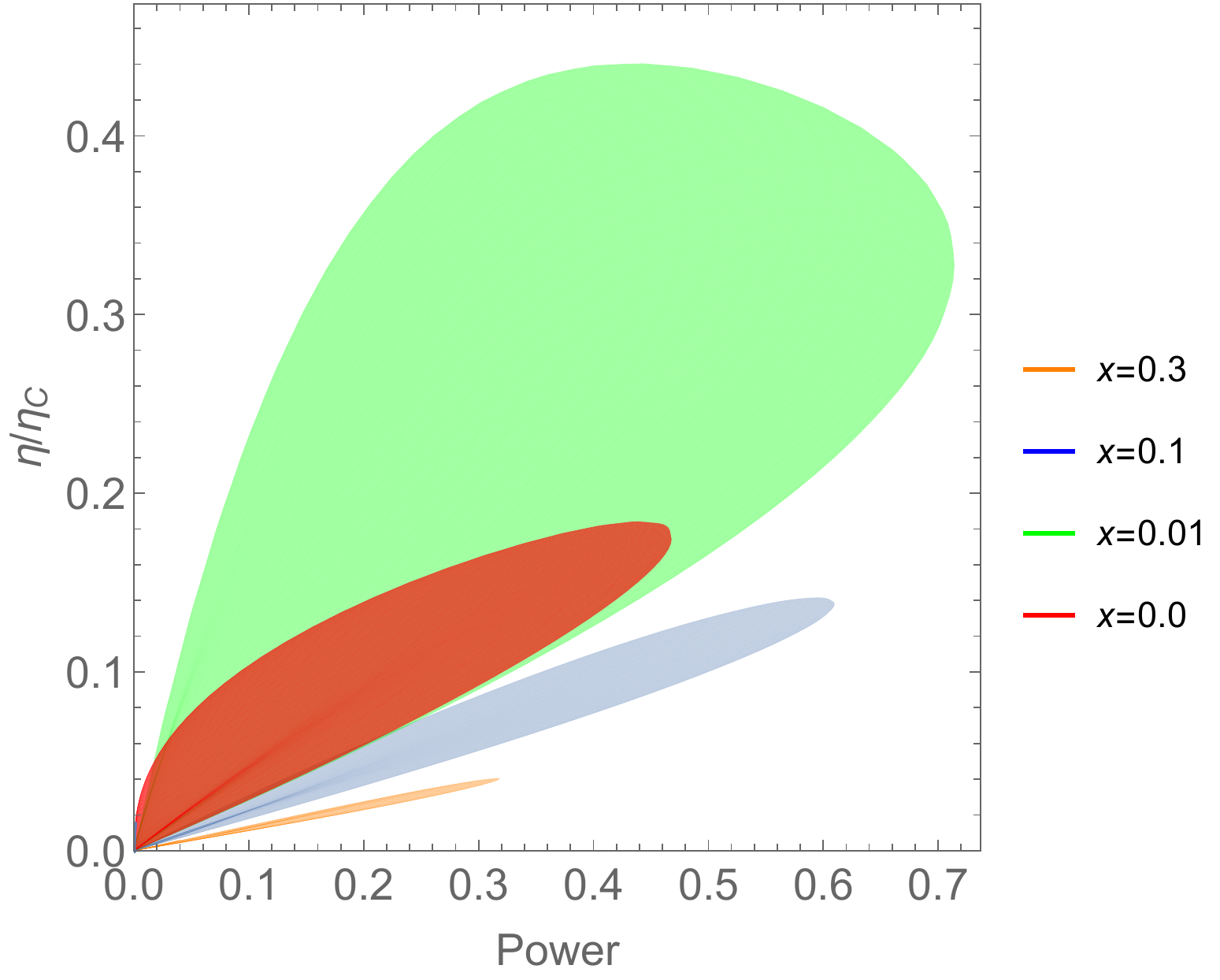}}
\caption{Linear-response estimate of the power-efficiency diagram at
  (a) $k_{B}T = 0.2\gamma$ and (b) $k_{B}T =\gamma$, by changing $x$:
  $x=0.0$ (red), $x=0.01$ (green), $x=0.1$ (blue) and $x=0.3$
  (orange). }
\label{fig:PL-etaL}
\end{figure}

Figures \ref{fig:Pmax-Beta5} and \ref{fig:Pmax-Beta1} show the density
plot of the linear-response estimate of the maximal output power
$\mathcal{P}_{\max}$ in the unit of
$\mathcal{P}_{\Delta T} = (k_{B} \Delta T)^{2}/4h$.  The quantity is
nothing but $hGS^{2}/k_{B}^{2} = \mathcal{K}_{1}^{2}/\mathcal{K}_{0}$,
discussed in Appendix~\ref{sec:linear-response} [see
Eq.~\eqref{eq:power-factor}].  We find the maximal achievable power
$\mathcal{P}_{\max}$ is less sensitive to the value of
$k_{B}T/\gamma$.  Comparing Figs.~\ref{fig:ZT-Beta5} and
\ref{fig:Pmax-Beta5}, or \ref{fig:ZT-Beta1} and \ref{fig:Pmax-Beta1},
we see that the optimal sets of parameters $(x, \epsilon_{d} - \mu)$
to maximize $(ZT)$ or $\mathcal{P}_{\max}$ differs, but they are not
so far apart when $x$ is finite.  We can recapitulate the
linear-response performance by drawing the power-efficiency diagram in
Fig.~\ref{fig:PL-etaL}.  We see that introducing $x$ helps drastically
enhance it in both cases $k_{B}T=0.2\gamma$ and $k_{B}T=\gamma$.

\subsubsection{Nonlinear responses}

We now examine how nonlinear thermoelectric performance can be
possibly improved by utilizing the Fano effect or the parameter $x$.
The analysis of linear-response quantities tells us what parameter
range we should look at to improve it.  Seeing
Figs.~\ref{fig:ZT-Beta5} and \ref{fig:Pmax-Beta5}, we choose to
examine mainly the setting $k_{B}T_{L} =2 k_{B}T_{R} = 0.2\gamma$ with
$x=0.01$ and $x=0.1$, where the corresponding Carnot efficiency is
$\eta_{C} = 0.5$.

\begin{figure}
  \centering
\subfloat[$\eta/\eta_{C}$ at $x=0.01$.\label{fig:eta-Beta5-xi01-etaC50}]{
  \includegraphics[width=0.45\linewidth]{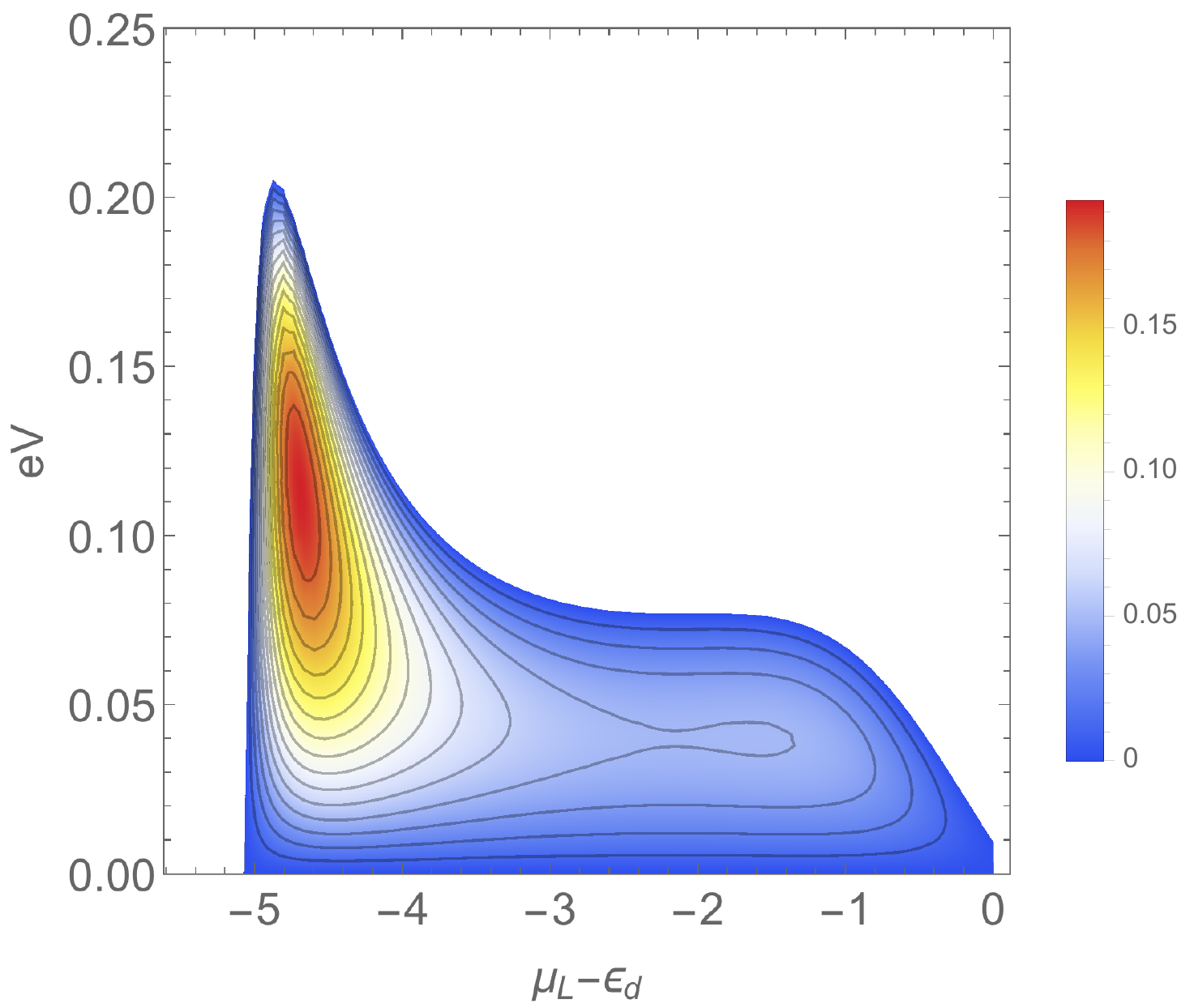}} \quad
\subfloat[$\eta/\eta_{C}$ at $x=0.1$.\label{fig:eta-Beta5-xi10-etaC50}]{
  \includegraphics[width=0.45\linewidth]{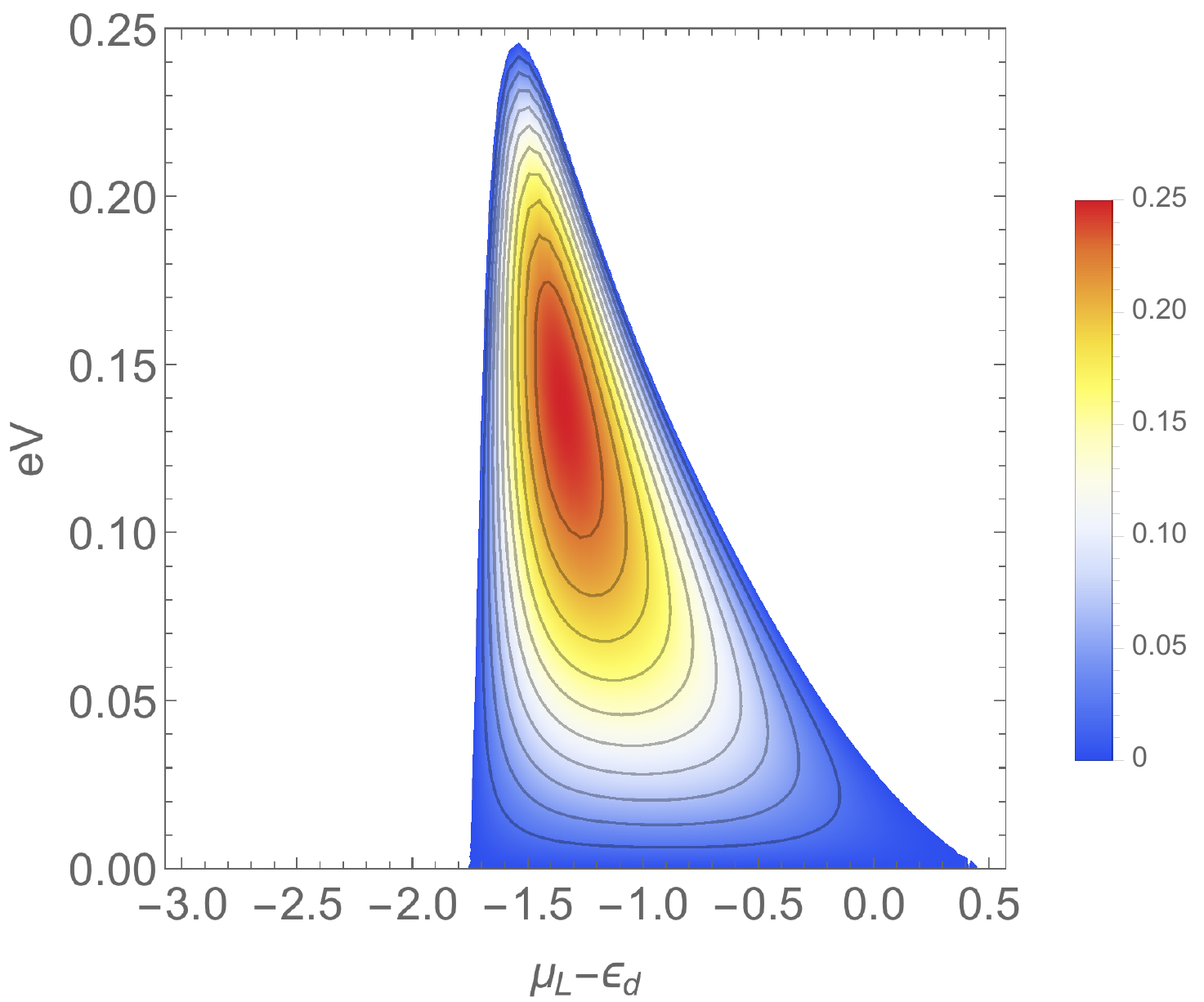}} \\
\subfloat[$\mathcal{P}$ at $x=0.01$.\label{fig:P-Beta5-xi01-etaC50}]{
  \includegraphics[width=0.45\linewidth]{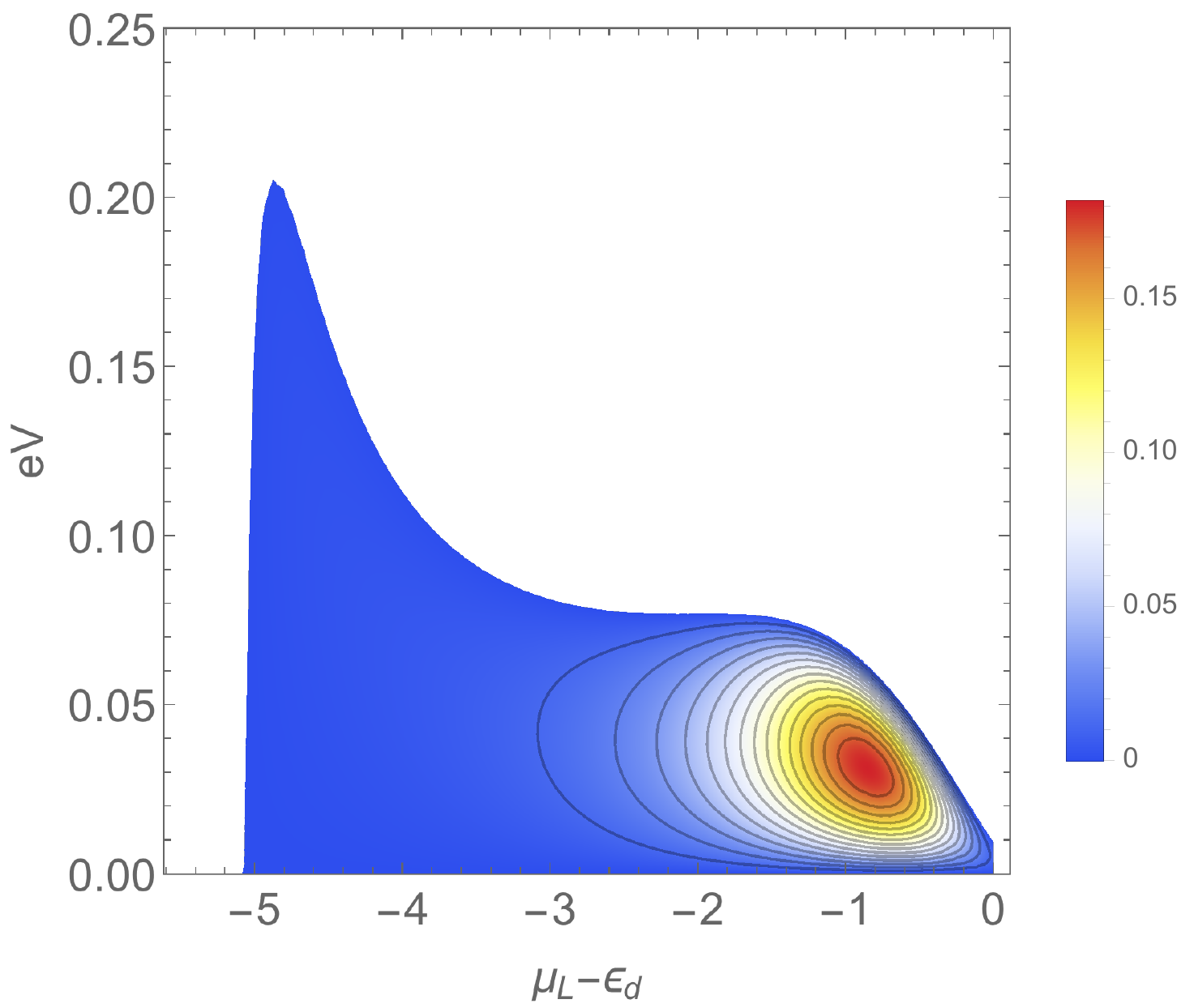}} \quad
\subfloat[$\mathcal{P}$ at $x=0.1$.\label{fig:P-Beta5-xi10-etaC50}]{
  \includegraphics[width=0.45\linewidth]{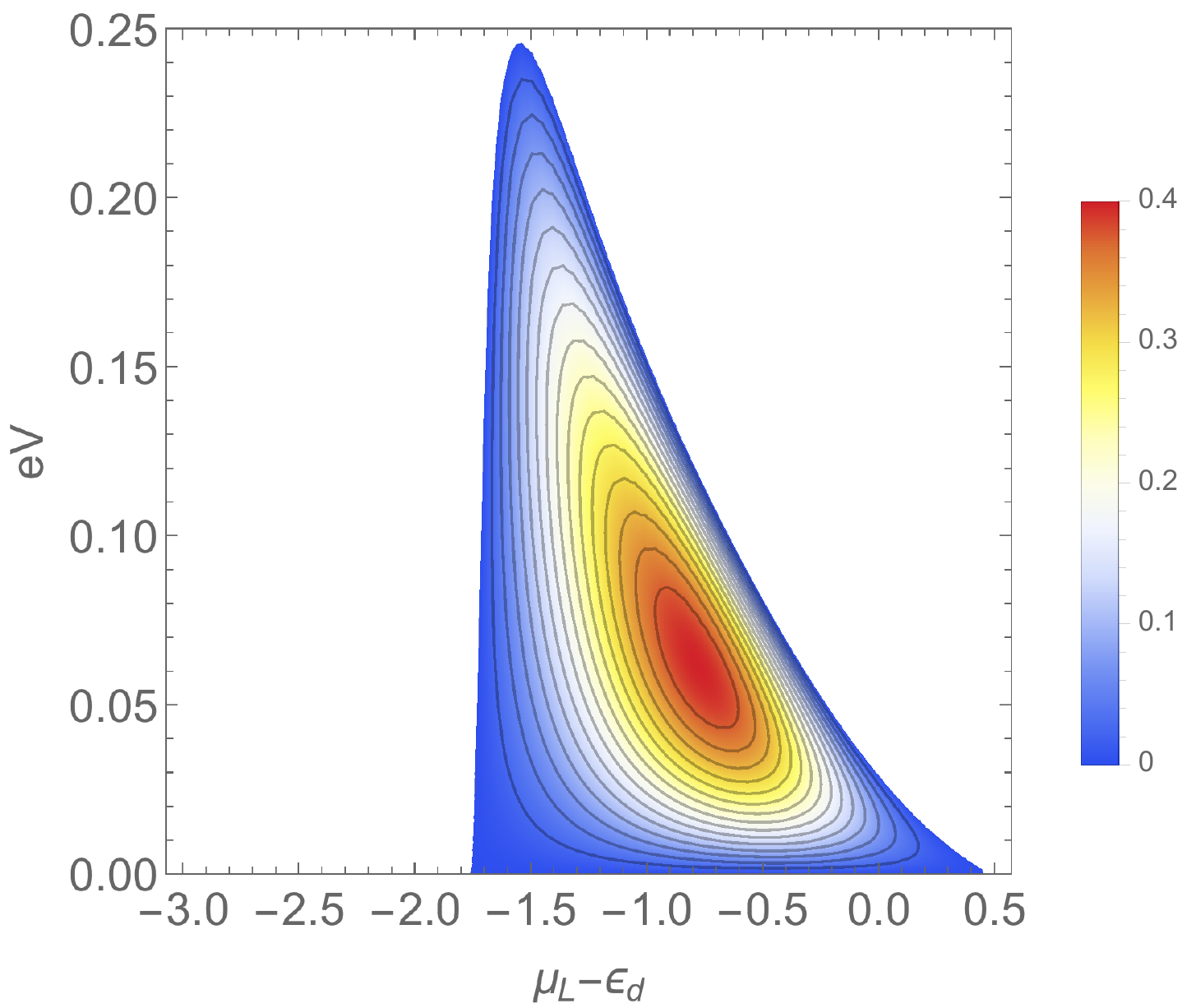}}
\caption{Evolution of nonlinear efficiencies $\eta/\eta_{C}$ and the
  output powers $\mathcal{P}$ by changing gate and bias voltages. The
  setting is $k_{B}T_{L} = 2 k_{B} T_{R} = 0.2\gamma$ (with
  $\eta_{C}=0.5$). We normalize $\mathcal{P}$ by
  $\mathcal{P}_{\Delta T}$.  }
\label{fig:NL-eta-power-etaC50}
\end{figure}
\begin{figure}
  \centering
\subfloat[$x=0.01$\label{fig:P-eta-Beta5-xi01-etaC50}]{
  \includegraphics[width=0.45\linewidth]{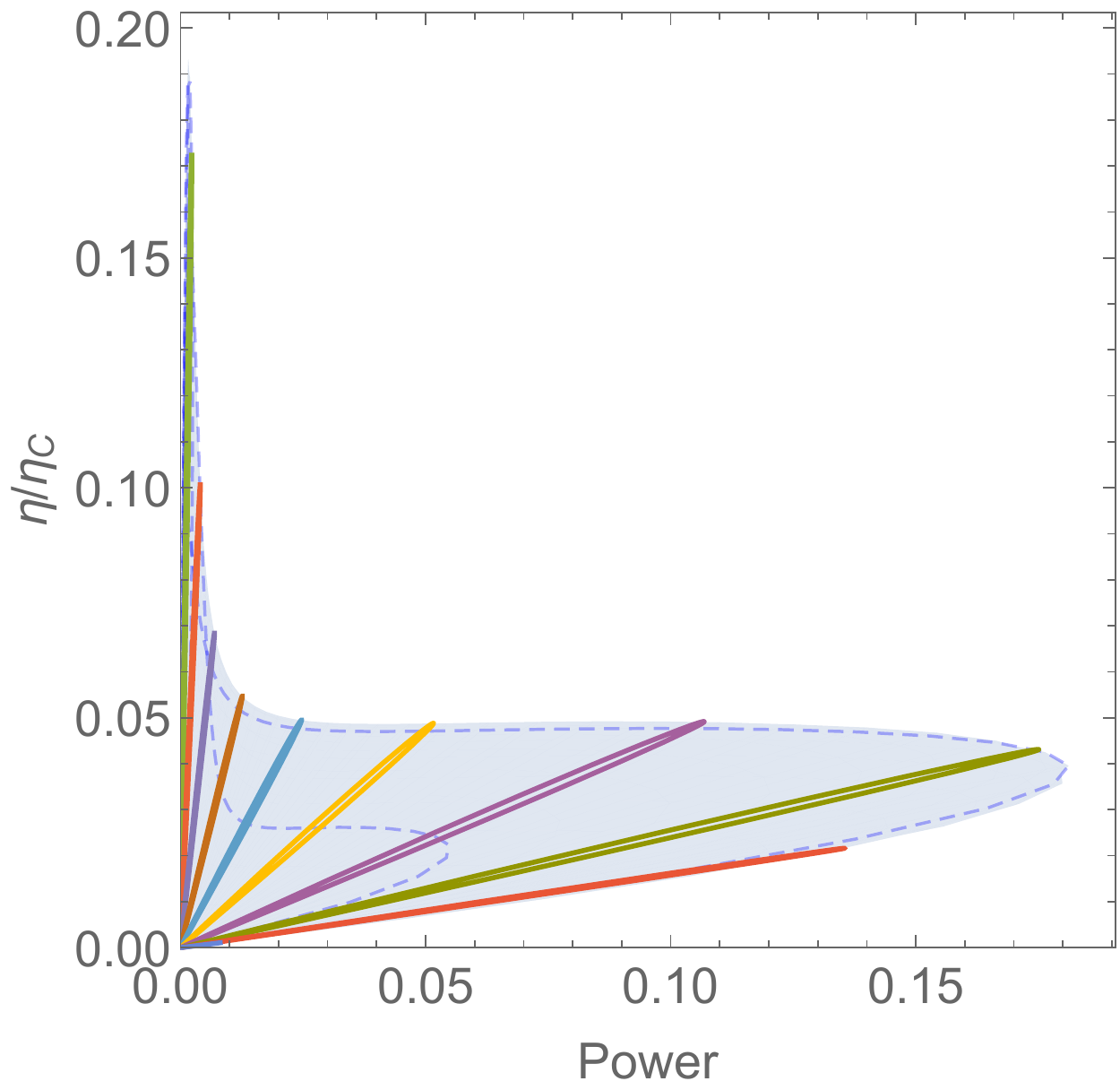}} \quad
\subfloat[$x=0.1$\label{fig:P-eta-Beta5-xi10-etaC50}]{
  \includegraphics[width=0.45\linewidth]{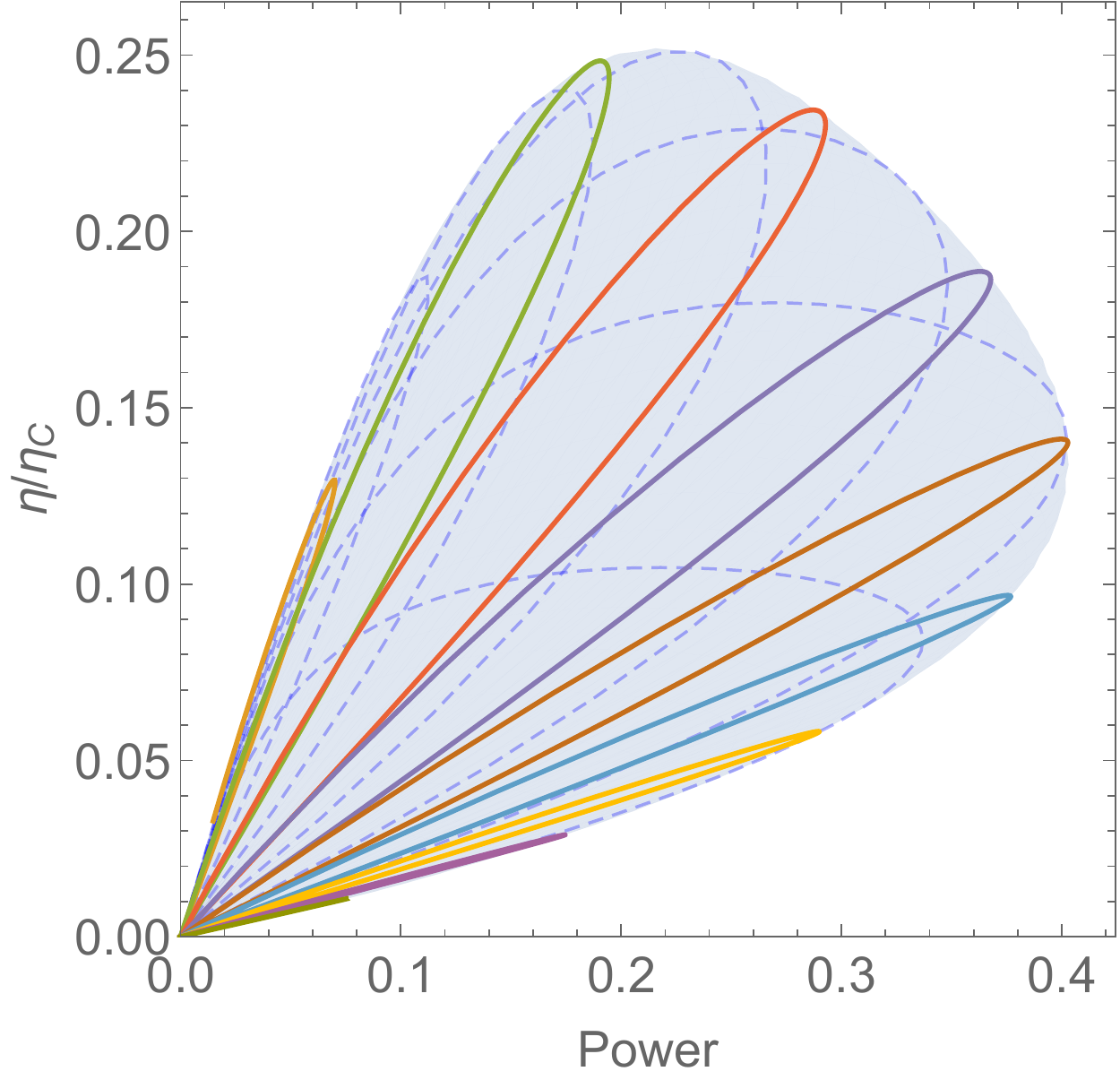}}
  \caption{Power-efficiency diagrams of
    Fig.~\ref{fig:NL-eta-power-etaC50}. Color lines correspond 
    to the evolution by changing bias voltage with a fixed
    gate voltage, and dashed
    lines correspond to the one by changing the gate voltage with a
    fixed bias voltage.  } 
\label{fig:P-eta-diagram}
\end{figure}

Figures \ref{fig:NL-eta-power-etaC50} show density plots of nonlinear
efficiencies $\eta$ (modulo $\eta_{C}$) and output powers
$\mathcal{P}$ (modulo $\mathcal{P}_{\Delta T}$) for $x=0.01$ (a, c) and $x=0.1$ (b,d).  We find the
maximal values of the efficiencies reaches $0.20\eta_{C}$ (for
$x=0.01$) or $0.25\eta_{C}$ (for $x=0.1$), exceeding their
linear-response estimates.  For the 
case of $x=0.01$, one sees that 
the settings of gate and bias voltages for maximizing either the
efficiency or the output power are irreconcilable.  For instance, at
the gate voltage achieving the highest efficiency, its output power 
almost vanishes, as was argued in Ref.~\cite{Nakpathomkun10}.  For
the case of $x=0.1$, however, the two conditions are more compatible,
and both a higher $\eta$ and a larger $\mathcal{P}$ are achieved in
comparing with $x=0.01$.  The power-efficiency diagrams in 
Fig.~\ref{fig:P-eta-diagram} clearly show this difference in their
behaviors.  We see that for the case of $x=0.1$, the efficiency and the output
power are well balanced for a finite range of the gate voltage
$\mu_{L}$. 

\begin{figure}
  \centering
\subfloat[$T_{R} = 0.5T_{L}$ ($\eta_{C} = 0.50$)\label{fig:P-eta-Beta5-etaC50}]{
  \includegraphics[width=0.45\linewidth]{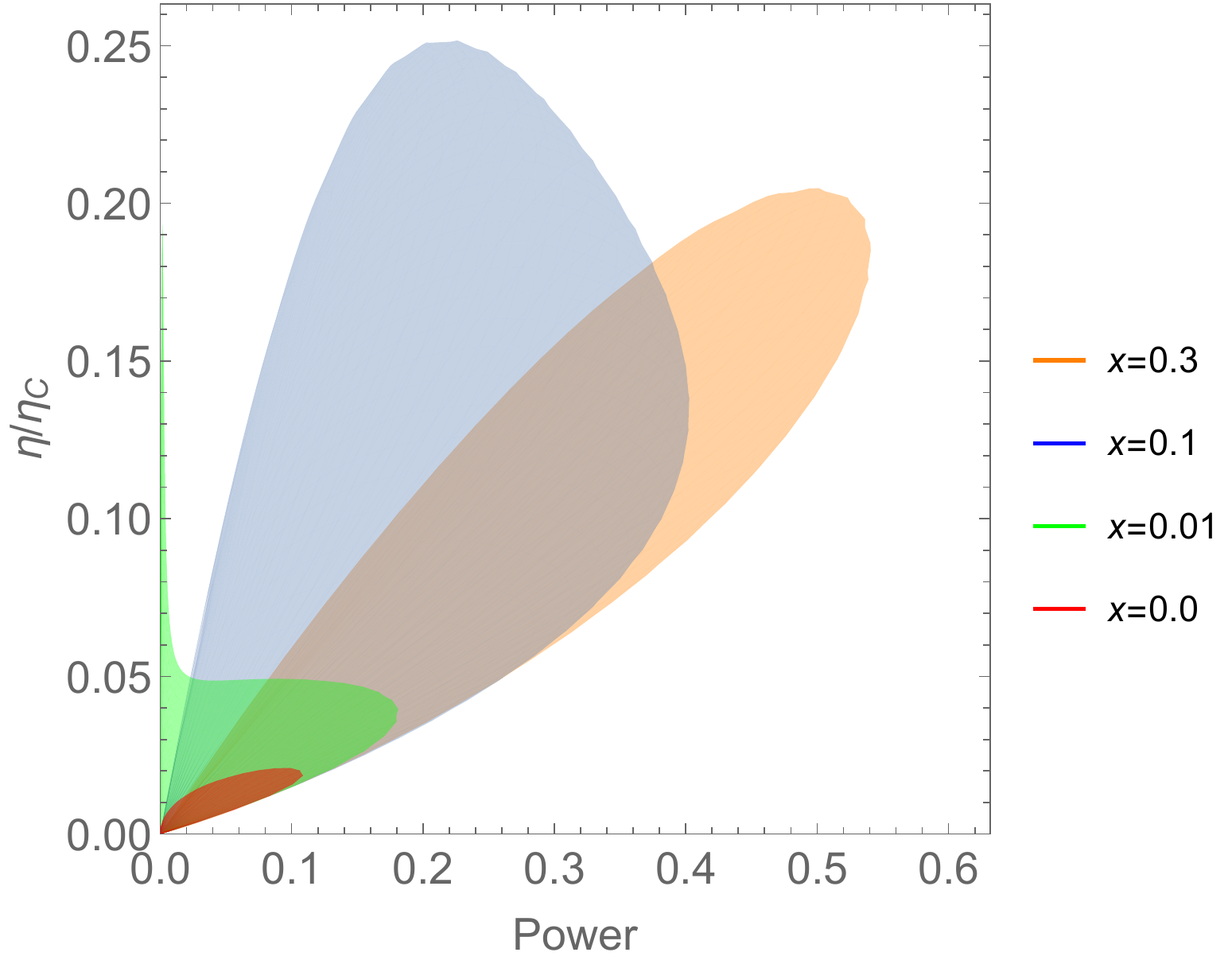}} \quad
\subfloat[$T_{R} = 0.05T_{L}$ ($\eta_{C} = 0.95$)\label{fig:P-eta-Beta5-etaC95}]{
  \includegraphics[width=0.45\linewidth]{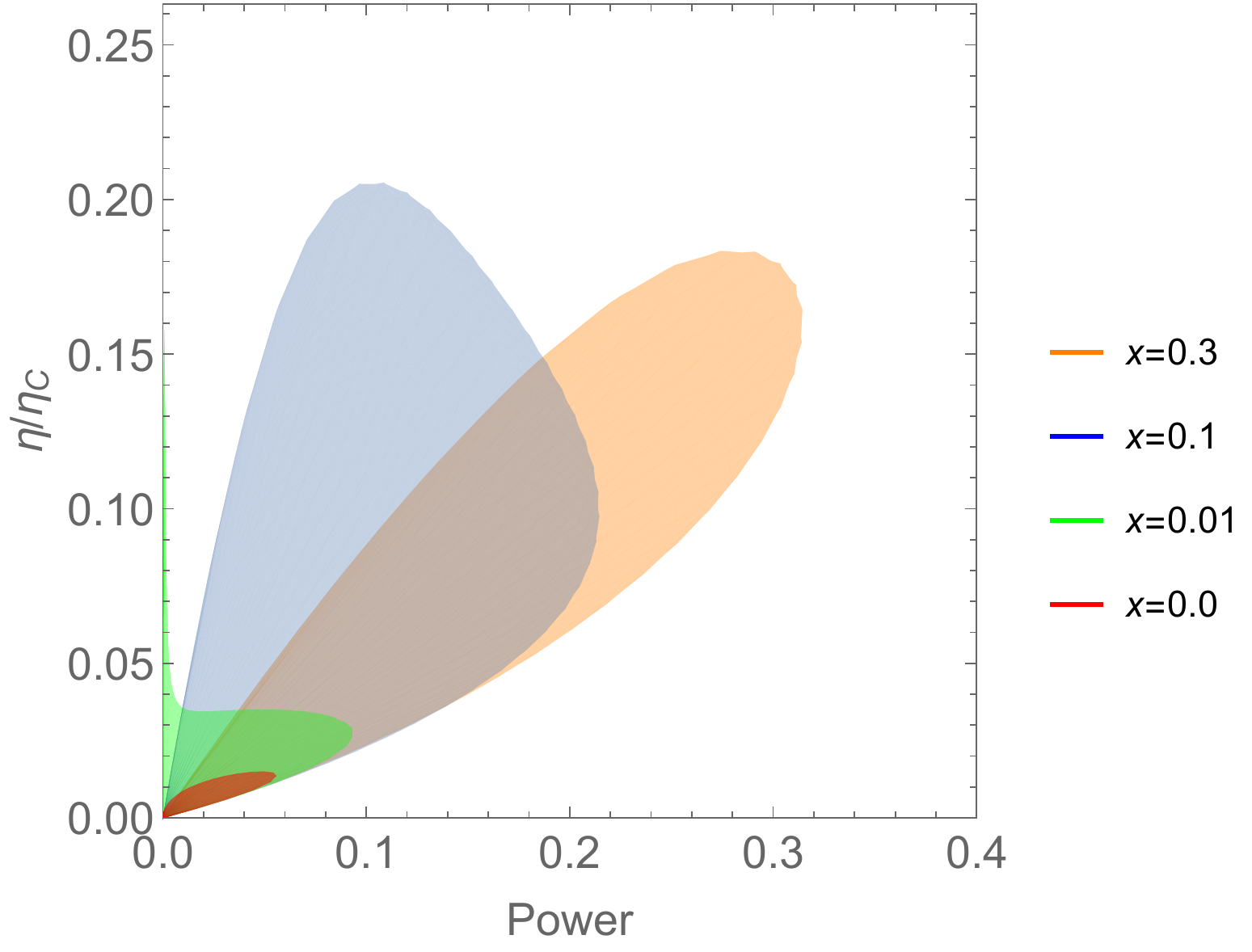}}
\caption{Comparison of the power-efficiency diagrams at different
  $k_{B}T_{R}$. Other parameters are the same with
  Figs.~\ref{fig:NL-eta-power-etaC50} and \ref{fig:P-eta-diagram}. 
  The power output is normalized by $\mathcal{P}_{\Delta T}$
  for both cases. }
\label{fig:P-eta-Beta5}
\end{figure}

To attain even higher efficiency and output power, one may place a
system subject to a larger temperature difference, because the
linear-response theory predicts the efficiency proportional to
$\eta_{C}$ and the power output, to $\eta_{C}^{2}$, though nonlinear
effects may well suppress such behavior. We expect, however, the
present thermoelectric enhancement is likely to survive for a large
temperature difference because it is a universal mechanism due to the
Fano resonance effect (see Sec.~\ref{sec:phenomenology}).
Figures~\ref{fig:P-eta-Beta5-etaC50} and \ref{fig:P-eta-Beta5-etaC95}
compare the power-efficiency diagrams between the two settings:
(a)~$k_{B}T_{L}=2k_{B}T_{R} = 0.2\gamma$ (with $\eta_{C} = 0.5$), and
(b)~$k_{B}T_{L}=20k_{B}T_{R} = 0.2\gamma$ (with $\eta_{C}=0.95$).  By
normalizing the efficiency $\eta$ by $\eta_{C}$, and the output power
$\mathcal{P}$ by $\mathcal{P}_{\Delta T}$, we can directly
compare those diagrams with Fig.~\ref{fig:PL-etaL-Beta5}.
We first notice that the linear-response estimate reasonably captures
the overall tendency for each $x$ in this fully nonlinear regime.
There is a noticeable saturation in the highest efficiency and a
severe reduction of the output power, though.  We can attribute those
to nonlinear thermal effects, which we will discuss further in
Sec.~\ref{sec:criteria}.
Comparing with the thermoelectric performance at $x=0$, the
enhancement effect due to finite $x$ is significant. For the case of
$\eta_{C}=0.5$, the power-efficiency improves from 
$(\mathcal{P}_{\max}, \eta_{\max}) = (0.11\mathcal{P}_{\Delta T},
0.02\eta_{C})$ at $x=0$ to
$(0.40\mathcal{P}_{\Delta T}, 0.25\eta_{C})$ at $x=0.1$, or
$(0.54\mathcal{P}_{\Delta T}, 0.20\eta_{C})$ at $x=0.3$; for the case
of $\eta_{C}=0.95$, from 
$(\mathcal{P}_{\max}, \eta_{\max}) = (0.06\mathcal{P}_{\Delta T},
0.015 \eta_{C})$ at $x=0$ to
$(0.21\mathcal{P}_{\Delta T}, 0.20\eta_{C})$ at $x=0.1$, or
$(0.31\mathcal{P}_{\Delta T}, 0.18\eta_{C})$ at $x=0.3$.  The
efficiency improves more than 10 times, while the output power
gets amplified nearly 5 times.

\subsection{Quantum dot with interaction}

\begin{figure}
  \centering
\subfloat[$x=0$\label{fig:T-Beta5-xi00}]{
  \includegraphics[width=0.4\linewidth]{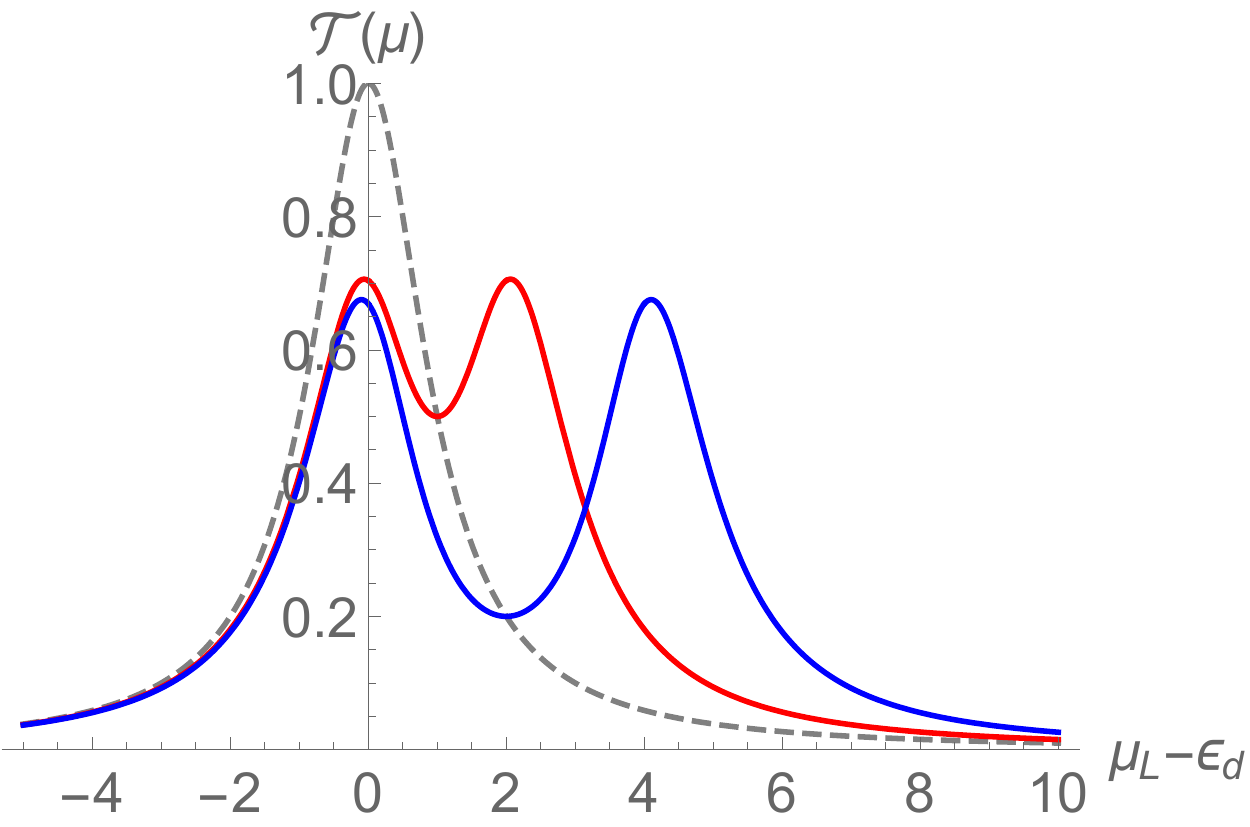}} \quad
\subfloat[$x=0.1$\label{fig:T-Beta5-xi10}]{
  \includegraphics[width=0.4\linewidth]{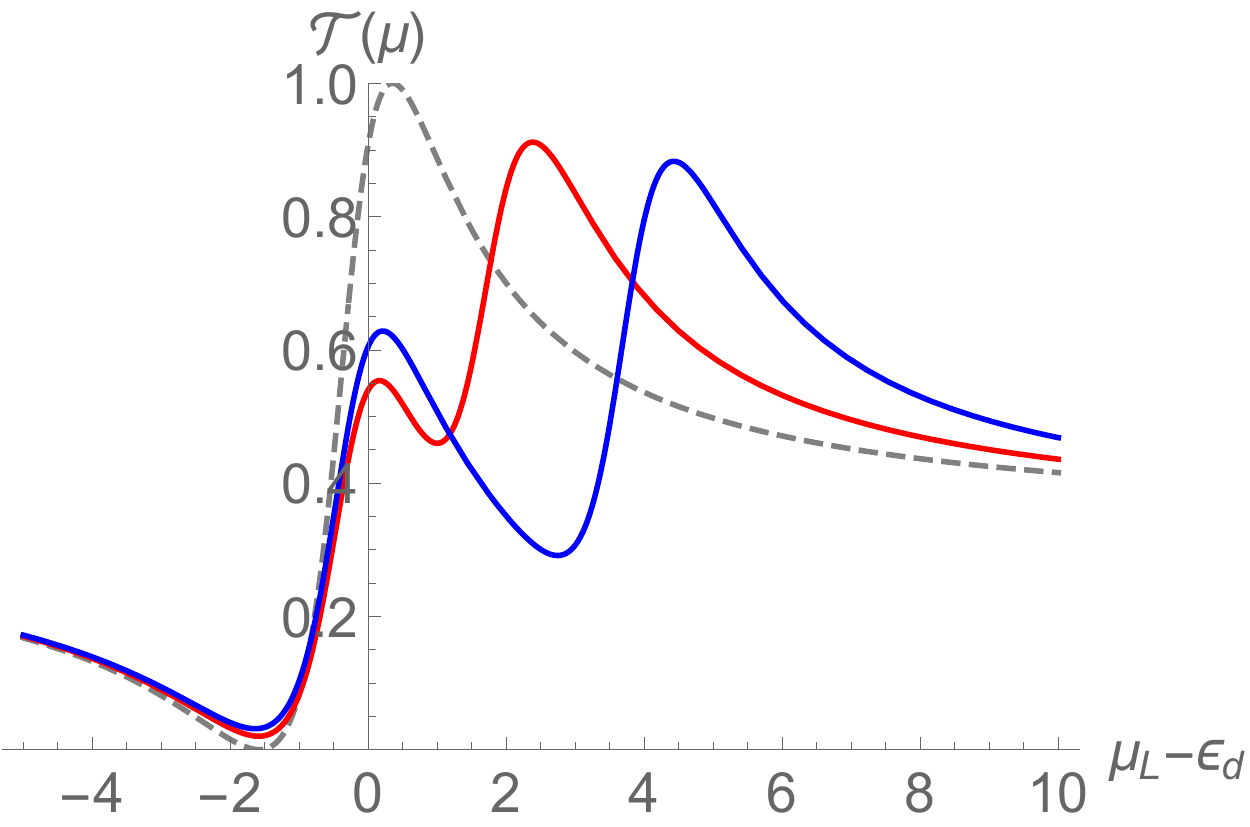}}
  \caption{Effective transmission $\mathcal{T}(\mu)$ for an
    interacting dot as a function of $\mu-\epsilon_{d}$ by changing
    $U$: $U/\gamma = 0$ (dashed gray lines), $2$ (red lines), and $4$
    (blue line). The results are shown for (a) $x = 0$ and (b) $x=0.1$.}
\label{fig:effective-T-int}
\end{figure}

As argued in Sec.~\ref{sec:GR-with-int}, we incorporate the strong
interaction on the dot in the effective transmission by
Eq.~\eqref{eq:effective-T-with-int}.  Figure~\ref{fig:effective-T-int}
illustrates how $\mathcal{T}(\varepsilon)$ describes the Coulomb
blockade peaks around $\mu - \epsilon_{d} = E_{d}$ and $E_{d}+U$
(Fig.~\ref{fig:T-Beta5-xi00}), and its deformations by the Fano
resonances (Fig.~\ref{fig:T-Beta5-xi10}).  Since nonlinear flows
become a superposition of two Fano-type contributions, as in
Eqs.~\eqref{eq:particle-flow-with-int} and
\eqref{eq:energy-flow-with-int}, we can still apply much of the
previous argument given for a noninteracting dot to an interacting
dot. That means we expect enhanced thermoelectricity by the Fano
resonance in an interacting dot as well.

\begin{figure}
  \centering
\subfloat[$\eta/\eta_{C}$ at
$U=2\gamma$\label{fig:etaInt-U2-Beta5-xi10-etaC50}]{ 
  \includegraphics[width=0.45\linewidth]{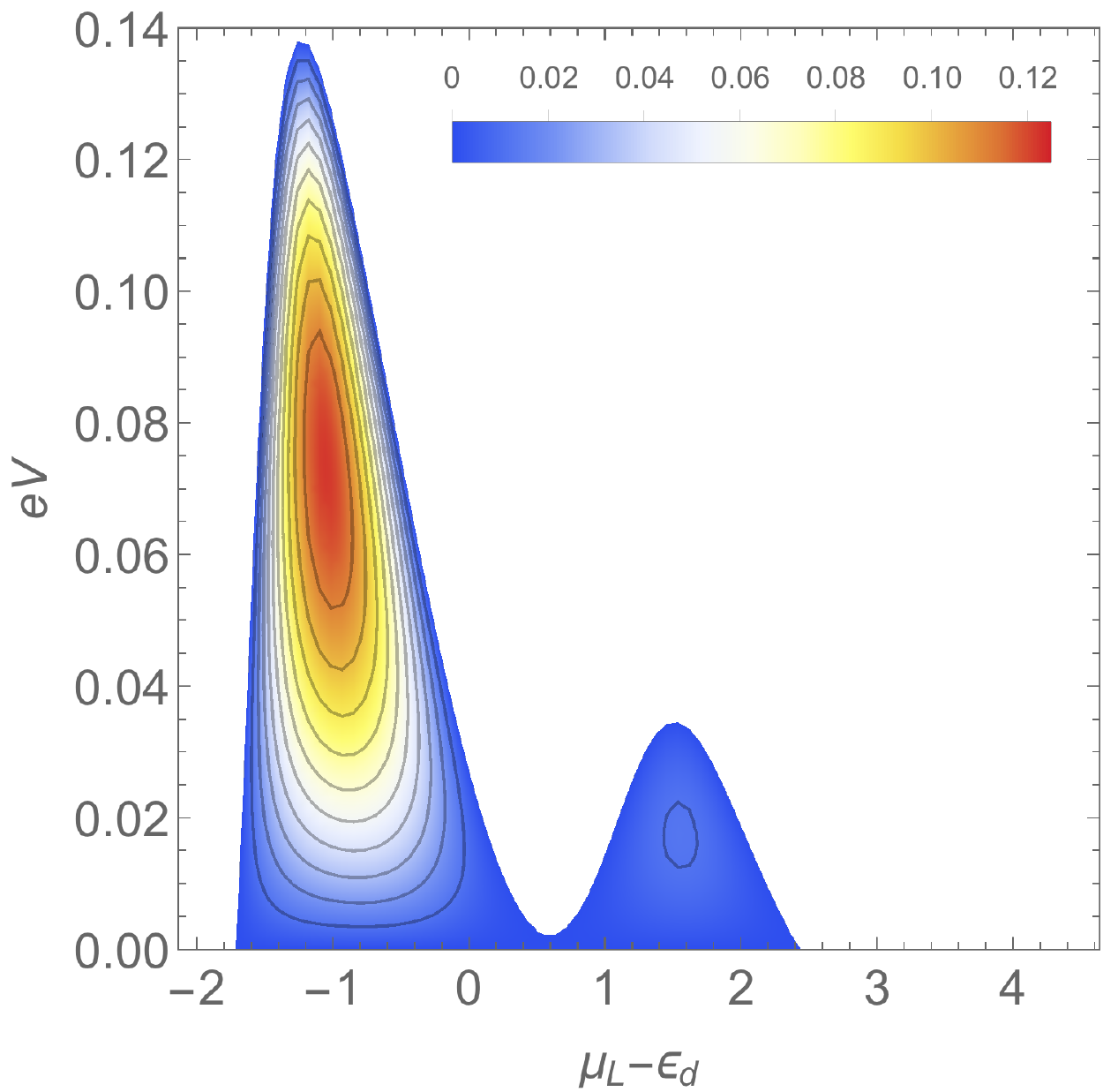}} \quad
\subfloat[$\eta/\eta_{C}$ at
$U=4\gamma$\label{fig:etaInt-U4-Beta5-xi10-etaC50}]{
  \includegraphics[width=0.45\linewidth]{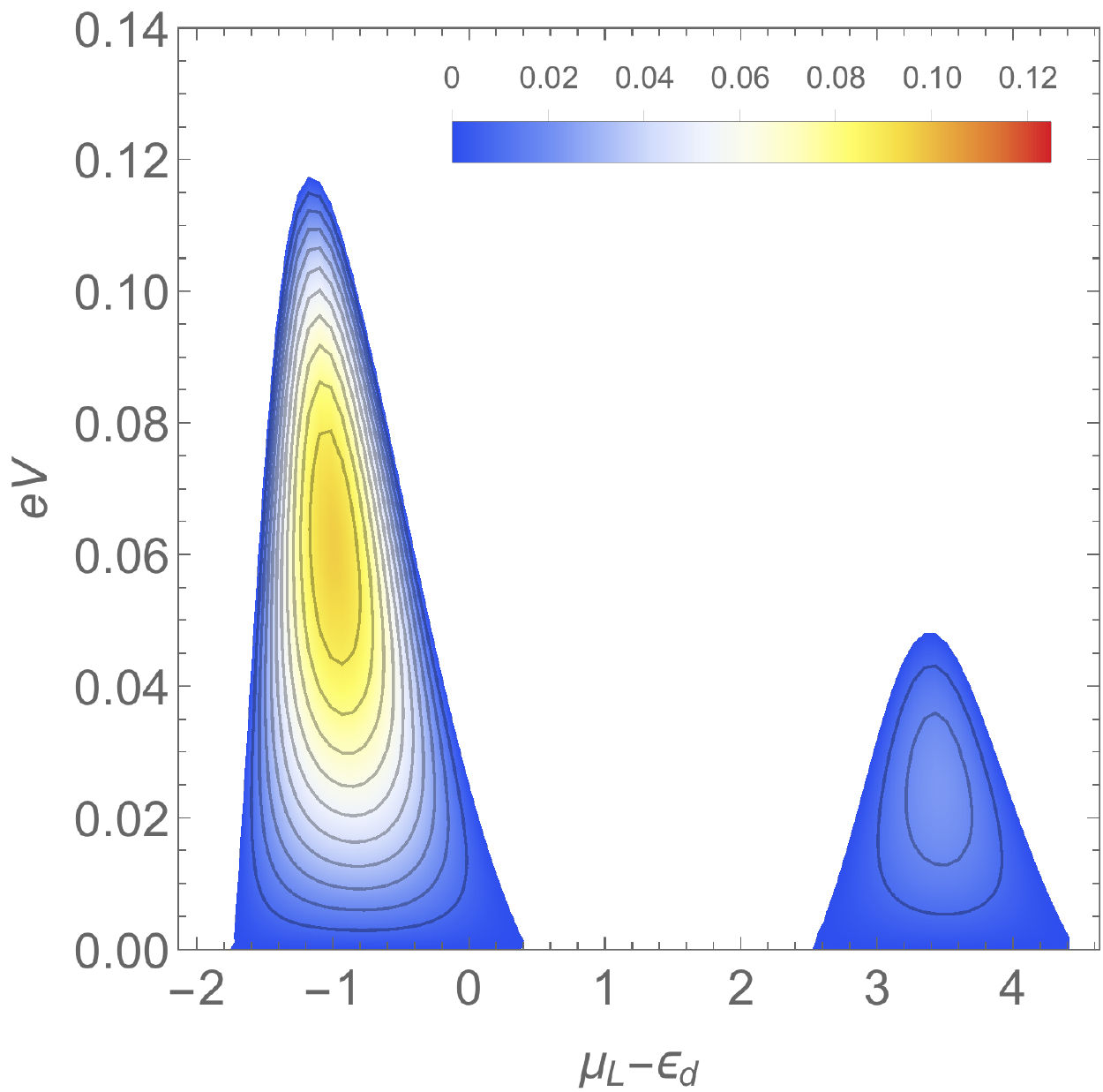}}
\\
\subfloat[$\mathcal{P}$ at $U=2\gamma$\label{fig:PInt-U2-Beta5-xi10-etaC50}]{
  \includegraphics[width=0.45\linewidth]{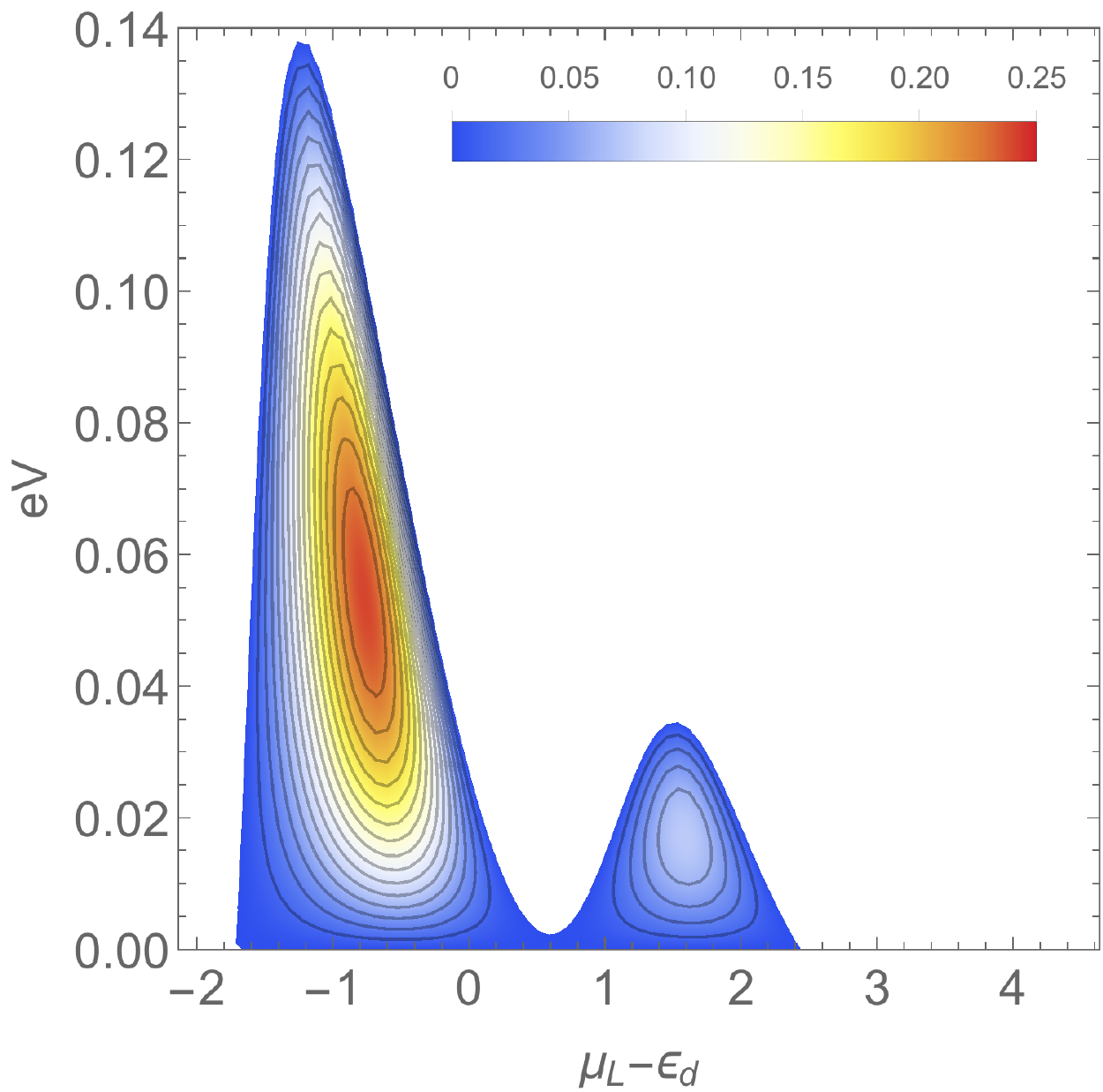}} \quad
\subfloat[$\mathcal{P}$ at $U=2\gamma$\label{fig:PInt-U4-Beta5-xi10-etaC50}]{
  \includegraphics[width=0.45\linewidth]{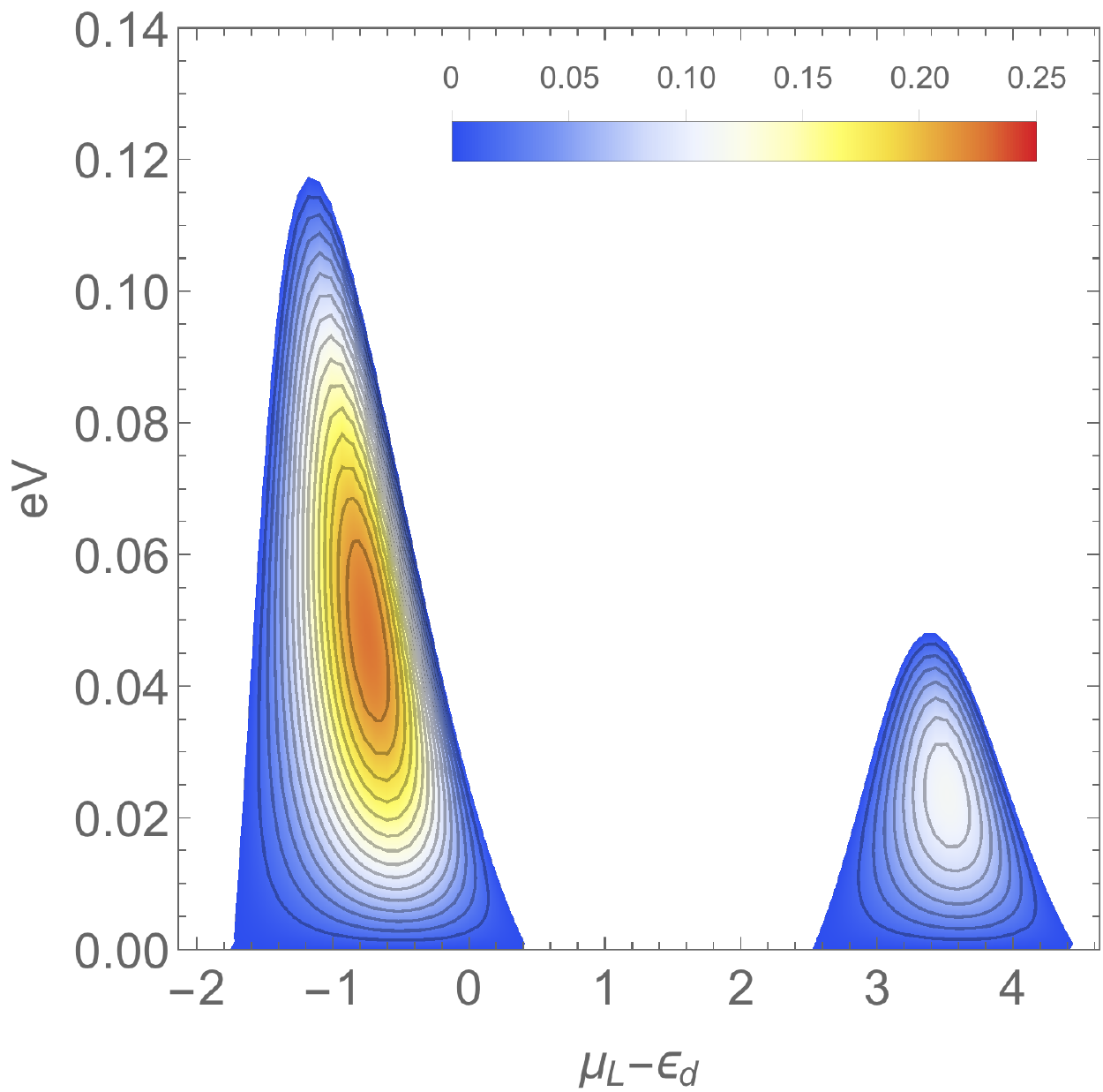}}
\caption{The evolution of the efficiency $\eta/\eta_{C}$ and the power output
  $\mathcal{P}$ at $x=0.1$ by changing $U$ at
  $k_{B}T_{L} = 2 k_{B} T_{R} = 0.2\gamma$ with $\eta_{C}=0.5$. The
  power $\mathcal{P}$ is normalized by
  $\mathcal{P}_{\Delta T}$.  }
  \label{fig:etaInt-PInt-density-plot}
\end{figure}

\begin{figure}
  \centering
\subfloat[$U=2\gamma$\label{fig:P-eta-U2-Beta5-etaC50}]{
  \includegraphics[height=0.25\linewidth]{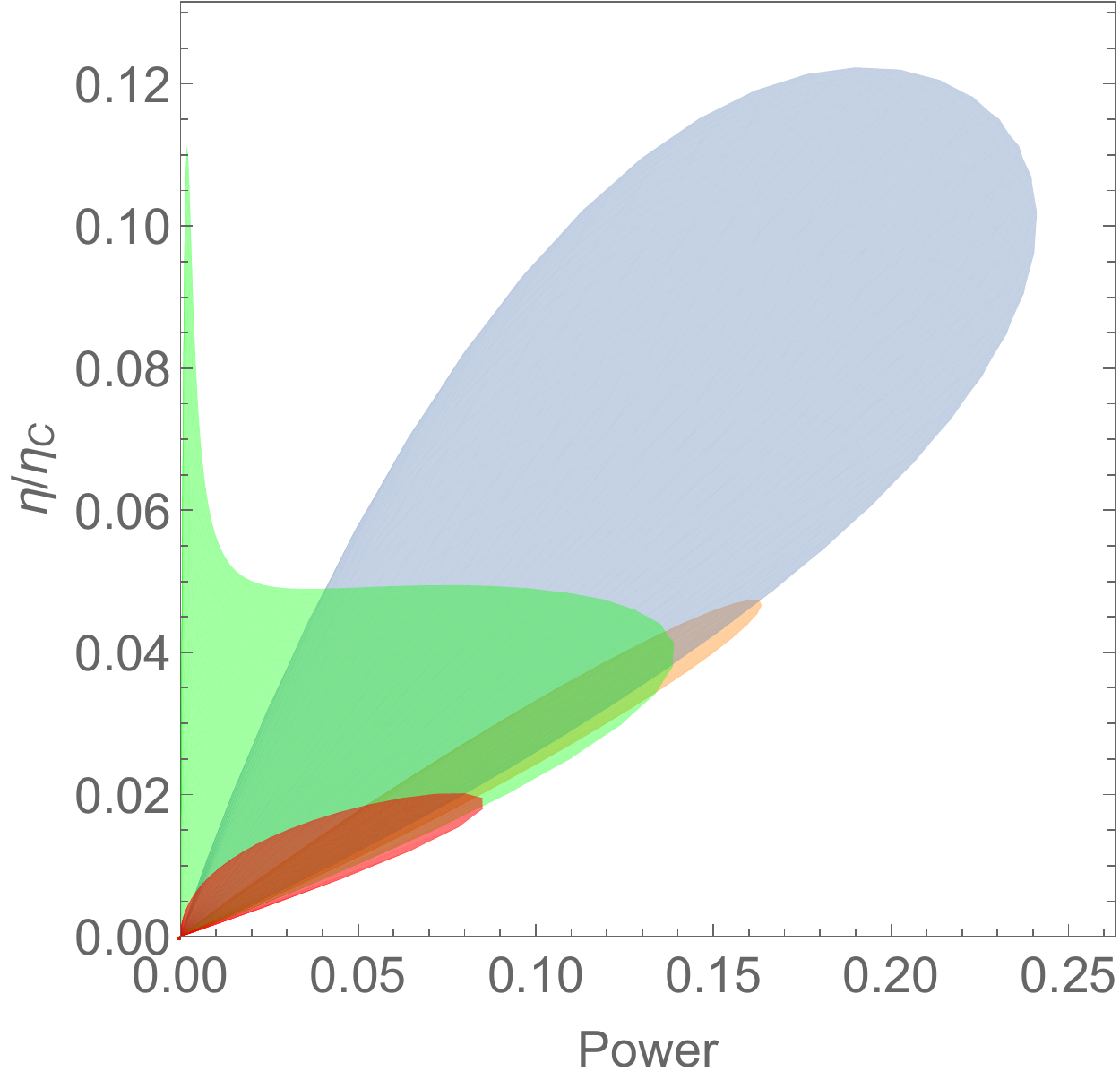}} \quad
\subfloat[$U=4\gamma$\label{fig:P-eta-U4-Beta5-etaC50}]{
  \includegraphics[height=0.25\linewidth]{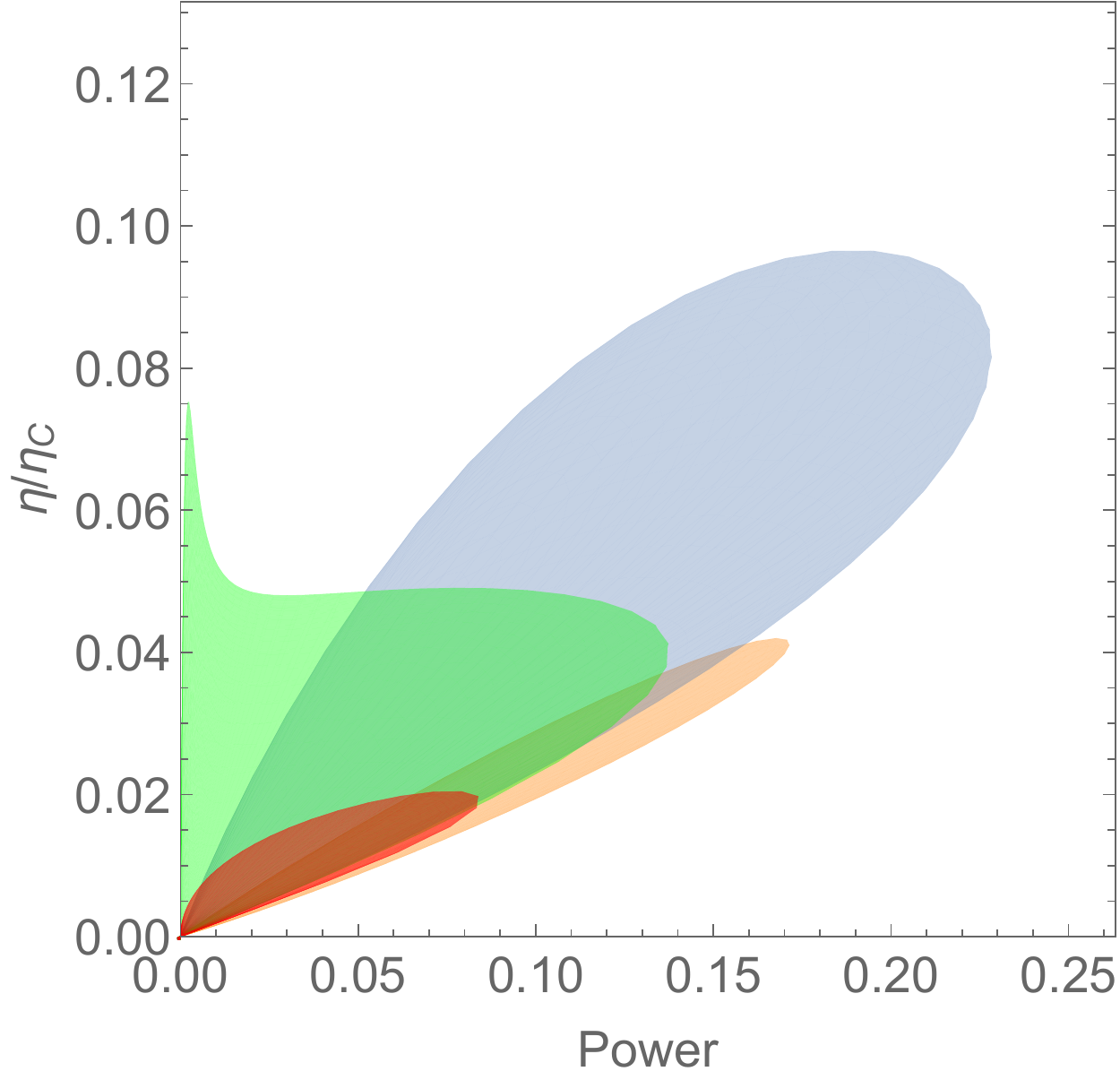}} \quad
\subfloat[$U=8\gamma$\label{fig:P-eta-U8-Beta5-etaC50}]{
  \includegraphics[height=0.25\linewidth]{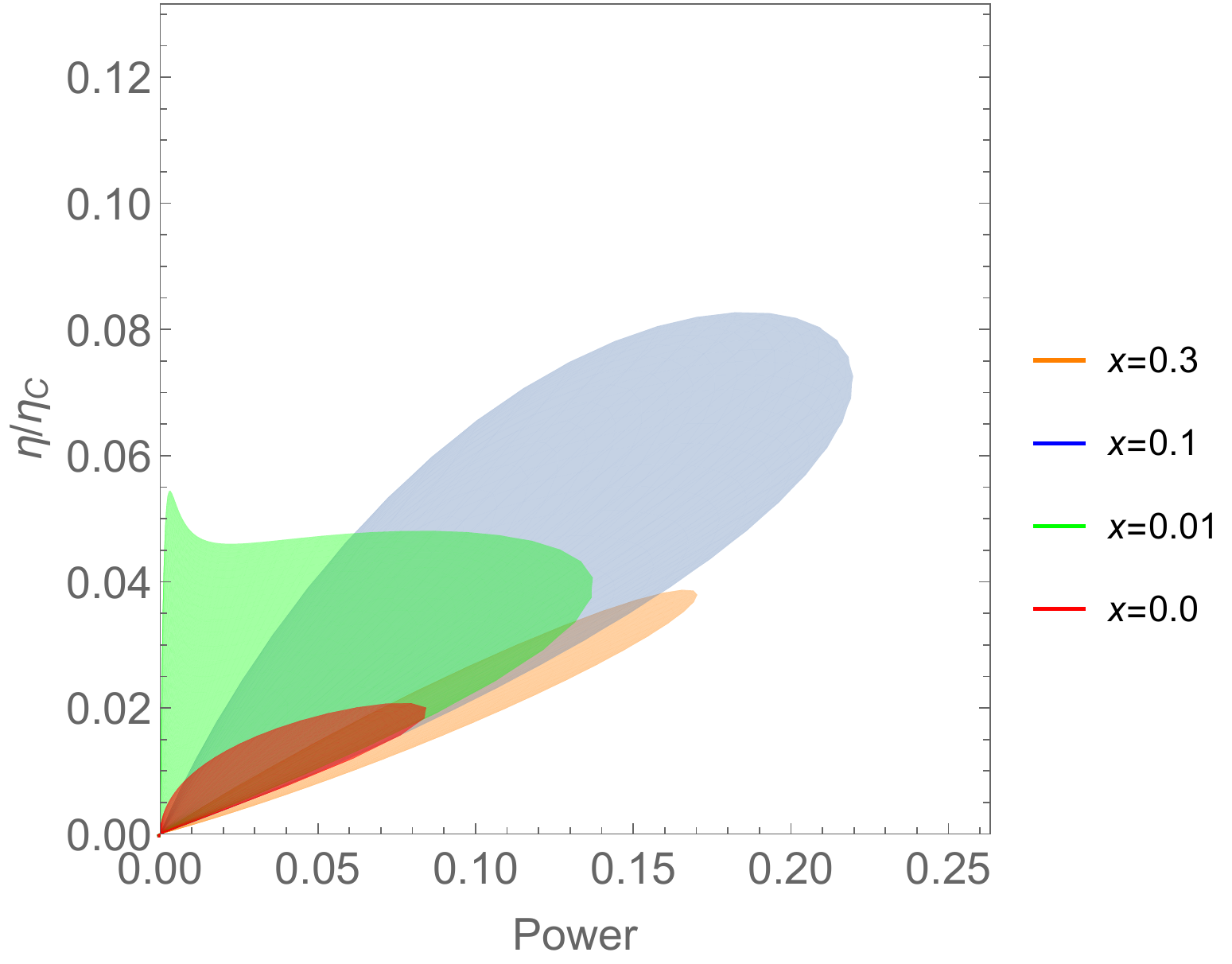}}
\caption{Power-efficiency diagram by changing $U/\gamma=2$, $4$, $8$.
  Other parameters are the same as in
  Fig.~\ref{fig:etaInt-PInt-density-plot}. Shaded regions correspond
  to the evolution for $x=0$ (red), $x=0.01$ (green),
  $x=0.1$ (blue), and $x=0.3$ (orange). }
  \label{fig:P-eta-U-Beta5-etaC50}
\end{figure}

Figures~\ref{fig:etaInt-PInt-density-plot} show the density plots of
the efficiency and the output power as a function of gate and bias
voltages by changing $U/\gamma=2, 4$.  For a better comparison, we use
the same color scheme for different values of $U$.  As one sees in
Fig.~\ref{fig:etaInt-PInt-density-plot}, the Fano resonance dominantly
affects only one of the two Coulomb blockade peaks to enhance both the
efficiency and the output power there (see also
Fig.~\ref{fig:P-eta-U-Beta5-etaC50}).  Simultaneously, we notice that
the highest efficiency decreases with increasing $U$, suggesting
strong interaction is rather detrimental to efficiency.  The
power-efficiency diagrams (Fig.~\ref{fig:P-eta-U-Beta5-etaC50})
clarify this point. For a fixed $U$, we see both the efficiency and
the output power greatly enhanced by introducing finite $x$, as in a
noninteracting dot. With $U=2\gamma$, we have
$(\mathcal{P}_{\max}, \eta_{\max}) = (0.08\mathcal{P}_{\Delta T},
0.02\eta_{C})$ to $( 0.24\mathcal{P}_{\Delta T}, 0.12\eta_{C})$ at
$x=0.1$ (Fig.~\ref{fig:P-eta-U2-Beta5-etaC50}). However, the
efficiency at $x=0.1$, which is most enhanced, gets suppressed by
increasing $U$.  One may understand it from the form of the effective
transmission~\eqref{eq:effective-T-with-int}.  Since the transmission
peak is split into two and and the average occupation around the Fano
resonance gets smaller by increasing $U$, the Fano enhancement plays a
less prominent role with a larger $U$.  Such interaction-induced
suppression is also seen in the output power but its reduction is much
more gradual.

\section{Assessing nonlinear thermoelectricity}
\label{sec:criteria}

In taking advantage of quantum control to achieve better
thermoelectricity, it is important to find an optimal set of relevant
parameters. With a fixed $x$, we find it crucial to adjust the gate
voltage for attaining a full enhancement due to the Fano effect.  
To assess its performance in the nonlinear
transport regime, we would be better off with a quantity that can
characterize nonlinear thermoelectric performance without delving into
a detailed analysis of nonlinear flows.  The figure of merit $ZT$
loses its authenticity beyond the linear
transport~\cite{Meair13,Azema14}.  We here address this issue 
with some speculations, based on our results.

\begin{figure}
  \centering
\subfloat[Particle flow $I_{L}$\label{fig:IV-Beta5-xi10-etaC50}]{
  \includegraphics[width=0.4\linewidth]{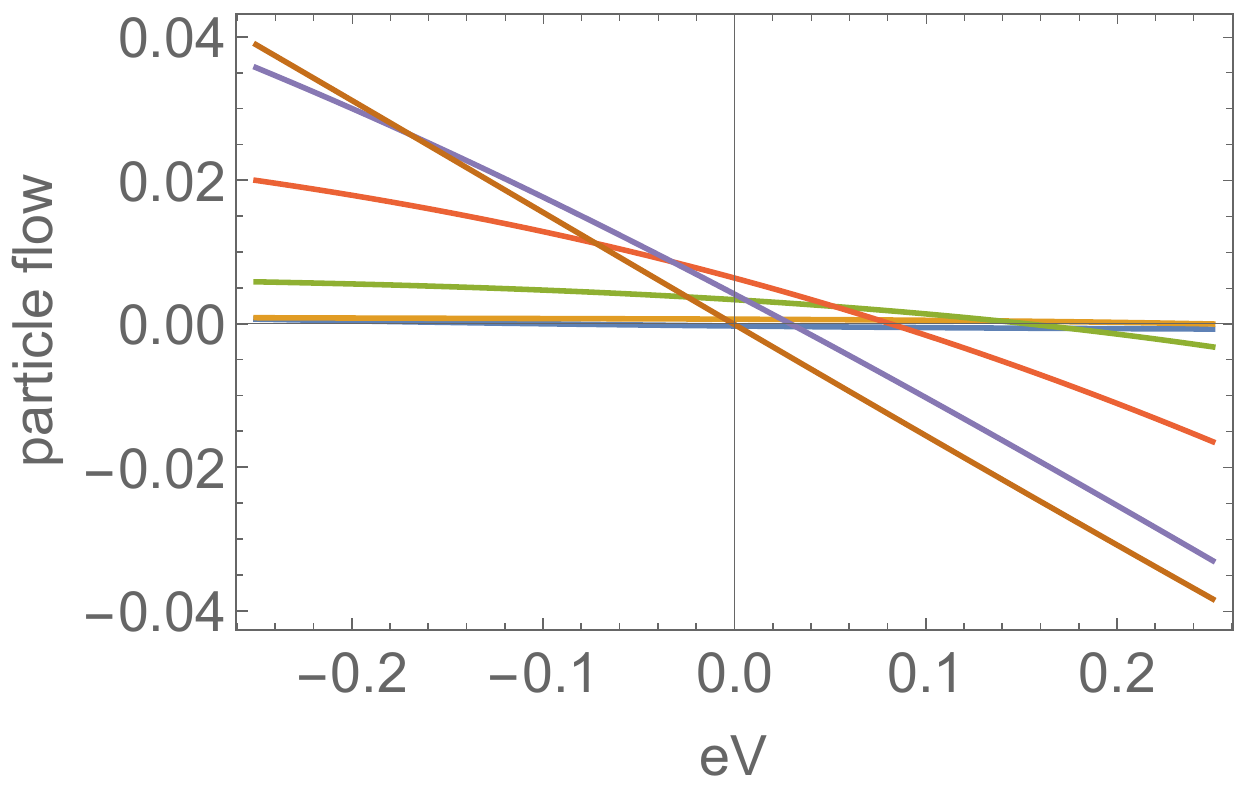}} \quad
\subfloat[Heat flow $J_{L}^{Q}$\label{fig:JV-Beta5-xi10-etaC50}]{
  \includegraphics[width=0.4\linewidth]{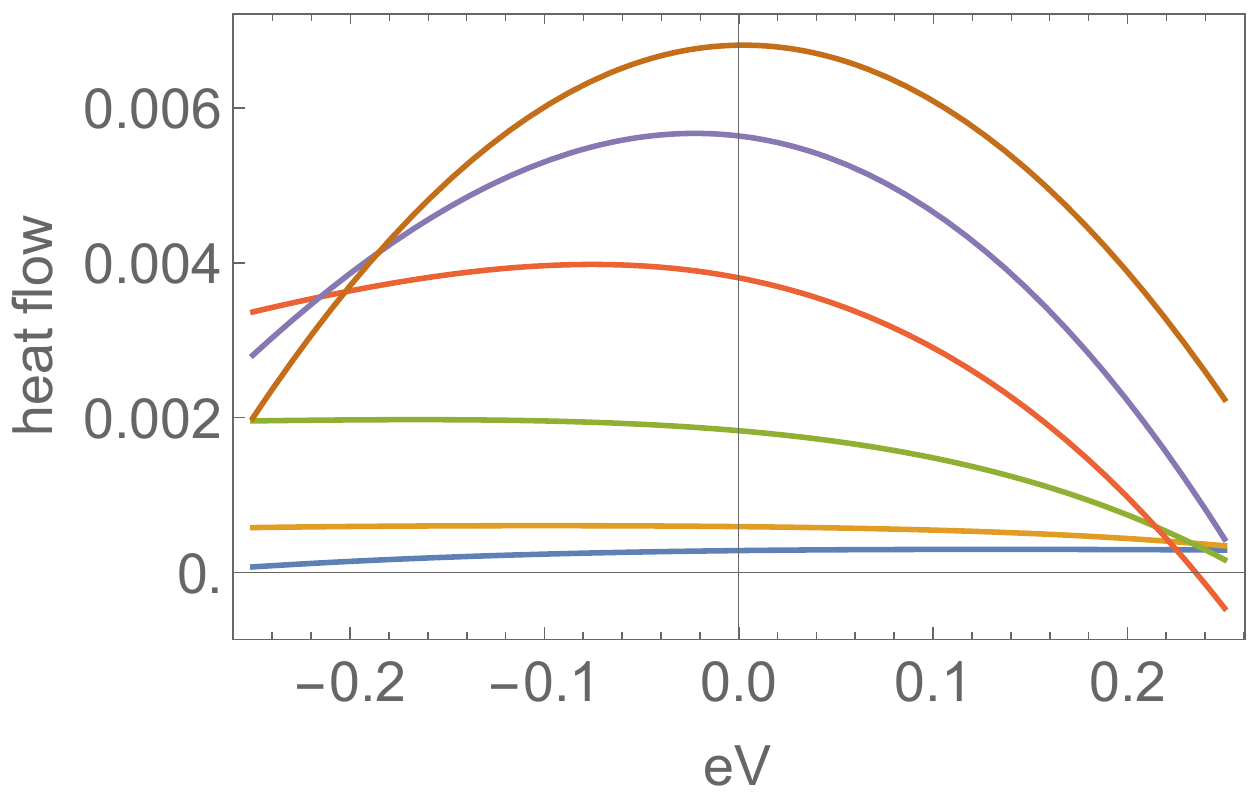}}
\caption{Nonlinear flows of the particle and the heat at $x=0.1$ as a
  function of bias voltage. Parameters are the same as in
  Fig.~\ref{fig:NL-eta-power-etaC50}. Different colored lines
  correspond to different values of the gate voltage from
  $\mu_{L}-\varepsilon_{d}=-2.0\gamma$ to $+0.5\gamma$.}
\label{fig:NL-flows-by-eV}
\end{figure}

To be concrete, we take the results of a noninteracting dot with
$x=0.01$ and $0.1$ (Fig.~\ref{fig:NL-eta-power-etaC50}), where we
evaluated nonlinear flows exactly.  Figure~\ref{fig:NL-flows-by-eV}
shows the corresponding bias voltage characteristic of the particle
and heat flows at $x=0.1$.  One finds that the particle flow depends almost
linearly while the heat flow, highly nonlinearly.  We intend to
exploit the former characteristic of the particle flow in asses
nonlinear thermoelectricity.

\subsection{Output power}

Since the particle flow is almost linear regarding the bias voltage,
we can use much of the linear response theory to investigate the
output power $\mathcal{P}$.  For instance, we immediately see that
$\mathcal{P}$ takes its maximum regarding the bias voltage around half
the stopping potential,
$\Delta \mu \equiv \mu_{R}-\mu_{L} = \Delta \mu_{\text{stop}}/2$ as in
the linear response theory.  What is more important in applications is
the location of the optimal gate voltage to maximize the power.  With
other parameters fixed, we can roughly estimate it by the
low-temperature expansion:
$\mathcal{P} \propto
[\mathcal{T}'(\bar{\mu})]^{2}/\mathcal{T}(\bar{\mu})$ (with
$\mu_{L}<\bar{\mu} < \mu_{R}$). For a real $q$, we can find an
analytical solution to maximize $\mathcal{P}$ at 
\begin{align}
& \bar{\mu} = E_{d} - \tilde{q}\, \Gamma; \quad \tilde{q} =
\tfrac{\sqrt{9+8q^{2}} -3}{4q}. 
\end{align}
By choosing $\bar{\mu}=\mu_{L}$, this provides an estimate of the
optimal gate voltage $\mu_{L}-\epsilon_{d} \approx -0.66 \gamma$ in
Fig.~\ref{fig:P-Beta5-xi01-etaC50}, or 
$\mu_{L}-\epsilon_{d} \approx -0.61\gamma$ in
Fig.~\ref{fig:P-Beta5-xi10-etaC50}.  We see that these estimates give a
reasonable agreement in both cases.

\subsection{Nonlinear efficiency}

We now consider how we find the optimal gate voltage to achieve
the highest nonlinear efficiency. As was discussed in
Sec.~\ref{sec:phenomenology}, we expect the enhancement for the
efficiency to occur near the Fano node $E_{d} - \Gamma \Re q$ (the
green dashed line in Fig.~\ref{fig:density-plot-linear}), which gives
$-5.0 \gamma$ for $x=0.01$ (Fig.~\ref{fig:eta-Beta5-xi01-etaC50}), or
$\mu-\epsilon_{d} \approx -1.58 \gamma$ for $x=0.1$
(Fig.~\ref{fig:eta-Beta5-xi10-etaC50}).  Since one can deduce the
value of $q$ from the observed transmission profile, this helps us
locate the optimal gate voltage.

\begin{figure}
  \centering
\subfloat[$\gamma_{L}=\gamma_{R}=0.5\gamma$
  \label{fig:S-Beta5-xi10-a100-etaC50}]{
  \includegraphics[width=0.45\linewidth]{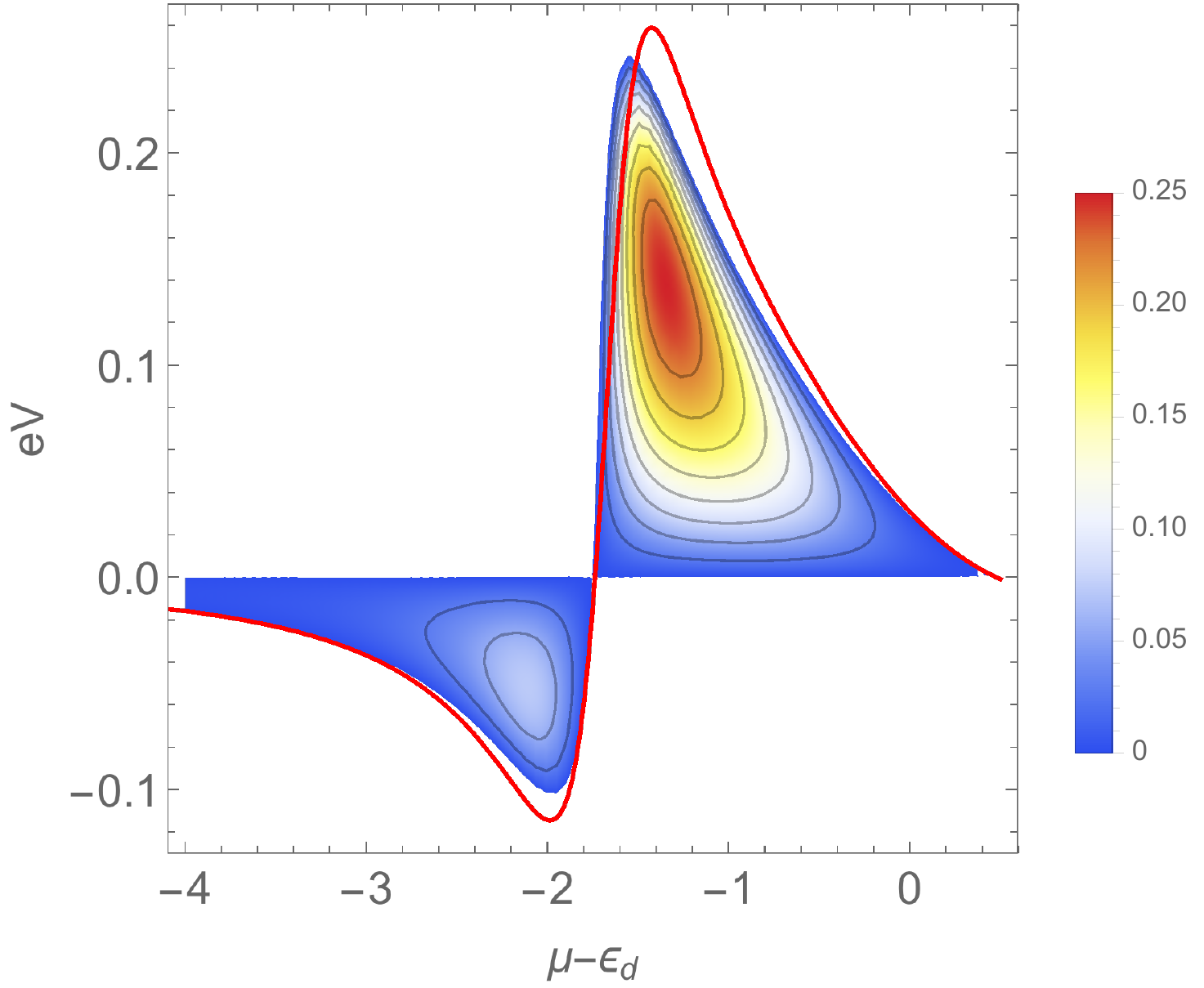}} \quad
\subfloat[$\gamma_{L}=0.05\gamma$, $\gamma_{R}=0.95\gamma$
  \label{fig:S-Beta5-xi10-a019-etaC50}]{
  \includegraphics[width=0.45\linewidth]{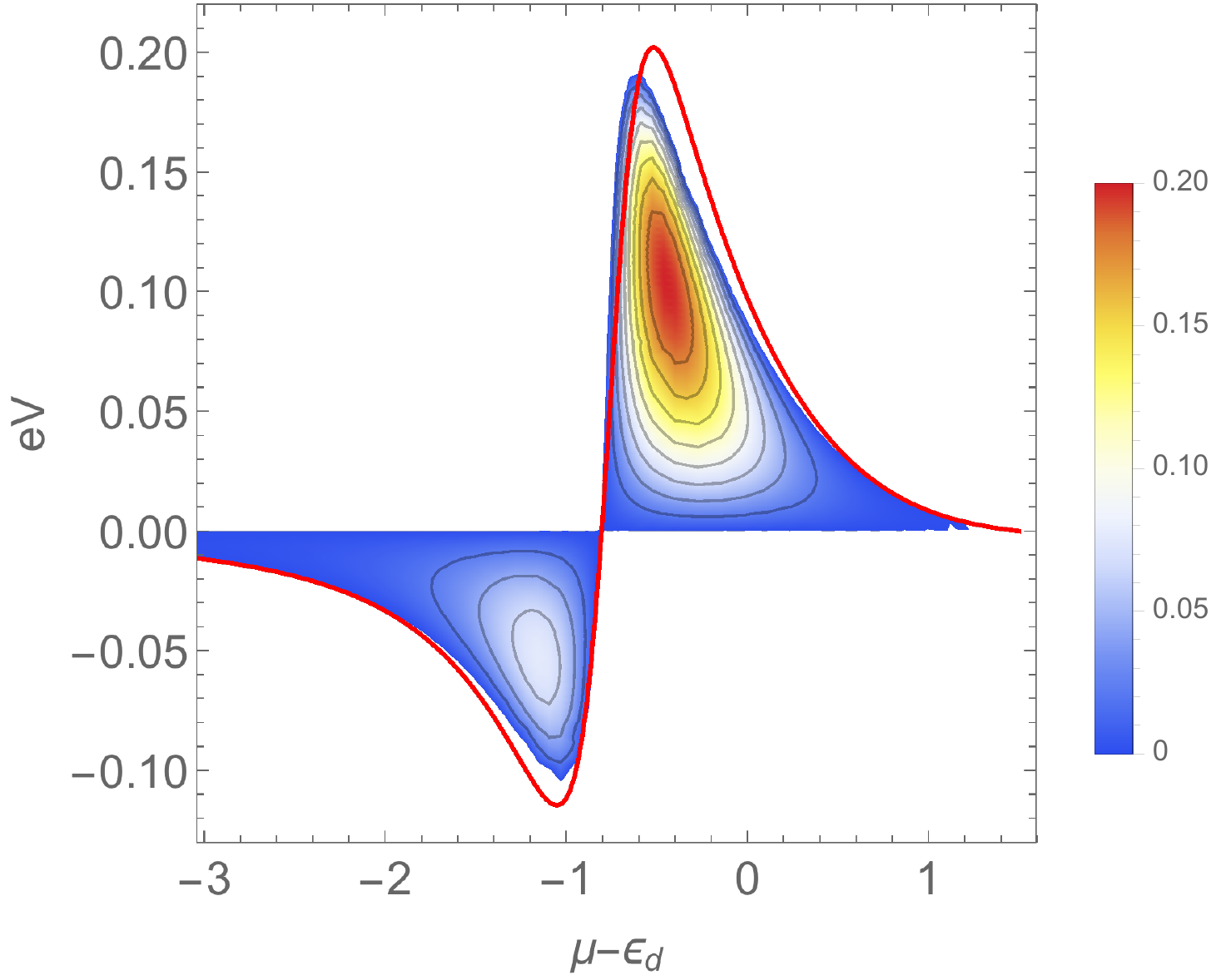}}
\caption{Comparison of the linear thermopower at $T_{\text{op}}$ with
  (a) nonlinear efficiency of Fig.~\ref{fig:eta-Beta5-xi10-etaC50}, and
  (b) the case where the dot-reservoir couplings are highly asymmetric.}
\label{fig:estimate-S}
\end{figure}

We can make a more quantitative argument and speculate about how it
connects with a linear response quantity by using the almost linear
characteristic of $I_{L}$.  As one notices immediately in
Figs.~\ref{fig:eta-Beta5-xi01-etaC50}
and~\ref{fig:eta-Beta5-xi10-etaC50}, the gate voltage optimal for
efficiency almost coincides with what maximize the boundary line.  The
latter boundary line defines the stopping potential
$\Delta \mu_{\text{stop}}$ when the particle flow vanishes.  This
implies that the (dimensionless) nonlinear thermopower
$\Delta \mu_{\text{stop}}/k\Delta T$ may well characterize the
nonlinear efficiency.  Furthermore, as was demonstrated numerically
for the single-impurity Anderson model~\cite{Azema14}, one can somehow
predict nonlinear thermopower well by using the linear-response
estimate at the ``operating temperature''
$T_{\text{op}} = (T_{L} + T_{R})/2$.  Therefore, we may conjecture
that we can use the linear-response estimate of the thermopower at
$T_{\text{op}}$ to assess the nonlinear efficiency.
Figure~\ref{fig:S-Beta5-xi10-a100-etaC50} supports this speculation:
the linear-response estimate of $\Delta \mu_{\text{stop}}$ at
$T_{\text{op}}$ (red line) is compared with the nonlinear efficiency
of Fig.~\ref{fig:eta-Beta5-xi10-etaC50}.  We see that the maximum of
the red line detects well the location of the gate voltage optimal for
efficiency.

One way to understand this unexpected role of $T_{\text{op}}$ is due
to a low-temperature expansion of $I_{L}$.  The lowest-order of the
temperature correction takes a form of
$ (\pi^{2} \mathcal{T}'(\bar{\mu}) k_{B}^{2} /6) (T_{L}^{2}-T_{R}^{2})
\propto T_{\text{op}} \Delta T$ (with
$\mu_{L} \le \bar{\mu} \le \mu_{R}$). The argument suggests that the
using $T_{\text{op}}$ may not be limited for the symmetric dot-bath
couplings case considered in Ref.~\cite{Azema14}.  In
Fig.~\ref{fig:S-Beta5-xi10-a019-etaC50}, we support the view by taking
a look at a dot with highly asymmetric couplings,
$\gamma_{L}=0.05\gamma$ and $\gamma_{R}=0.95\gamma$ with
$\alpha=0.19$.  In spite of high asymmetry, we see the linear-response
estimate at $T_{\text{op}}$ can tell reasonably well the location of
the gate voltage optimal for nonlinear efficiency.

\subsection{Power-efficiency diagram}

\begin{figure}
  \centering
\subfloat[$\eta_{C}=0.50$
  \label{fig:LR-P-eta-Beta5-etaC50}]{
  \includegraphics[width=0.45\linewidth]{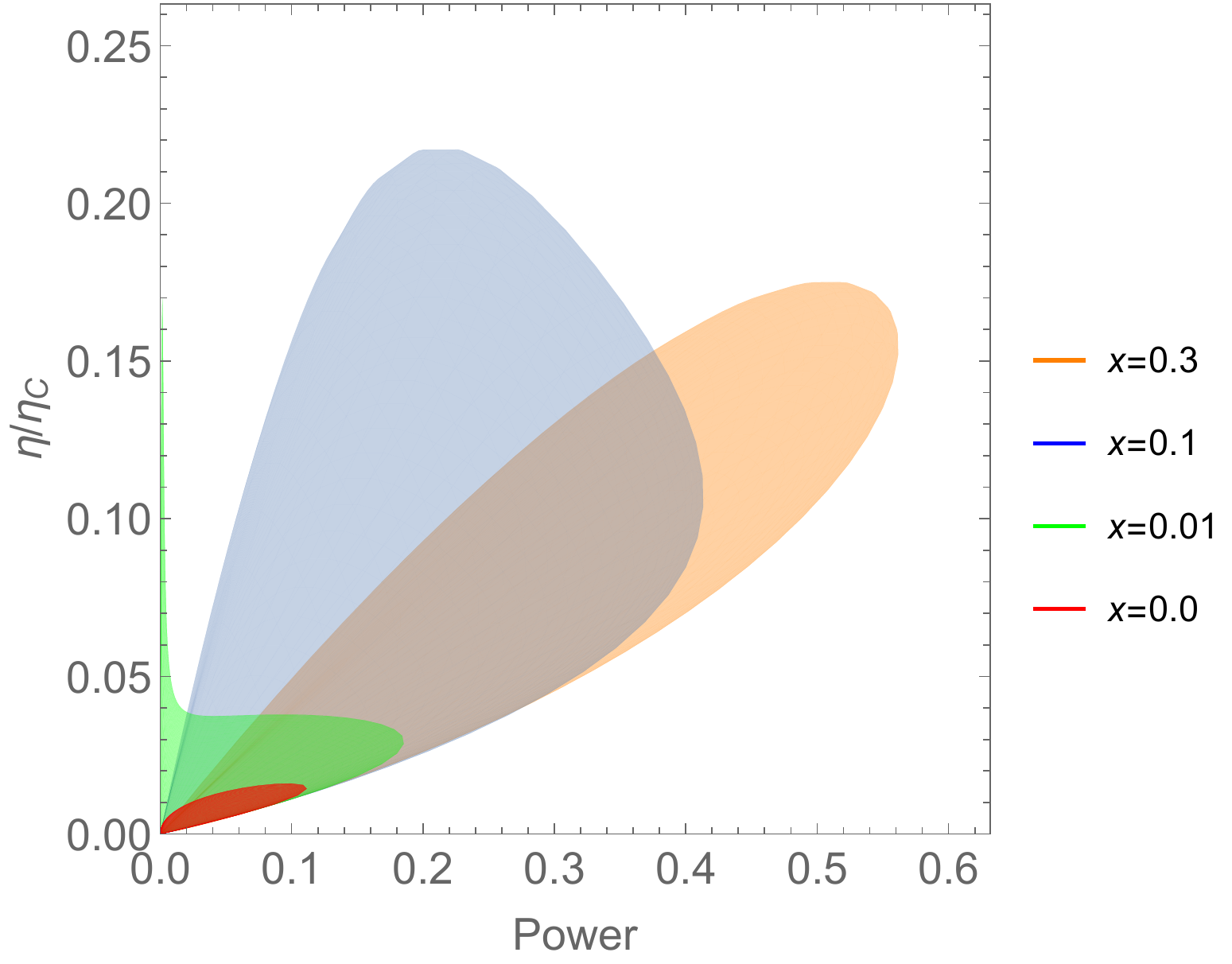}} \quad
\subfloat[$\eta_{C}=0.95$
  \label{fig:LR-P-eta-Beta5-etaC95}]{
  \includegraphics[width=0.45\linewidth]{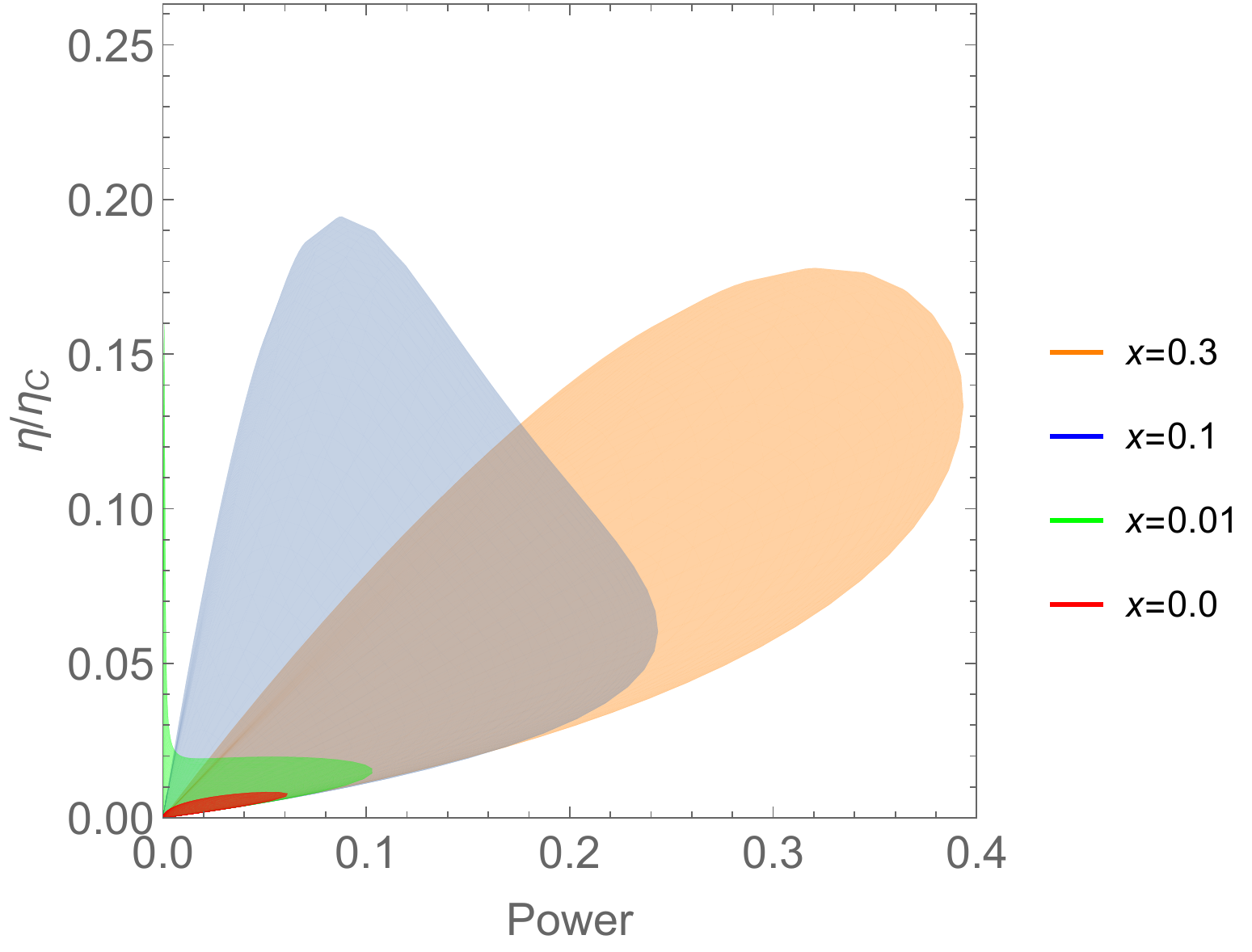}}
\caption{Linear-response estimate of the power-efficiency diagram
  using $T_{\text{op}}$, for (a) $\eta_{C}=0.50$ and (b)
  $\eta_{C}=0.95$. Parameters are chosen to correspond with
  Fig.~\ref{fig:P-eta-Beta5}.}
\label{fig:LR-P-eta-Beta5}
\end{figure}

Figure~\ref{fig:LR-P-eta-Beta5} illustrates how using the operating
temperature $T_{\text{op}}=T_{L}(1-\eta_{C}/2)$ helps improve a
linear-response estimate of the power-efficiency diagram for the
parameters chosen in Fig.~\ref{fig:P-eta-Beta5}: (a)
$k_{B}T_{L}=2 k_{B}T_{R} = 0.2\gamma$ (with $\eta_{C}=0.5$) and
$k_{B}T_{L}=20 k_{B}T_{R} = 0.2\gamma$ (with $\eta_{C}=0.95$).
Comparing with Fig.~\ref{fig:PL-etaL-Beta5} (with $\eta_{C}=0$), we
see the linear-response estimate with using $T_{\text{op}}$
(Fig.~\ref{fig:LR-P-eta-Beta5}) much improved in predicting the
behavior of Fig.~\ref{fig:P-eta-Beta5}, especially regarding the
output power. We find this way of the linear-response estimate of
nonlinear thermoelectricity useful for finding an optimal
setting of nonlinear thermoelectric performance, even in the fully
nonlinear regime $\eta_{C} = 0.95$. 

\section{Summary}
\label{sec:summary}

We have developed a theory for enhancing the thermoelectric
performance (efficiency and power) when a nanostructure acts as a heat
engine. We have demonstrated that even if the temperature is much
smaller than the resonant width, which is unfavorable for good
thermoelectricity, one can still achieve a reasonably good
thermoelectric performance by regulating quantum coherence via the
Fano effect.  Such thermoelectric enhancement stays effective in fully
nonlinear regimes.  We have shown that such thermoelectric enhancement
is viable in fully nonlinear responses.  With appropriate parameters,
the efficiency improves up to 10 times and the power, nearly 5 times
(Figs.~\ref{fig:P-eta-Beta5} and \ref{fig:P-eta-U-Beta5-etaC50}). We
have also estimated the optimal gate voltage that maximizes nonlinear
efficiency or power.  Furthermore, we argued the significance of the
linear thermopower at $T_{\text{op}}$ along assessing nonlinear
efficiency (Fig.~\ref{fig:LR-P-eta-Beta5}).  We believe quantum
control by the Fano effect is a promising, universal approach that
helps thermoelectric materials exhibit an even better thermoelectric
performance, as well as turn mediocre materials into thermoelectric.

\begin{acknowledgements}
  The author appreciates the help of S. Imai through helpful discussion. The author
  gratefully acknowledges financial support from JSPS KAKENHI Grant
  No.~JP19K03682.
\end{acknowledgements}



\appendix

\section{Integral formula}
\label{sec:integral}

We use the following integral formula to evaluate the energy integrals that
appear in nonlinear flows:
\begin{align}
& \int^{\infty}_{-\infty} \frac{d\varepsilon}{(\varepsilon - E_{d} +
  i\Gamma)[e^{\beta(\varepsilon - \mu)}+1]}
\notag \\ & \quad
=  \int^{\mu}_{-\infty} \frac{d\varepsilon}{\varepsilon - E_{d} +
  i\Gamma}
+ \psi(\tfrac{1}{2}+z) - \log z,
\label{eq:integral-formula}
\end{align}
where $\psi(\tfrac{1}{2}+z)$ is Euler's digamma function and the
parameter $z$ is defined in Eq.~\eqref{def:z-by-beta-mu}.  One can
derive Eq.~\eqref{eq:integral-formula} in various ways, \textit{e.g.},
by summing up the Sommerfeld expansion up to infinite order (see also
Ref.~\onlinecite{Taniguchi18} Appendix D).  The first term on the
right-hand side corresponds to the zero-temperature contribution. It
diverges logarithmically but will be canceled out in evaluating flows.

\section{Linear response quantities}
\label{sec:linear-response}

In this appendix, we collect the results of the linear-response
theory and connect them with the Onsager coefficients.  They can be
obtained by expanding our nonlinear results regarding small bias and
thermal gradient. 
Let us introduce the dimensionless Onsager coefficients
$\mathcal{K}_{n}$ (for $n=0,1,2$), which are defined by
\begin{align}
& \mathcal{K}_{n} = \beta^{n}  \int
d\varepsilon\, \mathcal{T} (\varepsilon) (\varepsilon -
\mu)^{n}\, [-f'(\varepsilon)].
\end{align}
Within the linear response theory, one can express flows of particle
and heat as
\begin{align}
& h\begin{pmatrix}  I_{L} \\ \beta J^{Q}_{L} \end{pmatrix}
= \begin{pmatrix}
\mathcal{K}_{0} &  \mathcal{K}_{1} \\ 
  \mathcal{K}_{1} & \mathcal{K}_{2}
\end{pmatrix}
\begin{pmatrix}
  \mu_{LR} \\ k_{B} T_{LR}
\end{pmatrix},
\end{align}
if $\mu_{LR} = \mu_{L} - \mu_{R}$ and $T_{LR} = T_{L} - T_{R}$ are
small.  We can readily evaluate $\mathcal{K}_{n}$ as
\begin{align}
& \mathcal{K}_{0} 
= \mathcal{T}_{0} + \frac{\beta \Gamma}{2\pi} \Re \left[
  \mathcal{T}_{q} \, \psi'(\tfrac{1}{2}+z) \right], \\
& \mathcal{K}_{1} = - \beta \Gamma \Im \left[
  \mathcal{T}_{q} \left( 1-z \psi'(\tfrac{1}{2}+z) \right) \right], \\
& \mathcal{K}_{2} 
= \frac{\pi^{2}}{3} \mathcal{T}_{0} + 2\pi \beta
\Gamma \Re \left[ \mathcal{T}_{q} \left( z-z^{2}
    \psi'(\tfrac{1}{2}+z) \right) \right]. 
\end{align}
We note that we straightforwardly restore the
results of the Sommerfeld expansion by seeing 
$\mathcal{K}_{0} \approx \mathcal{T}(\mu)$,
$\mathcal{K}_{1} \approx \frac{\pi^{2}}{3}(k_{B} T) \mathcal{T}'(\mu)$ and
$\mathcal{K}_{2} \approx \frac{\pi^{2}}{3} \mathcal{T}(\mu)$  at low
temperatures. 

In terms of these dimensionless Onsager coefficients
$\mathcal{K}_{n}$, standard linear-response quantities are given by
\begin{align}
& G = \frac{e^{2}}{2\pi \hbar} \mathcal{K}_{0}; &&
K = \frac{k_{B}^{2} T}{2\pi \hbar} \left( \mathcal{K}_{2} -
  \frac{\mathcal{K}_{1}^{2}}{\mathcal{K}_{0}} \right), \\
& S = -\frac{k_{B}}{e} \frac{\mathcal{K}_{1}}{\mathcal{K}_{0}}; &&
\Pi 
= -\frac{k_{B}T}{e} \frac{\mathcal{K}_{1}}{\mathcal{K}_{0}}. 
\end{align}
Therefore the figure of merit $ZT$ becomes
\begin{align}
& ZT = \frac{GT}{K} S^{2}
= \frac{\mathcal{K}_{1}^{2}}{\mathcal{K}_{0} \mathcal{K}_{2} -
  \mathcal{K}_{1}^{2}}.
\label{eq:ZT}
\end{align}

When one uses the above linear response theory with choosing
$\Delta \mu = \mu_{RL}>0$ and $\Delta T = T_{LR} > 0$, one finds the
stopping bias potential (or the open circuit potential) that makes the
particle flow vanish is given by 
\begin{align}
& \Delta \mu_{\text{stop}} = -e\, S
\Delta T = \frac{\mathcal{K}_{1}}{ \mathcal{K}_{0}} k_{B} \Delta T
\end{align}
According to Eq.~\eqref{def:P}, the output power $\mathcal{P}$ becomes
\begin{align}
&  \mathcal{P} = I \Delta \mu
= \left( - \mathcal{K}_{0} \Delta \mu + \mathcal{K}_{1} k_{B} \Delta T
\right) \cdot \frac{\Delta \mu}{h}, 
\end{align}
which we can maximize at $\Delta \mu = \Delta \mu_{\text{stop}}/2$ as
\begin{align}
& \mathcal{P}_{\max} = \frac{G S^{2}}{4} (\Delta T)^{2} 
= \frac{(k_{B}\Delta T)^{2}}{4h}\cdot 
\frac{\mathcal{K}_{1}^{2}}{\mathcal{K}_{0}}.
\label{eq:power-factor}
\end{align}
By introducing $v=\Delta \mu/\Delta \mu_{\text{stop}}$, one
can simply write the ratio as
$\mathcal{P}/\mathcal{P}_{\max} = 4v (1-v)$. 

The linear-response efficiency $\eta_{L}$ is also defined by
Eq.~\eqref{def:eta}.  After some manipulation, we can write it
as
\begin{align}
& \frac{\eta_{L}}{\eta_{C}} = \frac{v(1-v)}{1 + (ZT)^{-1}-v}. 
\end{align}
Accordingly, the efficiency at the maximal power ($v=1/2$) is equal to
\begin{align}
& \eta_{L} (\mathcal{P}_{\max}) = \frac{\eta_{C}}{2}\cdot \frac{ZT}{ZT+2}. 
\end{align}
We can maximize $\eta_{L}$ by changing $0<v<1$.  The maximal value
$\eta_{L}^{\max}$ is given by Eq.~\eqref{eq:linear-eta}, when one
chooses $v = ( 1+ \eta_{L}^{\max} /\eta_{C})/2$.


\bibliographystyle{apsrev4-2} 
\bibliography{../ref} 

\begin{thebibliography}{74}%
\makeatletter
\providecommand \@ifxundefined [1]{%
 \@ifx{#1\undefined}
}%
\providecommand \@ifnum [1]{%
 \ifnum #1\expandafter \@firstoftwo
 \else \expandafter \@secondoftwo
 \fi
}%
\providecommand \@ifx [1]{%
 \ifx #1\expandafter \@firstoftwo
 \else \expandafter \@secondoftwo
 \fi
}%
\providecommand \natexlab [1]{#1}%
\providecommand \enquote  [1]{``#1''}%
\providecommand \bibnamefont  [1]{#1}%
\providecommand \bibfnamefont [1]{#1}%
\providecommand \citenamefont [1]{#1}%
\providecommand \href@noop [0]{\@secondoftwo}%
\providecommand \href [0]{\begingroup \@sanitize@url \@href}%
\providecommand \@href[1]{\@@startlink{#1}\@@href}%
\providecommand \@@href[1]{\endgroup#1\@@endlink}%
\providecommand \@sanitize@url [0]{\catcode `\\12\catcode `\$12\catcode
  `\&12\catcode `\#12\catcode `\^12\catcode `\_12\catcode `\%12\relax}%
\providecommand \@@startlink[1]{}%
\providecommand \@@endlink[0]{}%
\providecommand \url  [0]{\begingroup\@sanitize@url \@url }%
\providecommand \@url [1]{\endgroup\@href {#1}{\urlprefix }}%
\providecommand \urlprefix  [0]{URL }%
\providecommand \Eprint [0]{\href }%
\providecommand \doibase [0]{https://doi.org/}%
\providecommand \selectlanguage [0]{\@gobble}%
\providecommand \bibinfo  [0]{\@secondoftwo}%
\providecommand \bibfield  [0]{\@secondoftwo}%
\providecommand \translation [1]{[#1]}%
\providecommand \BibitemOpen [0]{}%
\providecommand \bibitemStop [0]{}%
\providecommand \bibitemNoStop [0]{.\EOS\space}%
\providecommand \EOS [0]{\spacefactor3000\relax}%
\providecommand \BibitemShut  [1]{\csname bibitem#1\endcsname}%
\let\auto@bib@innerbib\@empty
\bibitem [{\citenamefont {Rowe}(2006)}]{RoweBook06}%
  \BibitemOpen
  \bibinfo {editor} {\bibfnamefont {D.~M.}\ \bibnamefont {Rowe}},\ ed.,\ \href
  {https://doi.org/10.1201/9781420038903} {\emph {\bibinfo {title}
  {Thermoelectrics Handbook, Macro to Nano}}}\ (\bibinfo  {publisher} {Taylor
  \& Francis},\ \bibinfo {address} {Boca Raton},\ \bibinfo {year}
  {2006})\BibitemShut {NoStop}%
\bibitem [{\citenamefont {Dubi}\ and\ \citenamefont
  {Di~Ventra}(2011)}]{Dubi11}%
  \BibitemOpen
  \bibfield  {author} {\bibinfo {author} {\bibfnamefont {Y.}~\bibnamefont
  {Dubi}}\ and\ \bibinfo {author} {\bibfnamefont {M.}~\bibnamefont
  {Di~Ventra}},\ }\href {https://doi.org/10.1103/RevModPhys.83.131} {\bibfield
  {journal} {\bibinfo  {journal} {Rev. Mod. Phys.}\ }\textbf {\bibinfo {volume}
  {83}},\ \bibinfo {pages} {131} (\bibinfo {year} {2011})}\BibitemShut
  {NoStop}%
\bibitem [{\citenamefont {Wang}\ and\ \citenamefont {Wang}(2014)}]{WangBook14}%
  \BibitemOpen
  \bibinfo {editor} {\bibfnamefont {X.}~\bibnamefont {Wang}}\ and\ \bibinfo
  {editor} {\bibfnamefont {Z.~M.}\ \bibnamefont {Wang}},\ eds.,\ \href
  {https://doi.org/10.1007/978-3-319-02012-9} {\emph {\bibinfo {title}
  {Nanoscale Thermoelectrics}}},\ \bibinfo {series} {Lecture Notes in Nanoscale
  Science and Technology}, Vol.~\bibinfo {volume} {16}\ (\bibinfo  {publisher}
  {Springer},\ \bibinfo {address} {Cham},\ \bibinfo {year} {2014})\BibitemShut
  {NoStop}%
\bibitem [{\citenamefont {Zlati\'{c}}\ and\ \citenamefont
  {Monnier}(2014)}]{ZlaticBook14}%
  \BibitemOpen
  \bibfield  {author} {\bibinfo {author} {\bibfnamefont {V.}~\bibnamefont
  {Zlati\'{c}}}\ and\ \bibinfo {author} {\bibfnamefont {R.}~\bibnamefont
  {Monnier}},\ }\href
  {https://doi.org/10.1093/acprof:oso/9780198705413.001.0001} {\emph {\bibinfo
  {title} {Modern Theory of Thermoelectricity}}}\ (\bibinfo  {publisher}
  {Oxford University Press},\ \bibinfo {address} {Oxford},\ \bibinfo {year}
  {2014})\BibitemShut {NoStop}%
\bibitem [{\citenamefont {Goldsmid}(2016)}]{GoldsmidBook16}%
  \BibitemOpen
  \bibfield  {author} {\bibinfo {author} {\bibfnamefont {H.~J.}\ \bibnamefont
  {Goldsmid}},\ }\href {https://doi.org/10.1007/978-3-642-00716-3} {\emph
  {\bibinfo {title} {Introduction to Thermoelectricity}}},\ \bibinfo {series}
  {Springer Series in Materials Science}, Vol.\ \bibinfo {volume} {121}\
  (\bibinfo  {publisher} {Springer},\ \bibinfo {address} {Berlin},\ \bibinfo
  {year} {2016})\BibitemShut {NoStop}%
\bibitem [{\citenamefont {Majumdar}(2004)}]{Majumdar04}%
  \BibitemOpen
  \bibfield  {author} {\bibinfo {author} {\bibfnamefont {A.}~\bibnamefont
  {Majumdar}},\ }\href {https://doi.org/10.1126/science.1093164} {\bibfield
  {journal} {\bibinfo  {journal} {Science}\ }\textbf {\bibinfo {volume}
  {303}},\ \bibinfo {pages} {777} (\bibinfo {year} {2004})}\BibitemShut
  {NoStop}%
\bibitem [{\citenamefont {Vining}(2009)}]{Vining09}%
  \BibitemOpen
  \bibfield  {author} {\bibinfo {author} {\bibfnamefont {C.~B.}\ \bibnamefont
  {Vining}},\ }\href {https://doi.org/10.1038/nmat2361} {\bibfield  {journal}
  {\bibinfo  {journal} {Nature Materials}\ }\textbf {\bibinfo {volume} {8}},\
  \bibinfo {pages} {83} (\bibinfo {year} {2009})}\BibitemShut {NoStop}%
\bibitem [{\citenamefont {He}\ and\ \citenamefont {Tritt}(2017)}]{He17b}%
  \BibitemOpen
  \bibfield  {author} {\bibinfo {author} {\bibfnamefont {J.}~\bibnamefont
  {He}}\ and\ \bibinfo {author} {\bibfnamefont {T.~M.}\ \bibnamefont {Tritt}},\
  }\href {https://doi.org/10.1126/science.aak9997} {\bibfield  {journal}
  {\bibinfo  {journal} {Science}\ }\textbf {\bibinfo {volume} {357}},\ \bibinfo
  {pages} {eaak9997} (\bibinfo {year} {2017})}\BibitemShut {NoStop}%
\bibitem [{\citenamefont {Dresselhaus}\ \emph {et~al.}(2007)\citenamefont
  {Dresselhaus}, \citenamefont {Chen}, \citenamefont {Tang}, \citenamefont
  {Yang}, \citenamefont {Lee}, \citenamefont {Wang}, \citenamefont {Ren},
  \citenamefont {Fleurial},\ and\ \citenamefont {Gogna}}]{Dresselhaus07}%
  \BibitemOpen
  \bibfield  {author} {\bibinfo {author} {\bibfnamefont {M.~S.}\ \bibnamefont
  {Dresselhaus}}, \bibinfo {author} {\bibfnamefont {G.}~\bibnamefont {Chen}},
  \bibinfo {author} {\bibfnamefont {M.~Y.}\ \bibnamefont {Tang}}, \bibinfo
  {author} {\bibfnamefont {R.~G.}\ \bibnamefont {Yang}}, \bibinfo {author}
  {\bibfnamefont {H.}~\bibnamefont {Lee}}, \bibinfo {author} {\bibfnamefont
  {D.~Z.}\ \bibnamefont {Wang}}, \bibinfo {author} {\bibfnamefont {Z.~F.}\
  \bibnamefont {Ren}}, \bibinfo {author} {\bibfnamefont {J.~P.}\ \bibnamefont
  {Fleurial}},\ and\ \bibinfo {author} {\bibfnamefont {P.}~\bibnamefont
  {Gogna}},\ }\href {https://doi.org/10.1002/adma.200600527} {\bibfield
  {journal} {\bibinfo  {journal} {Advanced Materials}\ }\textbf {\bibinfo
  {volume} {19}},\ \bibinfo {pages} {1043} (\bibinfo {year}
  {2007})}\BibitemShut {NoStop}%
\bibitem [{\citenamefont {Humphrey}\ \emph {et~al.}(2002)\citenamefont
  {Humphrey}, \citenamefont {Newbury}, \citenamefont {Taylor},\ and\
  \citenamefont {Linke}}]{Humphrey02}%
  \BibitemOpen
  \bibfield  {author} {\bibinfo {author} {\bibfnamefont {T.~E.}\ \bibnamefont
  {Humphrey}}, \bibinfo {author} {\bibfnamefont {R.}~\bibnamefont {Newbury}},
  \bibinfo {author} {\bibfnamefont {R.~P.}\ \bibnamefont {Taylor}},\ and\
  \bibinfo {author} {\bibfnamefont {H.}~\bibnamefont {Linke}},\ }\href
  {https://doi.org/10.1103/PhysRevLett.89.116801} {\bibfield  {journal}
  {\bibinfo  {journal} {Phys. Rev. Lett.}\ }\textbf {\bibinfo {volume} {89}},\
  \bibinfo {pages} {116801} (\bibinfo {year} {2002})}\BibitemShut {NoStop}%
\bibitem [{\citenamefont {Zhou}\ \emph {et~al.}(2015)\citenamefont {Zhou},
  \citenamefont {Yang},\ and\ \citenamefont {Zhang}}]{Zhou15}%
  \BibitemOpen
  \bibfield  {author} {\bibinfo {author} {\bibfnamefont {J.}~\bibnamefont
  {Zhou}}, \bibinfo {author} {\bibfnamefont {Y.}~\bibnamefont {Yang}},\ and\
  \bibinfo {author} {\bibfnamefont {C.-y.}\ \bibnamefont {Zhang}},\ }\href
  {https://doi.org/10.1021/acs.chemrev.5b00049} {\bibfield  {journal} {\bibinfo
   {journal} {Chemical Reviews}\ }\textbf {\bibinfo {volume} {115}},\ \bibinfo
  {pages} {11669} (\bibinfo {year} {2015})}\BibitemShut {NoStop}%
\bibitem [{\citenamefont {Hicks}\ and\ \citenamefont
  {Dresselhaus}(1993{\natexlab{a}})}]{Hicks93}%
  \BibitemOpen
  \bibfield  {author} {\bibinfo {author} {\bibfnamefont {L.~D.}\ \bibnamefont
  {Hicks}}\ and\ \bibinfo {author} {\bibfnamefont {M.~S.}\ \bibnamefont
  {Dresselhaus}},\ }\href {https://doi.org/10.1103/PhysRevB.47.12727}
  {\bibfield  {journal} {\bibinfo  {journal} {Phys. Rev. B}\ }\textbf {\bibinfo
  {volume} {47}},\ \bibinfo {pages} {12727} (\bibinfo {year}
  {1993}{\natexlab{a}})}\BibitemShut {NoStop}%
\bibitem [{\citenamefont {Hicks}\ and\ \citenamefont
  {Dresselhaus}(1993{\natexlab{b}})}]{Hicks93b}%
  \BibitemOpen
  \bibfield  {author} {\bibinfo {author} {\bibfnamefont {L.~D.}\ \bibnamefont
  {Hicks}}\ and\ \bibinfo {author} {\bibfnamefont {M.~S.}\ \bibnamefont
  {Dresselhaus}},\ }\href {https://doi.org/10.1103/PhysRevB.47.16631}
  {\bibfield  {journal} {\bibinfo  {journal} {Phys. Rev. B}\ }\textbf {\bibinfo
  {volume} {47}},\ \bibinfo {pages} {16631} (\bibinfo {year}
  {1993}{\natexlab{b}})}\BibitemShut {NoStop}%
\bibitem [{\citenamefont {Mahan}\ and\ \citenamefont {Sofo}(1996)}]{Mahan96}%
  \BibitemOpen
  \bibfield  {author} {\bibinfo {author} {\bibfnamefont {G.~D.}\ \bibnamefont
  {Mahan}}\ and\ \bibinfo {author} {\bibfnamefont {J.~O.}\ \bibnamefont
  {Sofo}},\ }\href {https://doi.org/10.1073/pnas.93.15.7436} {\bibfield
  {journal} {\bibinfo  {journal} {Proceedings of the National Academy of
  Sciences}\ }\textbf {\bibinfo {volume} {93}},\ \bibinfo {pages} {7436}
  (\bibinfo {year} {1996})}\BibitemShut {NoStop}%
\bibitem [{\citenamefont {Finch}\ \emph {et~al.}(2009)\citenamefont {Finch},
  \citenamefont {Garc\'{\i}a-Su\'{a}rez},\ and\ \citenamefont
  {Lambert}}]{Finch09}%
  \BibitemOpen
  \bibfield  {author} {\bibinfo {author} {\bibfnamefont {C.~M.}\ \bibnamefont
  {Finch}}, \bibinfo {author} {\bibfnamefont {V.~M.}\ \bibnamefont
  {Garc\'{\i}a-Su\'{a}rez}},\ and\ \bibinfo {author} {\bibfnamefont {C.~J.}\
  \bibnamefont {Lambert}},\ }\href {https://doi.org/10.1103/PhysRevB.79.033405}
  {\bibfield  {journal} {\bibinfo  {journal} {Phys. Rev. B}\ }\textbf {\bibinfo
  {volume} {79}},\ \bibinfo {pages} {033405} (\bibinfo {year}
  {2009})}\BibitemShut {NoStop}%
\bibitem [{\citenamefont {G\'{o}mez-Silva}\ \emph {et~al.}(2012)\citenamefont
  {G\'{o}mez-Silva}, \citenamefont {\'{A}valos Ovando}, \citenamefont
  {Ladr\'{o}n~de Guevara},\ and\ \citenamefont {Orellana}}]{Gomez-Silva12}%
  \BibitemOpen
  \bibfield  {author} {\bibinfo {author} {\bibfnamefont {G.}~\bibnamefont
  {G\'{o}mez-Silva}}, \bibinfo {author} {\bibfnamefont {O.}~\bibnamefont
  {\'{A}valos Ovando}}, \bibinfo {author} {\bibfnamefont {M.~L.}\ \bibnamefont
  {Ladr\'{o}n~de Guevara}},\ and\ \bibinfo {author} {\bibfnamefont {P.~A.}\
  \bibnamefont {Orellana}},\ }\href {https://doi.org/10.1063/1.3689817}
  {\bibfield  {journal} {\bibinfo  {journal} {Journal of Applied Physics}\
  }\textbf {\bibinfo {volume} {111}},\ \bibinfo {pages} {053704} (\bibinfo
  {year} {2012})}\BibitemShut {NoStop}%
\bibitem [{\citenamefont {Trocha}\ and\ \citenamefont
  {Barna\'{s}}(2012)}]{Trocha12}%
  \BibitemOpen
  \bibfield  {author} {\bibinfo {author} {\bibfnamefont {P.}~\bibnamefont
  {Trocha}}\ and\ \bibinfo {author} {\bibfnamefont {J.}~\bibnamefont
  {Barna\'{s}}},\ }\href {https://doi.org/10.1103/PhysRevB.85.085408}
  {\bibfield  {journal} {\bibinfo  {journal} {Phys. Rev. B}\ }\textbf {\bibinfo
  {volume} {85}},\ \bibinfo {pages} {085408} (\bibinfo {year}
  {2012})}\BibitemShut {NoStop}%
\bibitem [{\citenamefont {Garc\'{\i}a-Su\'{a}rez}\ \emph
  {et~al.}(2013)\citenamefont {Garc\'{\i}a-Su\'{a}rez}, \citenamefont
  {Ferrad\'{a}s},\ and\ \citenamefont {Ferrer}}]{Garcia-Suarez13}%
  \BibitemOpen
  \bibfield  {author} {\bibinfo {author} {\bibfnamefont {V.~M.}\ \bibnamefont
  {Garc\'{\i}a-Su\'{a}rez}}, \bibinfo {author} {\bibfnamefont {R.}~\bibnamefont
  {Ferrad\'{a}s}},\ and\ \bibinfo {author} {\bibfnamefont {J.}~\bibnamefont
  {Ferrer}},\ }\href {https://doi.org/10.1103/PhysRevB.88.235417} {\bibfield
  {journal} {\bibinfo  {journal} {Phys. Rev. B}\ }\textbf {\bibinfo {volume}
  {88}},\ \bibinfo {pages} {235417} (\bibinfo {year} {2013})}\BibitemShut
  {NoStop}%
\bibitem [{\citenamefont {Bevilacqua}\ \emph {et~al.}(2016)\citenamefont
  {Bevilacqua}, \citenamefont {Grosso}, \citenamefont {Menichetti},\ and\
  \citenamefont {Pastori~Parravicini}}]{Bevilacqua16}%
  \BibitemOpen
  \bibfield  {author} {\bibinfo {author} {\bibfnamefont {G.}~\bibnamefont
  {Bevilacqua}}, \bibinfo {author} {\bibfnamefont {G.}~\bibnamefont {Grosso}},
  \bibinfo {author} {\bibfnamefont {G.}~\bibnamefont {Menichetti}},\ and\
  \bibinfo {author} {\bibfnamefont {G.}~\bibnamefont {Pastori~Parravicini}},\
  }\href {https://doi.org/10.1103/PhysRevB.94.245419} {\bibfield  {journal}
  {\bibinfo  {journal} {Phys. Rev. B}\ }\textbf {\bibinfo {volume} {94}},\
  \bibinfo {pages} {245419} (\bibinfo {year} {2016})}\BibitemShut {NoStop}%
\bibitem [{\citenamefont {W\'{o}jcik}\ and\ \citenamefont
  {Weymann}(2016)}]{Wojcik16}%
  \BibitemOpen
  \bibfield  {author} {\bibinfo {author} {\bibfnamefont {K.~P.}\ \bibnamefont
  {W\'{o}jcik}}\ and\ \bibinfo {author} {\bibfnamefont {I.}~\bibnamefont
  {Weymann}},\ }\href {https://doi.org/10.1103/PhysRevB.93.085428} {\bibfield
  {journal} {\bibinfo  {journal} {Phys. Rev. B}\ }\textbf {\bibinfo {volume}
  {93}},\ \bibinfo {pages} {085428} (\bibinfo {year} {2016})}\BibitemShut
  {NoStop}%
\bibitem [{\citenamefont {Menichetti}\ \emph {et~al.}(2018)\citenamefont
  {Menichetti}, \citenamefont {Grosso},\ and\ \citenamefont
  {Parravicini}}]{Menichetti18}%
  \BibitemOpen
  \bibfield  {author} {\bibinfo {author} {\bibfnamefont {G.}~\bibnamefont
  {Menichetti}}, \bibinfo {author} {\bibfnamefont {G.}~\bibnamefont {Grosso}},\
  and\ \bibinfo {author} {\bibfnamefont {G.~P.}\ \bibnamefont {Parravicini}},\
  }\href {https://doi.org/10.1088/2399-6528/aac423} {\bibfield  {journal}
  {\bibinfo  {journal} {Journal of Physics Communications}\ }\textbf {\bibinfo
  {volume} {2}},\ \bibinfo {pages} {055026} (\bibinfo {year}
  {2018})}\BibitemShut {NoStop}%
\bibitem [{\citenamefont {Cui}\ \emph {et~al.}(2017)\citenamefont {Cui},
  \citenamefont {Miao}, \citenamefont {Jiang}, \citenamefont {Meyhofer},\ and\
  \citenamefont {Reddy}}]{Cui17b}%
  \BibitemOpen
  \bibfield  {author} {\bibinfo {author} {\bibfnamefont {L.}~\bibnamefont
  {Cui}}, \bibinfo {author} {\bibfnamefont {R.}~\bibnamefont {Miao}}, \bibinfo
  {author} {\bibfnamefont {C.}~\bibnamefont {Jiang}}, \bibinfo {author}
  {\bibfnamefont {E.}~\bibnamefont {Meyhofer}},\ and\ \bibinfo {author}
  {\bibfnamefont {P.}~\bibnamefont {Reddy}},\ }\href
  {https://doi.org/10.1063/1.4976982} {\bibfield  {journal} {\bibinfo
  {journal} {The Journal of Chemical Physics}\ }\textbf {\bibinfo {volume}
  {146}},\ \bibinfo {pages} {092201} (\bibinfo {year} {2017})}\BibitemShut
  {NoStop}%
\bibitem [{\citenamefont {Karlstr\"{o}m}\ \emph {et~al.}(2011)\citenamefont
  {Karlstr\"{o}m}, \citenamefont {Linke}, \citenamefont {Karlstr\"{o}m},\ and\
  \citenamefont {Wacker}}]{Karlstrom11}%
  \BibitemOpen
  \bibfield  {author} {\bibinfo {author} {\bibfnamefont {O.}~\bibnamefont
  {Karlstr\"{o}m}}, \bibinfo {author} {\bibfnamefont {H.}~\bibnamefont
  {Linke}}, \bibinfo {author} {\bibfnamefont {G.}~\bibnamefont
  {Karlstr\"{o}m}},\ and\ \bibinfo {author} {\bibfnamefont {A.}~\bibnamefont
  {Wacker}},\ }\href {https://doi.org/10.1103/PhysRevB.84.113415} {\bibfield
  {journal} {\bibinfo  {journal} {Phys. Rev. B}\ }\textbf {\bibinfo {volume}
  {84}},\ \bibinfo {pages} {113415} (\bibinfo {year} {2011})}\BibitemShut
  {NoStop}%
\bibitem [{\citenamefont {Lambert}\ \emph {et~al.}(2016)\citenamefont
  {Lambert}, \citenamefont {Sadeghi},\ and\ \citenamefont
  {Al-Galiby}}]{Lambert16}%
  \BibitemOpen
  \bibfield  {author} {\bibinfo {author} {\bibfnamefont {C.~J.}\ \bibnamefont
  {Lambert}}, \bibinfo {author} {\bibfnamefont {H.}~\bibnamefont {Sadeghi}},\
  and\ \bibinfo {author} {\bibfnamefont {Q.~H.}\ \bibnamefont {Al-Galiby}},\
  }\href {https://doi.org/10.1016/j.crhy.2016.08.003} {\bibfield  {journal}
  {\bibinfo  {journal} {Comptes Rendus Physique}\ }\textbf {\bibinfo {volume}
  {17}},\ \bibinfo {pages} {1084} (\bibinfo {year} {2016})}\BibitemShut
  {NoStop}%
\bibitem [{\citenamefont {Bergfield}\ and\ \citenamefont
  {Stafford}(2009)}]{Bergfield09b}%
  \BibitemOpen
  \bibfield  {author} {\bibinfo {author} {\bibfnamefont {J.~P.}\ \bibnamefont
  {Bergfield}}\ and\ \bibinfo {author} {\bibfnamefont {C.~A.}\ \bibnamefont
  {Stafford}},\ }\href {https://doi.org/10.1021/nl901554s} {\bibfield
  {journal} {\bibinfo  {journal} {Nano Letters}\ }\textbf {\bibinfo {volume}
  {9}},\ \bibinfo {pages} {3072} (\bibinfo {year} {2009})}\BibitemShut
  {NoStop}%
\bibitem [{\citenamefont {Bergfield}\ \emph {et~al.}(2010)\citenamefont
  {Bergfield}, \citenamefont {Solis},\ and\ \citenamefont
  {Stafford}}]{Bergfield10}%
  \BibitemOpen
  \bibfield  {author} {\bibinfo {author} {\bibfnamefont {J.~P.}\ \bibnamefont
  {Bergfield}}, \bibinfo {author} {\bibfnamefont {M.~A.}\ \bibnamefont
  {Solis}},\ and\ \bibinfo {author} {\bibfnamefont {C.~A.}\ \bibnamefont
  {Stafford}},\ }\href {https://doi.org/10.1021/nn100490g} {\bibfield
  {journal} {\bibinfo  {journal} {ACS Nano}\ }\textbf {\bibinfo {volume} {4}},\
  \bibinfo {pages} {5314} (\bibinfo {year} {2010})}\BibitemShut {NoStop}%
\bibitem [{\citenamefont {Abbout}\ \emph {et~al.}(2013)\citenamefont {Abbout},
  \citenamefont {Ouerdane},\ and\ \citenamefont {Goupil}}]{Abbout13}%
  \BibitemOpen
  \bibfield  {author} {\bibinfo {author} {\bibfnamefont {A.}~\bibnamefont
  {Abbout}}, \bibinfo {author} {\bibfnamefont {H.}~\bibnamefont {Ouerdane}},\
  and\ \bibinfo {author} {\bibfnamefont {C.}~\bibnamefont {Goupil}},\ }\href
  {https://doi.org/10.1103/PhysRevB.87.155410} {\bibfield  {journal} {\bibinfo
  {journal} {Phys. Rev. B}\ }\textbf {\bibinfo {volume} {87}},\ \bibinfo
  {pages} {155410} (\bibinfo {year} {2013})}\BibitemShut {NoStop}%
\bibitem [{\citenamefont {Yamamoto}\ \emph {et~al.}(2017)\citenamefont
  {Yamamoto}, \citenamefont {Aharony}, \citenamefont {Entin-Wohlman},\ and\
  \citenamefont {Hatano}}]{Yamamoto17}%
  \BibitemOpen
  \bibfield  {author} {\bibinfo {author} {\bibfnamefont {K.}~\bibnamefont
  {Yamamoto}}, \bibinfo {author} {\bibfnamefont {A.}~\bibnamefont {Aharony}},
  \bibinfo {author} {\bibfnamefont {O.}~\bibnamefont {Entin-Wohlman}},\ and\
  \bibinfo {author} {\bibfnamefont {N.}~\bibnamefont {Hatano}},\ }\href
  {https://doi.org/10.1103/PhysRevB.96.155201} {\bibfield  {journal} {\bibinfo
  {journal} {Phys. Rev. B}\ }\textbf {\bibinfo {volume} {96}},\ \bibinfo
  {pages} {155201} (\bibinfo {year} {2017})},\ \Eprint
  {https://arxiv.org/abs/1707.08286} {1707.08286} \BibitemShut {NoStop}%
\bibitem [{\citenamefont {Meair}\ and\ \citenamefont
  {Jacquod}(2013)}]{Meair13}%
  \BibitemOpen
  \bibfield  {author} {\bibinfo {author} {\bibfnamefont {J.}~\bibnamefont
  {Meair}}\ and\ \bibinfo {author} {\bibfnamefont {P.}~\bibnamefont
  {Jacquod}},\ }\href {https://doi.org/10.1088/0953-8984/25/8/082201}
  {\bibfield  {journal} {\bibinfo  {journal} {Journal of Physics: Condensed
  Matter}\ }\textbf {\bibinfo {volume} {25}},\ \bibinfo {pages} {082201}
  (\bibinfo {year} {2013})}\BibitemShut {NoStop}%
\bibitem [{\citenamefont {Azema}\ \emph {et~al.}(2014)\citenamefont {Azema},
  \citenamefont {Lombardo},\ and\ \citenamefont {Dar\'{e}}}]{Azema14}%
  \BibitemOpen
  \bibfield  {author} {\bibinfo {author} {\bibfnamefont {J.}~\bibnamefont
  {Azema}}, \bibinfo {author} {\bibfnamefont {P.}~\bibnamefont {Lombardo}},\
  and\ \bibinfo {author} {\bibfnamefont {A.-M.}\ \bibnamefont {Dar\'{e}}},\
  }\href {https://doi.org/10.1103/PhysRevB.90.205437} {\bibfield  {journal}
  {\bibinfo  {journal} {Phys. Rev. B}\ }\textbf {\bibinfo {volume} {90}},\
  \bibinfo {pages} {205437} (\bibinfo {year} {2014})}\BibitemShut {NoStop}%
\bibitem [{\citenamefont {Miroshnichenko}\ \emph {et~al.}(2010)\citenamefont
  {Miroshnichenko}, \citenamefont {Flach},\ and\ \citenamefont
  {Kivshar}}]{Miroshnichenko10}%
  \BibitemOpen
  \bibfield  {author} {\bibinfo {author} {\bibfnamefont {A.~E.}\ \bibnamefont
  {Miroshnichenko}}, \bibinfo {author} {\bibfnamefont {S.}~\bibnamefont
  {Flach}},\ and\ \bibinfo {author} {\bibfnamefont {Y.~S.}\ \bibnamefont
  {Kivshar}},\ }\href {https://doi.org/10.1103/RevModPhys.82.2257} {\bibfield
  {journal} {\bibinfo  {journal} {Rev. Mod. Phys.}\ }\textbf {\bibinfo {volume}
  {82}},\ \bibinfo {pages} {2257} (\bibinfo {year} {2010})}\BibitemShut
  {NoStop}%
\bibitem [{\citenamefont {G\"{o}res}\ \emph {et~al.}(2000)\citenamefont
  {G\"{o}res}, \citenamefont {Goldhaber-Gordon}, \citenamefont {Heemeyer},
  \citenamefont {Kastner}, \citenamefont {Shtrikman}, \citenamefont {Mahalu},\
  and\ \citenamefont {Meirav}}]{Gores00}%
  \BibitemOpen
  \bibfield  {author} {\bibinfo {author} {\bibfnamefont {J.}~\bibnamefont
  {G\"{o}res}}, \bibinfo {author} {\bibfnamefont {D.}~\bibnamefont
  {Goldhaber-Gordon}}, \bibinfo {author} {\bibfnamefont {S.}~\bibnamefont
  {Heemeyer}}, \bibinfo {author} {\bibfnamefont {M.~A.}\ \bibnamefont
  {Kastner}}, \bibinfo {author} {\bibfnamefont {H.}~\bibnamefont {Shtrikman}},
  \bibinfo {author} {\bibfnamefont {D.}~\bibnamefont {Mahalu}},\ and\ \bibinfo
  {author} {\bibfnamefont {U.}~\bibnamefont {Meirav}},\ }\href
  {https://doi.org/10.1103/PhysRevB.62.2188} {\bibfield  {journal} {\bibinfo
  {journal} {Phys. Rev. B}\ }\textbf {\bibinfo {volume} {62}},\ \bibinfo
  {pages} {2188} (\bibinfo {year} {2000})}\BibitemShut {NoStop}%
\bibitem [{\citenamefont {Kobayashi}\ \emph {et~al.}(2002)\citenamefont
  {Kobayashi}, \citenamefont {Aikawa}, \citenamefont {Katsumoto},\ and\
  \citenamefont {Iye}}]{Kobayashi02}%
  \BibitemOpen
  \bibfield  {author} {\bibinfo {author} {\bibfnamefont {K.}~\bibnamefont
  {Kobayashi}}, \bibinfo {author} {\bibfnamefont {H.}~\bibnamefont {Aikawa}},
  \bibinfo {author} {\bibfnamefont {S.}~\bibnamefont {Katsumoto}},\ and\
  \bibinfo {author} {\bibfnamefont {Y.}~\bibnamefont {Iye}},\ }\href
  {https://doi.org/10.1103/PhysRevLett.88.256806} {\bibfield  {journal}
  {\bibinfo  {journal} {Phys. Rev. Lett.}\ }\textbf {\bibinfo {volume} {88}},\
  \bibinfo {pages} {256806} (\bibinfo {year} {2002})}\BibitemShut {NoStop}%
\bibitem [{\citenamefont {Kobayashi}\ \emph {et~al.}(2003)\citenamefont
  {Kobayashi}, \citenamefont {Aikawa}, \citenamefont {Katsumoto},\ and\
  \citenamefont {Iye}}]{Kobayashi03}%
  \BibitemOpen
  \bibfield  {author} {\bibinfo {author} {\bibfnamefont {K.}~\bibnamefont
  {Kobayashi}}, \bibinfo {author} {\bibfnamefont {H.}~\bibnamefont {Aikawa}},
  \bibinfo {author} {\bibfnamefont {S.}~\bibnamefont {Katsumoto}},\ and\
  \bibinfo {author} {\bibfnamefont {Y.}~\bibnamefont {Iye}},\ }\href
  {https://doi.org/10.1103/PhysRevB.68.235304} {\bibfield  {journal} {\bibinfo
  {journal} {Phys. Rev. B}\ }\textbf {\bibinfo {volume} {68}},\ \bibinfo
  {pages} {235304} (\bibinfo {year} {2003})}\BibitemShut {NoStop}%
\bibitem [{\citenamefont {Johnson}\ \emph {et~al.}(2004)\citenamefont
  {Johnson}, \citenamefont {Marcus}, \citenamefont {Hanson},\ and\
  \citenamefont {Gossard}}]{Johnson04}%
  \BibitemOpen
  \bibfield  {author} {\bibinfo {author} {\bibfnamefont {A.~C.}\ \bibnamefont
  {Johnson}}, \bibinfo {author} {\bibfnamefont {C.~M.}\ \bibnamefont {Marcus}},
  \bibinfo {author} {\bibfnamefont {M.~P.}\ \bibnamefont {Hanson}},\ and\
  \bibinfo {author} {\bibfnamefont {A.~C.}\ \bibnamefont {Gossard}},\ }\href
  {https://doi.org/10.1103/PhysRevLett.93.106803} {\bibfield  {journal}
  {\bibinfo  {journal} {Phys. Rev. Lett.}\ }\textbf {\bibinfo {volume} {93}},\
  \bibinfo {pages} {106803} (\bibinfo {year} {2004})}\BibitemShut {NoStop}%
\bibitem [{\citenamefont {Papadopoulos}\ \emph {et~al.}(2006)\citenamefont
  {Papadopoulos}, \citenamefont {Grace},\ and\ \citenamefont
  {Lambert}}]{Papadopoulos06}%
  \BibitemOpen
  \bibfield  {author} {\bibinfo {author} {\bibfnamefont {T.~A.}\ \bibnamefont
  {Papadopoulos}}, \bibinfo {author} {\bibfnamefont {I.~M.}\ \bibnamefont
  {Grace}},\ and\ \bibinfo {author} {\bibfnamefont {C.~J.}\ \bibnamefont
  {Lambert}},\ }\href {https://doi.org/10.1103/PhysRevB.74.193306} {\bibfield
  {journal} {\bibinfo  {journal} {Phys. Rev. B}\ }\textbf {\bibinfo {volume}
  {74}},\ \bibinfo {pages} {193306} (\bibinfo {year} {2006})}\BibitemShut
  {NoStop}%
\bibitem [{\citenamefont {Kim}\ \emph {et~al.}(2003)\citenamefont {Kim},
  \citenamefont {Kim}, \citenamefont {Lee}, \citenamefont {Park}, \citenamefont
  {So}, \citenamefont {Kim}, \citenamefont {Kang}, \citenamefont {Yoo},\ and\
  \citenamefont {Kim}}]{Kim03b}%
  \BibitemOpen
  \bibfield  {author} {\bibinfo {author} {\bibfnamefont {J.}~\bibnamefont
  {Kim}}, \bibinfo {author} {\bibfnamefont {J.-R.}\ \bibnamefont {Kim}},
  \bibinfo {author} {\bibfnamefont {J.-O.}\ \bibnamefont {Lee}}, \bibinfo
  {author} {\bibfnamefont {J.~W.}\ \bibnamefont {Park}}, \bibinfo {author}
  {\bibfnamefont {H.~M.}\ \bibnamefont {So}}, \bibinfo {author} {\bibfnamefont
  {N.}~\bibnamefont {Kim}}, \bibinfo {author} {\bibfnamefont {K.}~\bibnamefont
  {Kang}}, \bibinfo {author} {\bibfnamefont {K.-H.}\ \bibnamefont {Yoo}},\ and\
  \bibinfo {author} {\bibfnamefont {J.-J.}\ \bibnamefont {Kim}},\ }\href
  {https://doi.org/10.1103/PhysRevLett.90.166403} {\bibfield  {journal}
  {\bibinfo  {journal} {Phys. Rev. Lett.}\ }\textbf {\bibinfo {volume} {90}},\
  \bibinfo {pages} {166403} (\bibinfo {year} {2003})}\BibitemShut {NoStop}%
\bibitem [{\citenamefont {Kim}\ \emph {et~al.}(2005)\citenamefont {Kim},
  \citenamefont {Lee}, \citenamefont {Kim},\ and\ \citenamefont {Ihm}}]{Kim05}%
  \BibitemOpen
  \bibfield  {author} {\bibinfo {author} {\bibfnamefont {G.}~\bibnamefont
  {Kim}}, \bibinfo {author} {\bibfnamefont {S.~B.}\ \bibnamefont {Lee}},
  \bibinfo {author} {\bibfnamefont {T.-S.}\ \bibnamefont {Kim}},\ and\ \bibinfo
  {author} {\bibfnamefont {J.}~\bibnamefont {Ihm}},\ }\href
  {https://doi.org/10.1103/PhysRevB.71.205415} {\bibfield  {journal} {\bibinfo
  {journal} {Phys. Rev. B}\ }\textbf {\bibinfo {volume} {71}},\ \bibinfo
  {pages} {205415} (\bibinfo {year} {2005})}\BibitemShut {NoStop}%
\bibitem [{\citenamefont {Babi\'{c}}\ and\ \citenamefont
  {Sch\"{o}nenberger}(2004)}]{Babic04b}%
  \BibitemOpen
  \bibfield  {author} {\bibinfo {author} {\bibfnamefont {B.}~\bibnamefont
  {Babi\'{c}}}\ and\ \bibinfo {author} {\bibfnamefont {C.}~\bibnamefont
  {Sch\"{o}nenberger}},\ }\href {https://doi.org/10.1103/PhysRevB.70.195408}
  {\bibfield  {journal} {\bibinfo  {journal} {Phys. Rev. B}\ }\textbf {\bibinfo
  {volume} {70}},\ \bibinfo {pages} {195408} (\bibinfo {year}
  {2004})}\BibitemShut {NoStop}%
\bibitem [{\citenamefont {Gong}\ \emph {et~al.}(2013)\citenamefont {Gong},
  \citenamefont {Sui}, \citenamefont {Zhu}, \citenamefont {Yu},\ and\
  \citenamefont {Chen}}]{Gong13}%
  \BibitemOpen
  \bibfield  {author} {\bibinfo {author} {\bibfnamefont {W.~J.}\ \bibnamefont
  {Gong}}, \bibinfo {author} {\bibfnamefont {X.~Y.}\ \bibnamefont {Sui}},
  \bibinfo {author} {\bibfnamefont {L.}~\bibnamefont {Zhu}}, \bibinfo {author}
  {\bibfnamefont {G.~D.}\ \bibnamefont {Yu}},\ and\ \bibinfo {author}
  {\bibfnamefont {X.~H.}\ \bibnamefont {Chen}},\ }\href
  {https://doi.org/10.1209/0295-5075/103/18003} {\bibfield  {journal} {\bibinfo
   {journal} {EPL (Europhysics Letters)}\ }\textbf {\bibinfo {volume} {103}},\
  \bibinfo {pages} {18003} (\bibinfo {year} {2013})}\BibitemShut {NoStop}%
\bibitem [{\citenamefont {Briones-Torres}\ and\ \citenamefont
  {Rodr{\'\i}guez-Vargas}(2017)}]{Briones-Torres17}%
  \BibitemOpen
  \bibfield  {author} {\bibinfo {author} {\bibfnamefont {J.~A.}\ \bibnamefont
  {Briones-Torres}}\ and\ \bibinfo {author} {\bibfnamefont {I.}~\bibnamefont
  {Rodr{\'\i}guez-Vargas}},\ }\href
  {https://doi.org/10.1038/s41598-017-16838-9} {\bibfield  {journal} {\bibinfo
  {journal} {Scientific Reports}\ }\textbf {\bibinfo {volume} {7}},\ \bibinfo
  {pages} {16708} (\bibinfo {year} {2017})}\BibitemShut {NoStop}%
\bibitem [{\citenamefont {Zhou}\ \emph {et~al.}(2018)\citenamefont {Zhou},
  \citenamefont {Liu}, \citenamefont {Ban}, \citenamefont {Li}, \citenamefont
  {Huang}, \citenamefont {Xia}, \citenamefont {Wang},\ and\ \citenamefont
  {Zhan}}]{Zhou18b}%
  \BibitemOpen
  \bibfield  {author} {\bibinfo {author} {\bibfnamefont {C.}~\bibnamefont
  {Zhou}}, \bibinfo {author} {\bibfnamefont {G.}~\bibnamefont {Liu}}, \bibinfo
  {author} {\bibfnamefont {G.}~\bibnamefont {Ban}}, \bibinfo {author}
  {\bibfnamefont {S.}~\bibnamefont {Li}}, \bibinfo {author} {\bibfnamefont
  {Q.}~\bibnamefont {Huang}}, \bibinfo {author} {\bibfnamefont
  {J.}~\bibnamefont {Xia}}, \bibinfo {author} {\bibfnamefont {Y.}~\bibnamefont
  {Wang}},\ and\ \bibinfo {author} {\bibfnamefont {M.}~\bibnamefont {Zhan}},\
  }\href {https://doi.org/10.1063/1.5020576} {\bibfield  {journal} {\bibinfo
  {journal} {Applied Physics Letters}\ }\textbf {\bibinfo {volume} {112}},\
  \bibinfo {pages} {101904} (\bibinfo {year} {2018})}\BibitemShut {NoStop}%
\bibitem [{\citenamefont {Vazquez}\ \emph {et~al.}(2012)\citenamefont
  {Vazquez}, \citenamefont {Skouta}, \citenamefont {Schneebeli}, \citenamefont
  {Kamenetska}, \citenamefont {Breslow}, \citenamefont {Venkataraman},\ and\
  \citenamefont {Hybertsen}}]{Vazquez12}%
  \BibitemOpen
  \bibfield  {author} {\bibinfo {author} {\bibfnamefont {H.}~\bibnamefont
  {Vazquez}}, \bibinfo {author} {\bibfnamefont {R.}~\bibnamefont {Skouta}},
  \bibinfo {author} {\bibfnamefont {S.}~\bibnamefont {Schneebeli}}, \bibinfo
  {author} {\bibfnamefont {M.}~\bibnamefont {Kamenetska}}, \bibinfo {author}
  {\bibfnamefont {R.}~\bibnamefont {Breslow}}, \bibinfo {author} {\bibfnamefont
  {L.}~\bibnamefont {Venkataraman}},\ and\ \bibinfo {author} {\bibfnamefont
  {M.~S.}\ \bibnamefont {Hybertsen}},\ }\href
  {https://doi.org/10.1038/nnano.2012.147} {\bibfield  {journal} {\bibinfo
  {journal} {Nature Nanotechnology}\ }\textbf {\bibinfo {volume} {7}},\
  \bibinfo {pages} {663} (\bibinfo {year} {2012})}\BibitemShut {NoStop}%
\bibitem [{\citenamefont {Ballmann}\ \emph {et~al.}(2012)\citenamefont
  {Ballmann}, \citenamefont {H\"{a}rtle}, \citenamefont {Coto}, \citenamefont
  {Elbing}, \citenamefont {Mayor}, \citenamefont {Bryce}, \citenamefont
  {Thoss},\ and\ \citenamefont {Weber}}]{Ballmann12}%
  \BibitemOpen
  \bibfield  {author} {\bibinfo {author} {\bibfnamefont {S.}~\bibnamefont
  {Ballmann}}, \bibinfo {author} {\bibfnamefont {R.}~\bibnamefont
  {H\"{a}rtle}}, \bibinfo {author} {\bibfnamefont {P.~B.}\ \bibnamefont
  {Coto}}, \bibinfo {author} {\bibfnamefont {M.}~\bibnamefont {Elbing}},
  \bibinfo {author} {\bibfnamefont {M.}~\bibnamefont {Mayor}}, \bibinfo
  {author} {\bibfnamefont {M.~R.}\ \bibnamefont {Bryce}}, \bibinfo {author}
  {\bibfnamefont {M.}~\bibnamefont {Thoss}},\ and\ \bibinfo {author}
  {\bibfnamefont {H.~B.}\ \bibnamefont {Weber}},\ }\href
  {https://doi.org/10.1103/PhysRevLett.109.056801} {\bibfield  {journal}
  {\bibinfo  {journal} {Phys. Rev. Lett.}\ }\textbf {\bibinfo {volume} {109}},\
  \bibinfo {pages} {056801} (\bibinfo {year} {2012})}\BibitemShut {NoStop}%
\bibitem [{\citenamefont {Gu\'{e}don}\ \emph {et~al.}(2012)\citenamefont
  {Gu\'{e}don}, \citenamefont {Valkenier}, \citenamefont {Markussen},
  \citenamefont {Thygesen}, \citenamefont {Hummelen},\ and\ \citenamefont
  {van~der Molen}}]{Guedon12}%
  \BibitemOpen
  \bibfield  {author} {\bibinfo {author} {\bibfnamefont {C.~M.}\ \bibnamefont
  {Gu\'{e}don}}, \bibinfo {author} {\bibfnamefont {H.}~\bibnamefont
  {Valkenier}}, \bibinfo {author} {\bibfnamefont {T.}~\bibnamefont
  {Markussen}}, \bibinfo {author} {\bibfnamefont {K.~S.}\ \bibnamefont
  {Thygesen}}, \bibinfo {author} {\bibfnamefont {J.~C.}\ \bibnamefont
  {Hummelen}},\ and\ \bibinfo {author} {\bibfnamefont {S.~J.}\ \bibnamefont
  {van~der Molen}},\ }\href {https://doi.org/10.1038/nnano.2012.37} {\bibfield
  {journal} {\bibinfo  {journal} {Nature Nanotechnology}\ }\textbf {\bibinfo
  {volume} {7}},\ \bibinfo {pages} {305} (\bibinfo {year} {2012})}\BibitemShut
  {NoStop}%
\bibitem [{\citenamefont {Prins}\ \emph {et~al.}(2011)\citenamefont {Prins},
  \citenamefont {Barreiro}, \citenamefont {Ruitenberg}, \citenamefont
  {Seldenthuis}, \citenamefont {Aliaga-Alcalde}, \citenamefont {Vandersypen},\
  and\ \citenamefont {van~der Zant}}]{Prins11}%
  \BibitemOpen
  \bibfield  {author} {\bibinfo {author} {\bibfnamefont {F.}~\bibnamefont
  {Prins}}, \bibinfo {author} {\bibfnamefont {A.}~\bibnamefont {Barreiro}},
  \bibinfo {author} {\bibfnamefont {J.~W.}\ \bibnamefont {Ruitenberg}},
  \bibinfo {author} {\bibfnamefont {J.~S.}\ \bibnamefont {Seldenthuis}},
  \bibinfo {author} {\bibfnamefont {N.}~\bibnamefont {Aliaga-Alcalde}},
  \bibinfo {author} {\bibfnamefont {L.~M.~K.}\ \bibnamefont {Vandersypen}},\
  and\ \bibinfo {author} {\bibfnamefont {H.~S.~J.}\ \bibnamefont {van~der
  Zant}},\ }\href {https://doi.org/10.1021/nl202065x} {\bibfield  {journal}
  {\bibinfo  {journal} {Nano Letters}\ }\textbf {\bibinfo {volume} {11}},\
  \bibinfo {pages} {4607} (\bibinfo {year} {2011})}\BibitemShut {NoStop}%
\bibitem [{\citenamefont {Arroyo}\ \emph {et~al.}(2013)\citenamefont {Arroyo},
  \citenamefont {Tarkuc}, \citenamefont {Frisenda}, \citenamefont
  {Seldenthuis}, \citenamefont {Woerde}, \citenamefont {Eelkema}, \citenamefont
  {Grozema},\ and\ \citenamefont {van~der Zant}}]{Arroyo13}%
  \BibitemOpen
  \bibfield  {author} {\bibinfo {author} {\bibfnamefont {C.~R.}\ \bibnamefont
  {Arroyo}}, \bibinfo {author} {\bibfnamefont {S.}~\bibnamefont {Tarkuc}},
  \bibinfo {author} {\bibfnamefont {R.}~\bibnamefont {Frisenda}}, \bibinfo
  {author} {\bibfnamefont {J.~S.}\ \bibnamefont {Seldenthuis}}, \bibinfo
  {author} {\bibfnamefont {C.~H.~M.}\ \bibnamefont {Woerde}}, \bibinfo {author}
  {\bibfnamefont {R.}~\bibnamefont {Eelkema}}, \bibinfo {author} {\bibfnamefont
  {F.~C.}\ \bibnamefont {Grozema}},\ and\ \bibinfo {author} {\bibfnamefont
  {H.~S.~J.}\ \bibnamefont {van~der Zant}},\ }\href
  {https://doi.org/10.1002/anie.201207667} {\bibfield  {journal} {\bibinfo
  {journal} {Angewandte Chemie International Edition}\ }\textbf {\bibinfo
  {volume} {52}},\ \bibinfo {pages} {3152} (\bibinfo {year}
  {2013})}\BibitemShut {NoStop}%
\bibitem [{\citenamefont {Aradhya}\ \emph {et~al.}(2012)\citenamefont
  {Aradhya}, \citenamefont {Meisner}, \citenamefont {Krikorian}, \citenamefont
  {Ahn}, \citenamefont {Parameswaran}, \citenamefont {Steigerwald},
  \citenamefont {Nuckolls},\ and\ \citenamefont {Venkataraman}}]{Aradhya12}%
  \BibitemOpen
  \bibfield  {author} {\bibinfo {author} {\bibfnamefont {S.~V.}\ \bibnamefont
  {Aradhya}}, \bibinfo {author} {\bibfnamefont {J.~S.}\ \bibnamefont
  {Meisner}}, \bibinfo {author} {\bibfnamefont {M.}~\bibnamefont {Krikorian}},
  \bibinfo {author} {\bibfnamefont {S.}~\bibnamefont {Ahn}}, \bibinfo {author}
  {\bibfnamefont {R.}~\bibnamefont {Parameswaran}}, \bibinfo {author}
  {\bibfnamefont {M.~L.}\ \bibnamefont {Steigerwald}}, \bibinfo {author}
  {\bibfnamefont {C.}~\bibnamefont {Nuckolls}},\ and\ \bibinfo {author}
  {\bibfnamefont {L.}~\bibnamefont {Venkataraman}},\ }\href
  {https://doi.org/10.1021/nl2045815} {\bibfield  {journal} {\bibinfo
  {journal} {Nano Letters}\ }\textbf {\bibinfo {volume} {12}},\ \bibinfo
  {pages} {1643} (\bibinfo {year} {2012})}\BibitemShut {NoStop}%
\bibitem [{\citenamefont {Aradhya}\ and\ \citenamefont
  {Venkataraman}(2013)}]{Aradhya13}%
  \BibitemOpen
  \bibfield  {author} {\bibinfo {author} {\bibfnamefont {S.~V.}\ \bibnamefont
  {Aradhya}}\ and\ \bibinfo {author} {\bibfnamefont {L.}~\bibnamefont
  {Venkataraman}},\ }\href {https://doi.org/10.1038/nnano.2013.91} {\bibfield
  {journal} {\bibinfo  {journal} {Nature Nanotechnology}\ }\textbf {\bibinfo
  {volume} {8}},\ \bibinfo {pages} {399} (\bibinfo {year} {2013})}\BibitemShut
  {NoStop}%
\bibitem [{\citenamefont {Saiz-Bret\'{\i}n}\ \emph {et~al.}(2015)\citenamefont
  {Saiz-Bret\'{\i}n}, \citenamefont {Malyshev}, \citenamefont {Orellana},\ and\
  \citenamefont {Dom\'{\i}nguez-Adame}}]{Saiz-Bretin15}%
  \BibitemOpen
  \bibfield  {author} {\bibinfo {author} {\bibfnamefont {M.}~\bibnamefont
  {Saiz-Bret\'{\i}n}}, \bibinfo {author} {\bibfnamefont {A.~V.}\ \bibnamefont
  {Malyshev}}, \bibinfo {author} {\bibfnamefont {P.~A.}\ \bibnamefont
  {Orellana}},\ and\ \bibinfo {author} {\bibfnamefont {F.}~\bibnamefont
  {Dom\'{\i}nguez-Adame}},\ }\href {https://doi.org/10.1103/PhysRevB.91.085431}
  {\bibfield  {journal} {\bibinfo  {journal} {Phys. Rev. B}\ }\textbf {\bibinfo
  {volume} {91}},\ \bibinfo {pages} {085431} (\bibinfo {year}
  {2015})}\BibitemShut {NoStop}%
\bibitem [{\citenamefont {Cutler}\ and\ \citenamefont {Mott}(1969)}]{Cutler69}%
  \BibitemOpen
  \bibfield  {author} {\bibinfo {author} {\bibfnamefont {M.}~\bibnamefont
  {Cutler}}\ and\ \bibinfo {author} {\bibfnamefont {N.~F.}\ \bibnamefont
  {Mott}},\ }\href {https://doi.org/10.1103/PhysRev.181.1336} {\bibfield
  {journal} {\bibinfo  {journal} {Phys. Rev.}\ }\textbf {\bibinfo {volume}
  {181}},\ \bibinfo {pages} {1336} (\bibinfo {year} {1969})}\BibitemShut
  {NoStop}%
\bibitem [{\citenamefont {Fano}(1961)}]{Fano61}%
  \BibitemOpen
  \bibfield  {author} {\bibinfo {author} {\bibfnamefont {U.}~\bibnamefont
  {Fano}},\ }\href {https://doi.org/10.1103/PhysRev.124.1866} {\bibfield
  {journal} {\bibinfo  {journal} {Phys. Rev.}\ }\textbf {\bibinfo {volume}
  {124}},\ \bibinfo {pages} {1866} (\bibinfo {year} {1961})}\BibitemShut
  {NoStop}%
\bibitem [{\citenamefont {Kim}\ and\ \citenamefont {Hershfield}(2003)}]{Kim03}%
  \BibitemOpen
  \bibfield  {author} {\bibinfo {author} {\bibfnamefont {T.-S.}\ \bibnamefont
  {Kim}}\ and\ \bibinfo {author} {\bibfnamefont {S.}~\bibnamefont
  {Hershfield}},\ }\href
  {https://doi.org/http://dx.doi.org/10.1103/PhysRevB.67.165313} {\bibfield
  {journal} {\bibinfo  {journal} {Phys. Rev. B}\ }\textbf {\bibinfo {volume}
  {67}},\ \bibinfo {pages} {165313} (\bibinfo {year} {2003})}\BibitemShut
  {NoStop}%
\bibitem [{\citenamefont {L\'{o}pez}\ and\ \citenamefont
  {S\'{a}nchez}(2013)}]{Lopez13b}%
  \BibitemOpen
  \bibfield  {author} {\bibinfo {author} {\bibfnamefont {R.}~\bibnamefont
  {L\'{o}pez}}\ and\ \bibinfo {author} {\bibfnamefont {D.}~\bibnamefont
  {S\'{a}nchez}},\ }\href {https://doi.org/10.1103/PhysRevB.88.045129}
  {\bibfield  {journal} {\bibinfo  {journal} {Phys. Rev. B}\ }\textbf {\bibinfo
  {volume} {88}},\ \bibinfo {pages} {045129} (\bibinfo {year}
  {2013})}\BibitemShut {NoStop}%
\bibitem [{\citenamefont {Sierra}\ and\ \citenamefont
  {S\'{a}nchez}(2014)}]{Sierra14}%
  \BibitemOpen
  \bibfield  {author} {\bibinfo {author} {\bibfnamefont {M.~A.}\ \bibnamefont
  {Sierra}}\ and\ \bibinfo {author} {\bibfnamefont {D.}~\bibnamefont
  {S\'{a}nchez}},\ }\href {https://doi.org/10.1103/PhysRevB.90.115313}
  {\bibfield  {journal} {\bibinfo  {journal} {Phys. Rev. B}\ }\textbf {\bibinfo
  {volume} {90}},\ \bibinfo {pages} {115313} (\bibinfo {year}
  {2014})}\BibitemShut {NoStop}%
\bibitem [{\citenamefont {Whitney}(2014)}]{Whitney14}%
  \BibitemOpen
  \bibfield  {author} {\bibinfo {author} {\bibfnamefont {R.~S.}\ \bibnamefont
  {Whitney}},\ }\href {https://doi.org/10.1103/PhysRevLett.112.130601}
  {\bibfield  {journal} {\bibinfo  {journal} {Phys. Rev. Lett.}\ }\textbf
  {\bibinfo {volume} {112}},\ \bibinfo {pages} {130601} (\bibinfo {year}
  {2014})}\BibitemShut {NoStop}%
\bibitem [{\citenamefont {Hofstetter}\ \emph {et~al.}(2001)\citenamefont
  {Hofstetter}, \citenamefont {K\"{o}nig},\ and\ \citenamefont
  {Schoeller}}]{Hofstetter01}%
  \BibitemOpen
  \bibfield  {author} {\bibinfo {author} {\bibfnamefont {W.}~\bibnamefont
  {Hofstetter}}, \bibinfo {author} {\bibfnamefont {J.}~\bibnamefont
  {K\"{o}nig}},\ and\ \bibinfo {author} {\bibfnamefont {H.}~\bibnamefont
  {Schoeller}},\ }\href {https://doi.org/10.1103/PhysRevLett.87.156803}
  {\bibfield  {journal} {\bibinfo  {journal} {Phys. Rev. Lett.}\ }\textbf
  {\bibinfo {volume} {87}},\ \bibinfo {pages} {156803} (\bibinfo {year}
  {2001})}\BibitemShut {NoStop}%
\bibitem [{\citenamefont {Sun}\ \emph {et~al.}(2005)\citenamefont {Sun},
  \citenamefont {Wang},\ and\ \citenamefont {Guo}}]{Sun05b}%
  \BibitemOpen
  \bibfield  {author} {\bibinfo {author} {\bibfnamefont {Q.-f.}\ \bibnamefont
  {Sun}}, \bibinfo {author} {\bibfnamefont {J.}~\bibnamefont {Wang}},\ and\
  \bibinfo {author} {\bibfnamefont {H.}~\bibnamefont {Guo}},\ }\href
  {https://doi.org/10.1103/PhysRevB.71.165310} {\bibfield  {journal} {\bibinfo
  {journal} {Phys. Rev. B}\ }\textbf {\bibinfo {volume} {71}},\ \bibinfo
  {pages} {165310} (\bibinfo {year} {2005})}\BibitemShut {NoStop}%
\bibitem [{\citenamefont {Crisan}\ \emph {et~al.}(2009)\citenamefont {Crisan},
  \citenamefont {S\'{a}nchez}, \citenamefont {L\'{o}pez}, \citenamefont
  {Serra},\ and\ \citenamefont {Grosu}}]{Crisan09}%
  \BibitemOpen
  \bibfield  {author} {\bibinfo {author} {\bibfnamefont {M.}~\bibnamefont
  {Crisan}}, \bibinfo {author} {\bibfnamefont {D.}~\bibnamefont {S\'{a}nchez}},
  \bibinfo {author} {\bibfnamefont {R.}~\bibnamefont {L\'{o}pez}}, \bibinfo
  {author} {\bibfnamefont {L.}~\bibnamefont {Serra}},\ and\ \bibinfo {author}
  {\bibfnamefont {I.}~\bibnamefont {Grosu}},\ }\href
  {https://doi.org/10.1103/PhysRevB.79.125319} {\bibfield  {journal} {\bibinfo
  {journal} {Phys. Rev. B}\ }\textbf {\bibinfo {volume} {79}},\ \bibinfo
  {pages} {125319} (\bibinfo {year} {2009})}\BibitemShut {NoStop}%
\bibitem [{\citenamefont {Taniguchi}\ and\ \citenamefont
  {Isozaki}(2012)}]{Taniguchi12}%
  \BibitemOpen
  \bibfield  {author} {\bibinfo {author} {\bibfnamefont {N.}~\bibnamefont
  {Taniguchi}}\ and\ \bibinfo {author} {\bibfnamefont {K.}~\bibnamefont
  {Isozaki}},\ }\href {https://doi.org/10.1143/JPSJ.81.124708} {\bibfield
  {journal} {\bibinfo  {journal} {J. Phys. Soc. Japan}\ }\textbf {\bibinfo
  {volume} {81}},\ \bibinfo {pages} {124708} (\bibinfo {year}
  {2012})}\BibitemShut {NoStop}%
\bibitem [{\citenamefont {Meir}\ and\ \citenamefont {Wingreen}(1992)}]{Meir92}%
  \BibitemOpen
  \bibfield  {author} {\bibinfo {author} {\bibfnamefont {Y.}~\bibnamefont
  {Meir}}\ and\ \bibinfo {author} {\bibfnamefont {N.~S.}\ \bibnamefont
  {Wingreen}},\ }\href {https://doi.org/10.1103/PhysRevLett.68.2512} {\bibfield
   {journal} {\bibinfo  {journal} {Phys. Rev. Lett.}\ }\textbf {\bibinfo
  {volume} {68}},\ \bibinfo {pages} {2512} (\bibinfo {year}
  {1992})}\BibitemShut {NoStop}%
\bibitem [{\citenamefont {Haug}\ and\ \citenamefont
  {Jauho}(2008)}]{HaugBook08}%
  \BibitemOpen
  \bibfield  {author} {\bibinfo {author} {\bibfnamefont {H.}~\bibnamefont
  {Haug}}\ and\ \bibinfo {author} {\bibfnamefont {A.-P.}\ \bibnamefont
  {Jauho}},\ }\href {https://doi.org/10.1007/978-3-540-73564-9} {\emph
  {\bibinfo {title} {Quantum Kinetics in Transport and Optics of
  Semiconductors}}},\ \bibinfo {edition} {2nd}\ ed.,\ \bibinfo {series}
  {Springer Series in Solid-State Sciences}, Vol.\ \bibinfo {volume} {123}\
  (\bibinfo  {publisher} {Springer},\ \bibinfo {address} {Heidelberg},\
  \bibinfo {year} {2008})\BibitemShut {NoStop}%
\bibitem [{\citenamefont {Sivan}\ and\ \citenamefont {Imry}(1986)}]{Sivan86}%
  \BibitemOpen
  \bibfield  {author} {\bibinfo {author} {\bibfnamefont {U.}~\bibnamefont
  {Sivan}}\ and\ \bibinfo {author} {\bibfnamefont {Y.}~\bibnamefont {Imry}},\
  }\href {https://doi.org/10.1103/PhysRevB.33.551} {\bibfield  {journal}
  {\bibinfo  {journal} {Phys. Rev. B}\ }\textbf {\bibinfo {volume} {33}},\
  \bibinfo {pages} {551} (\bibinfo {year} {1986})}\BibitemShut {NoStop}%
\bibitem [{\citenamefont {Butcher}(1990)}]{Butcher90}%
  \BibitemOpen
  \bibfield  {author} {\bibinfo {author} {\bibfnamefont {P.~N.}\ \bibnamefont
  {Butcher}},\ }\href {https://doi.org/10.1088/0953-8984/2/22/008} {\bibfield
  {journal} {\bibinfo  {journal} {Journal of Physics: Condensed Matter}\
  }\textbf {\bibinfo {volume} {2}},\ \bibinfo {pages} {4869} (\bibinfo {year}
  {1990})}\BibitemShut {NoStop}%
\bibitem [{\citenamefont {S\'{a}nchez}\ and\ \citenamefont
  {L\'{o}pez}(2016)}]{Sanchez16}%
  \BibitemOpen
  \bibfield  {author} {\bibinfo {author} {\bibfnamefont {D.}~\bibnamefont
  {S\'{a}nchez}}\ and\ \bibinfo {author} {\bibfnamefont {R.}~\bibnamefont
  {L\'{o}pez}},\ }\href {https://doi.org/10.1016/j.crhy.2016.08.005} {\bibfield
   {journal} {\bibinfo  {journal} {Comptes Rendus Physique}\ }\textbf {\bibinfo
  {volume} {17}},\ \bibinfo {pages} {1060} (\bibinfo {year}
  {2016})}\BibitemShut {NoStop}%
\bibitem [{\citenamefont {Benenti}\ \emph {et~al.}(2017)\citenamefont
  {Benenti}, \citenamefont {Casati}, \citenamefont {Saito},\ and\ \citenamefont
  {Whitney}}]{Benenti17}%
  \BibitemOpen
  \bibfield  {author} {\bibinfo {author} {\bibfnamefont {G.}~\bibnamefont
  {Benenti}}, \bibinfo {author} {\bibfnamefont {G.}~\bibnamefont {Casati}},
  \bibinfo {author} {\bibfnamefont {K.}~\bibnamefont {Saito}},\ and\ \bibinfo
  {author} {\bibfnamefont {R.~S.}\ \bibnamefont {Whitney}},\ }\href
  {https://doi.org/10.1016/j.physrep.2017.05.008} {\bibfield  {journal}
  {\bibinfo  {journal} {Physics Reports}\ }\textbf {\bibinfo {volume} {694}},\
  \bibinfo {pages} {1} (\bibinfo {year} {2017})},\ \bibinfo {note}
  {arXiv:1608.05595}\BibitemShut {NoStop}%
\bibitem [{\citenamefont {Hershfield}\ \emph {et~al.}(1992)\citenamefont
  {Hershfield}, \citenamefont {Davies},\ and\ \citenamefont
  {Wilkins}}]{Hershfield92}%
  \BibitemOpen
  \bibfield  {author} {\bibinfo {author} {\bibfnamefont {S.}~\bibnamefont
  {Hershfield}}, \bibinfo {author} {\bibfnamefont {J.~H.}\ \bibnamefont
  {Davies}},\ and\ \bibinfo {author} {\bibfnamefont {J.~W.}\ \bibnamefont
  {Wilkins}},\ }\href {https://doi.org/10.1103/PhysRevB.46.7046} {\bibfield
  {journal} {\bibinfo  {journal} {Phys. Rev. B}\ }\textbf {\bibinfo {volume}
  {46}},\ \bibinfo {pages} {7046} (\bibinfo {year} {1992})}\BibitemShut
  {NoStop}%
\bibitem [{\citenamefont {Taniguchi}(2014)}]{Taniguchi14}%
  \BibitemOpen
  \bibfield  {author} {\bibinfo {author} {\bibfnamefont {N.}~\bibnamefont
  {Taniguchi}},\ }\href {https://doi.org/10.1103/PhysRevB.90.115421} {\bibfield
   {journal} {\bibinfo  {journal} {Phys. Rev. B}\ }\textbf {\bibinfo {volume}
  {90}},\ \bibinfo {pages} {115421} (\bibinfo {year} {2014})}\BibitemShut
  {NoStop}%
\bibitem [{\citenamefont {Taniguchi}(2017)}]{Taniguchi17}%
  \BibitemOpen
  \bibfield  {author} {\bibinfo {author} {\bibfnamefont {N.}~\bibnamefont
  {Taniguchi}},\ }\href {https://doi.org/10.1103/PhysRevA.96.042105} {\bibfield
   {journal} {\bibinfo  {journal} {Phys. Rev. A}\ }\textbf {\bibinfo {volume}
  {96}},\ \bibinfo {pages} {042105} (\bibinfo {year} {2017})}\BibitemShut
  {NoStop}%
\bibitem [{\citenamefont {Hubbard}(1963)}]{Hubbard63}%
  \BibitemOpen
  \bibfield  {author} {\bibinfo {author} {\bibfnamefont {J.}~\bibnamefont
  {Hubbard}},\ }\href {https://doi.org/10.1098/rspa.1963.0204} {\bibfield
  {journal} {\bibinfo  {journal} {Proceedings of the Royal Society of London.
  Series A. Mathematical and Physical Sciences}\ }\textbf {\bibinfo {volume}
  {276}},\ \bibinfo {pages} {238} (\bibinfo {year} {1963})}\BibitemShut
  {NoStop}%
\bibitem [{\citenamefont {Hewson}(1966)}]{Hewson66}%
  \BibitemOpen
  \bibfield  {author} {\bibinfo {author} {\bibfnamefont {A.~C.}\ \bibnamefont
  {Hewson}},\ }\href {https://doi.org/10.1103/PhysRev.144.420} {\bibfield
  {journal} {\bibinfo  {journal} {Phys. Rev.}\ }\textbf {\bibinfo {volume}
  {144}},\ \bibinfo {pages} {420} (\bibinfo {year} {1966})}\BibitemShut
  {NoStop}%
\bibitem [{\citenamefont {Meir}\ \emph {et~al.}(1991)\citenamefont {Meir},
  \citenamefont {Wingreen},\ and\ \citenamefont {Lee}}]{Meir91}%
  \BibitemOpen
  \bibfield  {author} {\bibinfo {author} {\bibfnamefont {Y.}~\bibnamefont
  {Meir}}, \bibinfo {author} {\bibfnamefont {N.~S.}\ \bibnamefont {Wingreen}},\
  and\ \bibinfo {author} {\bibfnamefont {P.~A.}\ \bibnamefont {Lee}},\ }\href
  {https://doi.org/10.1103/PhysRevLett.66.3048} {\bibfield  {journal} {\bibinfo
   {journal} {Phys. Rev. Lett.}\ }\textbf {\bibinfo {volume} {66}},\ \bibinfo
  {pages} {3048} (\bibinfo {year} {1991})}\BibitemShut {NoStop}%
\bibitem [{\citenamefont {Nakpathomkun}\ \emph {et~al.}(2010)\citenamefont
  {Nakpathomkun}, \citenamefont {Xu},\ and\ \citenamefont
  {Linke}}]{Nakpathomkun10}%
  \BibitemOpen
  \bibfield  {author} {\bibinfo {author} {\bibfnamefont {N.}~\bibnamefont
  {Nakpathomkun}}, \bibinfo {author} {\bibfnamefont {H.~Q.}\ \bibnamefont
  {Xu}},\ and\ \bibinfo {author} {\bibfnamefont {H.}~\bibnamefont {Linke}},\
  }\href {https://doi.org/10.1103/PhysRevB.82.235428} {\bibfield  {journal}
  {\bibinfo  {journal} {Phys. Rev. B}\ }\textbf {\bibinfo {volume} {82}},\
  \bibinfo {pages} {235428} (\bibinfo {year} {2010})}\BibitemShut {NoStop}%
\bibitem [{\citenamefont {Taniguchi}(2018)}]{Taniguchi18}%
  \BibitemOpen
  \bibfield  {author} {\bibinfo {author} {\bibfnamefont {N.}~\bibnamefont
  {Taniguchi}},\ }\href {https://doi.org/10.1103/PhysRevB.97.155404} {\bibfield
   {journal} {\bibinfo  {journal} {Phys. Rev. B}\ }\textbf {\bibinfo {volume}
  {97}},\ \bibinfo {pages} {155404} (\bibinfo {year} {2018})}\BibitemShut
  {NoStop}%
\end{thebibliography}%

\end{document}